\RequirePackage{lineno}
\documentclass[aps,prd,showpacs,byrevtex,twocolumn]{revtex4}
\usepackage{dcolumn}
\usepackage{bm}
\usepackage{rotating}
\usepackage{times}
\usepackage{amsmath}
\usepackage{graphicx}
\usepackage{overpic}
\usepackage{color}
\usepackage{multirow}
\usepackage{setspace}
\usepackage{lineno}
\usepackage{verbatim}
\usepackage{float}
\usepackage[T1]{fontenc}

\emergencystretch=\maxdimen
\graphicspath{{figures/}}

\newcommand{\BR}{{\cal B}}

\newcommand{\pspp}{\psi(3770)}

\newcommand{\EE}{e^+e^-}

\newcommand{\pp}{\pi^+\pi^-}

\newcommand{\ddb}{D\bar{D}}
\newcommand{\ddn}{D^0\bar{D}^0}
\newcommand{\ddc}{D^+D^-}

\newcommand{\ppjpsi}{\pi^+\pi^- J/\psi}

\begin{document}
\graphicspath{{figure/}}
\DeclareGraphicsExtensions{.eps,.png,.ps}

\title{\boldmath Observation of $e^{+}e^{-}\rightarrow \pi^{+}\pi^{-}\psi(3770)$ and $D_{1}(2420)^{0}\bar{D}^{0} + c.c.$}

\author{
M.~Ablikim$^{1}$, M.~N.~Achasov$^{10,d}$, P.~Adlarson$^{59}$, S. ~Ahmed$^{15}$, M.~Albrecht$^{4}$, M.~Alekseev$^{58A,58C}$, A.~Amoroso$^{58A,58C}$, F.~F.~An$^{1}$, Q.~An$^{55,43}$, Y.~Bai$^{42}$
% ...
\\
\vspace{0.2cm}
(BESIII Collaboration)\\
\vspace{0.2cm} {\it
$^{1}$ Institute of High Energy Physics, Beijing 100049, People's Republic of China\\
% ...
$^{4}$ Bochum Ruhr-University, D-44780 Bochum, Germany\\
% ...
$^{10}$ G.I. Budker Institute of Nuclear Physics SB RAS (BINP), Novosibirsk 630090, Russia\\
% ...
$^{15}$ Helmholtz Institute Mainz, Johann-Joachim-Becher-Weg 45, D-55099 Mainz, Germany\\
% ...
$^{42}$ Southeast University, Nanjing 211100, People's Republic of China\\
$^{43}$ State Key Laboratory of Particle Detection and Electronics, Beijing 100049, Hefei 230026, People's Republic of China\\
% ...
$^{55}$ University of Science and Technology of China, Hefei 230026, People's Republic of China\\
% ...
$^{58}$ (A)University of Turin, I-10125, Turin, Italy; (B)University of Eastern Piedmont, I-15121, Alessandria, Italy; (C)INFN, I-10125, Turin, Italy\\
$^{59}$ Uppsala University, Box 516, SE-75120 Uppsala, Sweden\\
% ...
$^{d}$ Also at the Novosibirsk State University, Novosibirsk, 630090, Russia\\
% ...
}
}

\begin{abstract}
Several intermediate states of the reaction channels $e^{+}e^{-} \rightarrow \pi^{+} \pi^{-} D^{0} \bar{D}^{0}$ and $e^{+}e^{-} \rightarrow \pi^{+} \pi^{-} D^{+}D^{-}$ are studied using the data samples collected with the BESIII detector at center-of-mass energies above 4.08 GeV.  For the first time in this final state, a $\psi(3770)$ signal is seen in the $D\bar{D}$ invariant mass spectrum, with a statistical significance of $5.2\sigma$ at $\sqrt{s} = 4.42$~GeV.  There is also evidence for this resonance at $\sqrt{s}$ = 4.26 and 4.36 GeV with statistical significance of 3.2$\sigma$ and 3.3$\sigma$, respectively. In addition, the Born cross section of $e^{+}e^{-}\to \pi^{+}\pi^{-}\psi(3770)$ is measured. The proposed heavy-quark-spin-symmetry partner of the $X(3872)$, the state $X_{2}(4013)$, is also searched for in the $D\bar{D}$ invariant mass spectra. No obvious signal is found. The upper limit of the Born cross section of the process $e^{+}e^{-}\to \rho^{0}X_{2}(4013)$ combined with the branching fraction is measured. Also, the processes $\EE\to D_{1}(2420)\bar{D} + c.c.$ are investigated. The neutral mode with $D_{1}(2420)^{0}\to D^{0}\pi^{+}\pi^{-}$  is reported with statistical significance of 7.4$\sigma$ at $\sqrt{s} = 4.42$~GeV  for the first time, and  evidence with statistical significance of 3.2$\sigma$ and 3.3$\sigma$ at $\sqrt{s}$ = 4.36 and 4.60 GeV is seen, respectively.  No evident signal for the process $\EE\to D_{1}(2420)^{0}\bar{D}^{0} + c.c., D_{1}(2420)^{0}\to D^{*+}\pi^{-}$ is reported. Evidence for $\EE\to D_{1}(2420)^{+}D^{-} +~c.c., D_{1}(2420)^{+}\to D^{+}\pi^{+}\pi^{-}$ is reported with statistical significance of 3.1$\sigma$ and 3.0$\sigma$ at $\sqrt{s}$ = 4.36 and 4.42 GeV, respectively.
\end{abstract}

\pacs{14.40.Rt, 13.20.Gd, 13.66.Bc, 13.40.Hq, 14.40.Pq}

\maketitle

\section{Introduction}

Heavy quarkonia have been studied for more than forty years
for testing and developing quantum chromodynamics (QCD). On the
one hand, some effective theories have been developed to describe
 quarkonium spectroscopy and transition dynamics~\cite{cet,HQET,NRQCD}; on the other
hand, many $XYZ$ particles were
discovered~\cite{XYZreview1,XYZreview2,XYZreview3}, and some of them are beyond the scope of potential models. The rich information gained from the $XYZ$ particles
may have opened a door through which quark confinement can be
understood~\cite{Heavy quarkonium,hiddencharm_chen}. To understand these $XYZ$ particles, it is of great importance to understand also the
properties of the conventional quarkonia.

In recent years, several new vector charmonium-like states, the
$Y(4260)$, $Y(4360)$, and $Y(4660)$, have been discovered via their
decays into hidden-charm final states such as $\pi^{+}\pi^{-}J/\psi$ or
$\pi^{+}\pi^{-}\psi(3686)$~\cite{BabarY4260,
BabarY4360, BELLEY4260, BELLEY4360, BESpipiJpsi}. The charged $Z_{c}(3900)$s and
similar structures have been observed in the  $\pi^{\pm}J/\psi$  and
$\pi^{\pm}\psi(3686)$ invariant mass spectra in the processes
$e^{+}e^{-}\to \pi^{+}\pi^{-}J/\psi$ and
$\pp\psi(3686)$, respectively at BESIII, Belle, and with CLEO-c data~\cite{zc3900, BELLEY4260, BELLEY4360, BESpipipsip, Xiao:2013iha}.
 A natural extension would be a search
for the process $\EE\to \pi^{+}\pi^{-}\psi(3770)$ and
for the corresponding charged resonance that decays to
$\pi^{\pm}\psi(3770)$.

The $\psi(3770)$ is generally assumed to be the $1^{3}D_{1}$
charmonium state with some admixture of the $2^{3}S_{1}$ state. One of the
$D$-wave spin-triplet charmonium states, the $\psi(1^{3}D_{2})$ or $X(3823)$, has recently been observed in $\EE\to
\pp\psi(1^{3}D_{2})$ at BESIII~\cite{X3823}. Therefore, the final states $\pp \pspp$ and $\pp
\psi(1^3D_3)$ should be produced at BESIII as well,
although so far there is no calculation on how large the production
rates could be. The $\psi(3770)$ decays dominantly to $D\bar{D}$, which
is also expected to be an important decay mode of the
$\psi(1^{3}D_{3})$. The predicted mass of the $\psi(1^{3}D_{3})$
is at 3849 MeV/$c^{2}$~\cite{Highercharmonia}, however, there is no prediction for the width.
Therefore, by studying the process $\EE\to
\pi\pi D\bar{D}$, one can also search for the $\psi(1^{3}D_{3})$.

The $X(3872)$ state was first observed by Belle~\cite{X3872}, and
confirmed subsequently by several other
experiments~\cite{CDFX3872,D0X3872,BABARX3872}. Even though it
clearly contains a $c\bar{c}$ pair, the $X(3872)$ does not fit
in the conventional charmonium spectrum. It could be interpreted as a
$D\bar{D}^{*}$ molecule with $J^{PC} = 1^{++}$~\cite{3872DDstar,3872JP}. Throughout this paper, the charged conjugate mode is implied unless it is stated otherwise. Within this
picture the existence of its heavy quark-spin-symmetry partner
$X_{2}(4013)$ ($J^{PC} = 2^{++}$), an $S$-wave $D^{*}\bar{D}^{*}$ bound
state, is predicted~\cite{PRD056004,PRD076006}. Its mass and width are predicted as about
4013~MeV/$c^2$ and $\sim$2-8~MeV, respectively. The $X_{2}(4013)$ is expected to decay dominantly
to $D\bar{D}$ or $D\bar{D}^{*}$ in $D$-wave. So it may also be produced in $e^{+}e^{-}\to \pi^{+}\pi^{-} D\bar{D}$. The
possible discovery of the $2^{++}$ charmonium-like state will provide a strong support for the interpretation that the $X(3872)$
is dominantly a $D\bar{D}^{*}$ hadronic
molecule~\cite{DecaywidthX4013}.

Amongst various models to interpret the
$Y(4260)$~\cite{BabarY4260,BELLEY4260}, the authors of
Ref.~\cite{Y4260asDD1} argue that the $Y(4260)$ as a relative
$S$-wave $D_{1}(2420)\bar{D}$ system is able to accommodate nearly all
the present observations of the $Y(4260)$. Especially its absence
in various open charm decay channels and the observation of the
$Z_{c}(3900)$ in $Y(4260)\to \ppjpsi$ support this interpretation. In this model, the coupling strength of
$D_{1}(2420)\bar{D}$ to $Y(4260)$ is a key piece of
information. Because of $D_{1}(2420)$ decays to $D\pi\pi$ or $D^{*}\pi$,
this can also be studied  via the $\pi\pi D\bar{D}$ final state.

In this paper, we report the observation of $e^{+}e^{-}\rightarrow \pi^{+}\pi^{-}\psi(3770)$ and $D_{1}(2420)^{0}\bar{D}^{0}$ based on data samples  collected with the BESIII detector from 2012 to 2014.  The Born cross sections of  $e^{+}e^{-}\rightarrow \pi^{+}\pi^{-}\psi(3770)$ at center-of-mass (c.m.) energies $\sqrt{s}$ above 4.08 GeV, $e^{+}e^{-}\rightarrow \rho^{0} X_{2}(4013)$ at $\sqrt{s}=4.36$, 4.42, and 4.60 GeV, and  $e^{+}e^{-}\rightarrow D_{1}\bar{D}$ above 4.30 GeV are measured. The energies of the data samples used in this analysis are  4.0854, 4.1886, 4.2077,      4.2171, 4.2263, 4.2417, 4.2580, 4.3079,  4.3583,  4.3874,  4.4156, 4.4671, 4.5271,  4.5745, 4.5995~GeV, respectively. To make the text easier to read, we use 4.09, 4.19, 4.21, 4.22, 4.23, 4.245, 4.26, 4.31, 4.36, 4.39, 4.42,  4.47,  4.53, 4.575 and  4.60~GeV in the following instead.

\section{THE EXPERIMENT AND DATA SETS}

The BESIII detector is a magnetic
spectrometer~\cite{Ablikim:2009aa} located at the Beijing Electron
Positron Collider (BEPCII)~\cite{Yu:IPAC2016-TUYA01}. The
cylindrical core of the BESIII detector consists of a helium-based
 multilayer drift chamber (MDC), a plastic scintillator time-of-flight
system (TOF), and a CsI(Tl) electromagnetic calorimeter (EMC),
which are all enclosed in a superconducting solenoidal magnet
providing a 1.0~T  magnetic field. The solenoid is supported by an
octagonal flux-return yoke with resistive plate counter muon
identifier modules interleaved with steel. The acceptance of
charged particles and photons is 93\% over $4\pi$ solid angle. The
charged-particle momentum resolution at $1~{\rm GeV}/c$ is
$0.5\%$, and the specific energy loss ($dE/dx$) resolution is $6\%$ for the electrons
from Bhabha scattering. The EMC measures photon energies with a
resolution of $2.5\%$ ($5\%$) at $1$~GeV in the barrel (end cap)
region. The time resolution of the TOF barrel part is 68~ps, while
that of the end cap part is 110~ps.

For this analysis, the data sets above 4.08 GeV recorded with the BESIII detector are used.
The c.m.\ energy and
the corresponding integrated luminosity of each data sample are listed in
Table~\ref{RCFDDbar}. The c.m.\ energy is measured using di-muon
events with a precision of 0.8 MeV~\cite{Ecms}. The integrated luminosity is determined by analyzing
large-angle
Bhabha scattering events. The uncertainty of the integrated luminosity is 1.0\%~\cite{luminosity}.

Simulated data samples produced with the {\sc geant4}-based~\cite{geant4} Monte Carlo (MC) package, which
includes the geometric description of the BESIII detector and the
detector response, are used to determine the detection efficiency
and to estimate the background contributions. The simulation includes the beam
energy spread and allows for the production of  initial state radiation (ISR) photons in the $e^+e^-$
annihilation process. Both effects are modeled within the generator package {\sc
kkmc}~\cite{ref:kkmc}.

\begin{figure*}[htbp]
  \centering
   \begin{overpic}[width=0.42\textwidth]{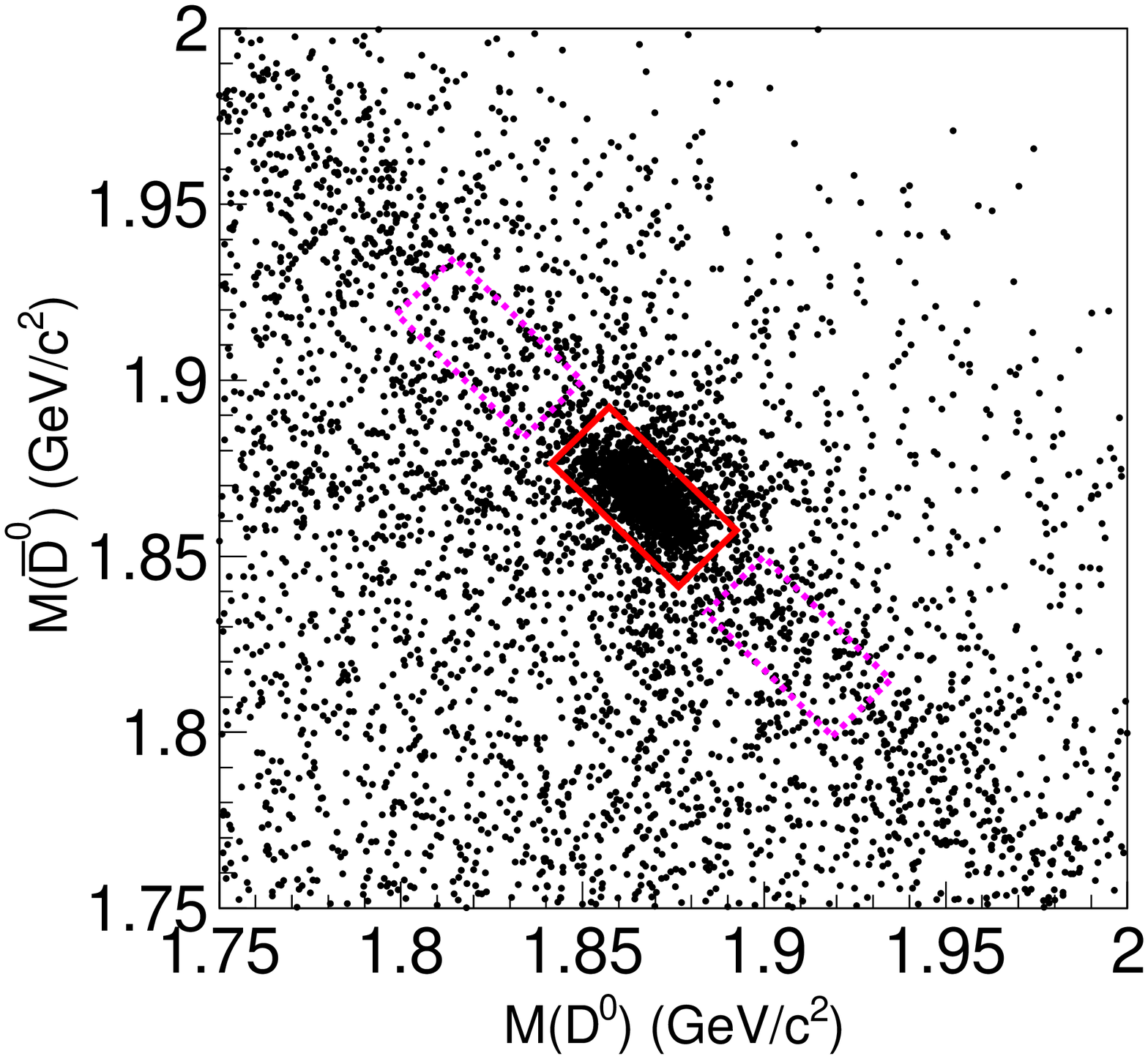}
   \put(80,80){(a)}
   \end{overpic}
   \begin{overpic}[width=0.42\textwidth]{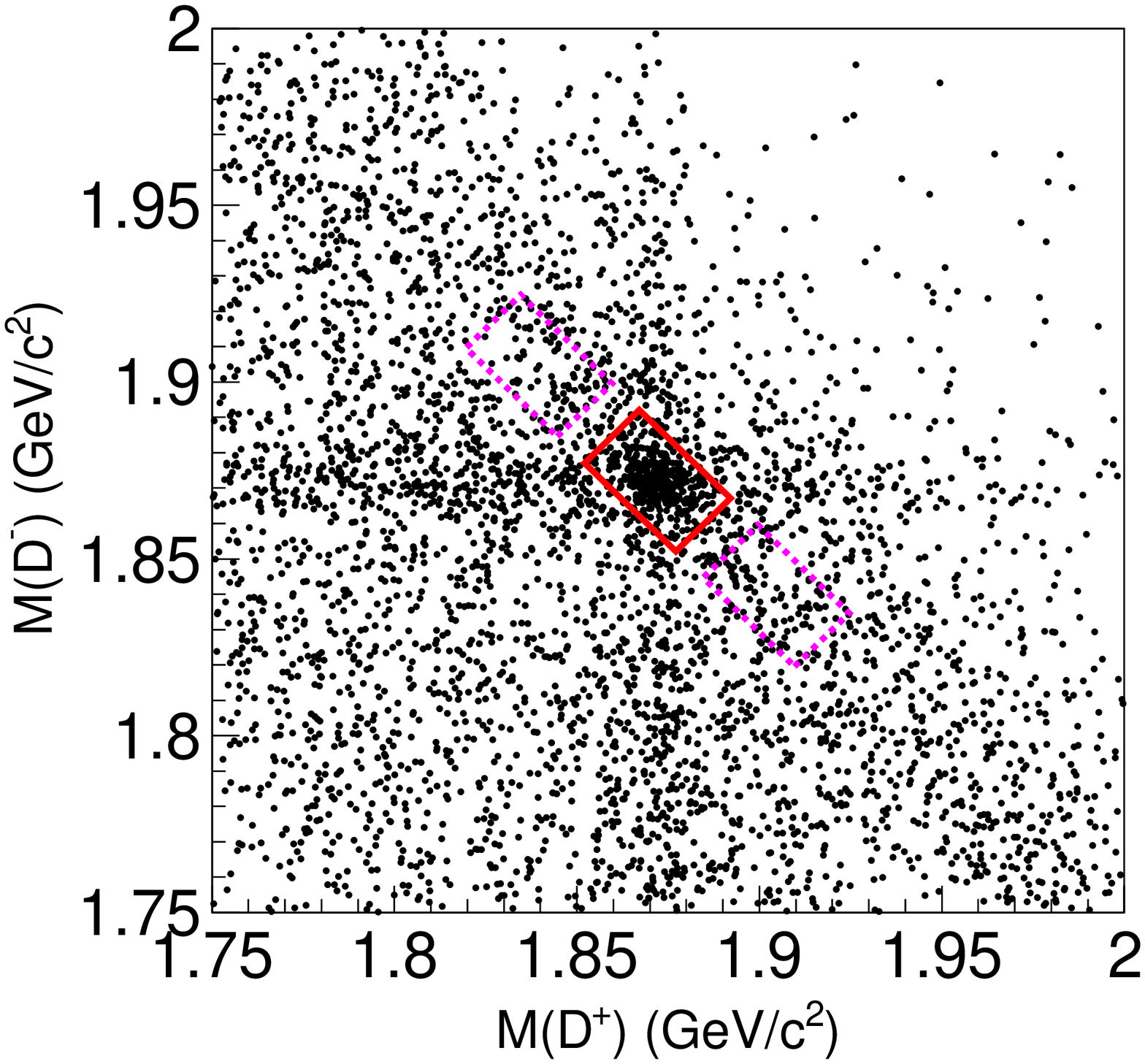}
   \put(80,80){(b)}
   \end{overpic}
\caption{Scatter plots of the invariant masses of $D^{0}$  versus  $\bar{D}^{0}$
meson candidates (a) and the invariant masses of $D^{+}$ versus  $D^{-}$ meson candidates (b) at
$\sqrt{s} = 4.42$~GeV. The rectangles show the signal regions (red solid lines)
and sideband regions (pink dotted lines).}
  \label{D0vsDbar}
\end{figure*}

For the optimization of the selection criteria, the following MC samples with 200,000 events for each process are produced at each c.m.\ energy:
$e^{+}e^{-}\to \pi^{+}\pi^{-}\psi(3770)$, with $\psi(3770) \to
D\bar{D}$; $e^{+}e^{-}\to \rho^{0}X_{2}(4013)$ with
$\rho^{0} \to \pi^{+}\pi^{-}$ and $X_{2}(4013)\to D\bar{D}$; $e^{+}e^{-}\to D_{1}(2420)^{0}\bar{D}^{0}$ with $D_{1}(2420)^{0}\rightarrow  D_{0}^{*}(2308)^{+}\pi^{-}  \to D^{0} \pi^{+}\pi^{-}$ ($D_{1} \to D\pi\pi$ decays through the quasi-two-body intermediate state $D_{0}^{*}(2308)$~\cite{D1_D2308}); $e^{+}e^{-}\to D_{1}(2420)^{0}\bar{D}^{0}$ with $D_{1}(2420)^{0}\to
D^{*+}\pi^{-}$; $e^{+}e^{-}\to D_{1}(2420)^{+}D^{-}$ with $D_{1}(2420)^{+}\rightarrow  D_{0}^{*}(2308)^{0} \pi^{+} \to  D^{+} \pi^{-}\pi^{+}$. The width of $X_{2}(4013)$ and $D_{0}^{*}(2308)$ are set to 8 and  276 MeV, respectively.

In order to estimate the potential background contributions, the BESIII official inclusive MC samples at  $\sqrt{s}=$ 4.23, 4.26, 4.36, 4.42, and 4.60~GeV are used. The inclusive MC samples consist of the production of open charm
processes, the ISR production of vector charmonium(-like) states, and the continuum processes incorporated in {\sc kkmc}~\cite{ref:kkmc}.
The known decay modes are modeled with {\sc
evtgen}~\cite{ref:evtgen} using branching fractions taken from the
Particle Data Group (PDG)~\cite{pdg}. The remaining unknown decays
of the charmonium states are generated with {\sc
lundcharm}~\cite{ref:lundcharm}. The generation of final state radiation (FSR) photons which are produced by
charged final state particles is incorporated by the usage of the {\sc
photos} package~\cite{photos}. The size of the MC samples is equivalent to the luminosity in data.

In addition, exclusive MC samples with 200,000 events each for the
processes $e^{+}e^{-}\to\pi^{+}\pi^{-}D^{0}\bar{D}^{0}$, $e^{+}e^{-}\to\pi^{+}\pi^{-}D^{+}D^{-}$, $e^{+}e^{-}\to D^{*+}D^{*-}$ with $D^{*+}\to D^{0}\pi^{+}$ and $D^{*-}\to \bar{D}^{0}\pi^{-}$, $e^{+}e^{-}\to D^{*+}\bar{D}^{0}\pi^{-}$ with $D^{*+}\to D^{0}\pi^{+}$, $e^{+}e^{-}\to D_{1}(2430)^{0}\bar{D}^{0}$ with $D_{1}(2430)^{0}\to D^{*+}\pi^{-}$, $e^{+}e^{-} \rightarrow D^{*}_{0}(2400)^{0}\bar{D}^{0}$ with $D^{*}_{0}(2400)^{0}\rightarrow D^{+}\pi^{-}$,   $e^{+}e^{-} \rightarrow D^{*}_{0}(2400)^{+}D^{-}$ with $D^{*}_{0}(2400)^{+}\rightarrow D^{0}\pi^{+}$  and $e^{+}e^{-}\to D^{*}_{2}(2460)^{0}\bar{D}^{0}$ with $D^{*}_{2}(2460)^{0}\to D^{*+}\pi^{-}$ are produced at each c.m.\ energy to study possible background contributions.

\section{EVENT SELECTION AND BACKGROUND ANALYSIS}

In this analysis the $D\bar{D}$ (denoting $\ddn$ and $\ddc$ in the following) pairs
are selected with both $D$ mesons fully reconstructed in a number of hadronic decay channels
(also called ``double $D$ tag'' in the following).
$D^{0}$ mesons are reconstructed in four decay modes ($K^{-}\pi^{+}$,
$K^{-}\pi^{+}\pi^{0}$, $K^{-}\pi^{+}\pi^{+}\pi^{-}$, and
$K^{-}\pi^{+}\pi^{+}\pi^{-}\pi^{0}$) and $D^{+}$ mesons in five decay
modes ($K^{-}\pi^{+}\pi^{+}$, $K^{-}\pi^{+}\pi^{+}\pi^{0}$,
$K^{0}_{S}\pi^{+}$, $K^{0}_{S}\pi^{+}\pi^{0}$, and
$K^{0}_{S}\pi^{+}\pi^{-}\pi^{+}$). The $\bar{D}^{0}$ and $D^{-}$ mesons are
reconstructed in the charge conjugate final states of the $D^0$ and
$D^+$ mesons, respectively. One $\pp$ pair is selected in addition to
the tracks from $D\bar{D}$ decays.

Charged tracks are reconstructed from MDC hits within a
polar-angle ($\theta$) acceptance range of $|\cos\theta| <0.93$ and required to pass within 10 cm of the interaction
point in the beam direction and within 1 cm in the plane perpendicular
to the beam. The TOF and $dE/dx$ information is combined for each
charged track to calculate the particle identification (PID)
probability $P_{i}$ ($i=\pi$, $K$) of each particle-type
hypothesis. $P_K>P_\pi$ is required for a kaon candidate and
$P_\pi>P_K$ is required for a pion candidate. Tracks used in
reconstructing $K^{0}_{S}$  are exempted from these
requirements.

Electromagnetic showers are reconstructed by clustering the energy deposits of the EMC crystals. Efficiency and energy resolution are improved by
including energy deposits in nearby TOF counters. A photon
candidate is defined as a shower with an energy deposit of at
least 25~MeV in the ``barrel" region ($|\cos\theta|<0.8$), or
at least 50~MeV in the ``end-cap" region
($0.86<|\cos\theta|<0.92$). Showers in the transition region between the barrel
and the end-cap are not well measured and are
rejected. An additional requirement on the EMC hit timing ($0\leq
T\leq700$~ns  relative to the event start time) suppresses electronic noise and energy
deposits unrelated to the event. To eliminate showers from bremsstrahlung photons which originated from charged particles, the angle between the shower and nearest
charged track is required to be greater than 20 degrees.

$\pi^{0}$ candidates are reconstructed from pairs of photon candidates with
an invariant mass in the range
$0.115<M_{\gamma\gamma}<0.150$~GeV/$c^{2}$. A one-constraint (1C)
kinematic fit with the mass of the $\pi^{0}$ constrained to the world average
value~\cite{pdg} is performed to improve the energy
resolution.

$K^{0}_{S}$ candidates are reconstructed from two oppositely
charged tracks which satisfy $|\cos\theta|<0.93$ for the polar
angle and the distance to the average beam position in beam
direction within 20~cm. For each pair of tracks, assuming they are $\pi^{+}$ and $\pi^{-}$ , a
vertex fit is performed and the resulting track parameters are
used to obtain the $\pi\pi$ invariant  mass which must be in the range
$0.487< M_{\pi\pi} < 0.511$~GeV/$c^{2}$. The $\chi^{2}$ from
the vertex fit is required to be smaller than 100.

The selected  $K^{\pm}$, $\pi^{\pm}$, $K^{0}_{S}$, and $\pi^{0}$
candidates are used to reconstruct $D$ meson candidates which are composed to $D^{0}\bar{D}^{0}$ and $D^{+}D^{-}$  meson pairs. If more
than one $D\bar{D}$ pair per event is found with both $D$ mesons decaying in the same way, the
pair with the average mass $\hat{M} = [M(D) + M(\bar{D})]/2$ closest
to the nominal mass of the $D$ meson~\cite{pdg} is chosen. In each event,
one negative and one positive charged $\pi$ are
required in addition. To reduce the background contribution and improve the mass resolution, a
four-constraint (4C) kinematic fit is performed.  The
total four-momentum of all selected charged tracks and good
photons from $\pi^{0}$ are constrained to that of the initial $e^{+}e^{-}$ system.
If the final state contains a $\pi^0$ or $K^{0}_S$ meson, its mass is constrained in the kinematic fit as well.
If there are multiple candidates in an
event, the one with the smallest $\chi^{2}$ of the kinematic fit is chosen.
To find the optimal $\chi^{2}$ criteria, the figure of merit $ {\rm FOM} = \frac{n_{s}}{\sqrt{n_{s}+n_{b}}}$ is maximized. Here  $n_{s}$ is the number of signal events from signal MC simulation and  $n_{b}$ is the number of backgrounds events from inclusive MC samples.
The $\chi^{2}$ is required to be less than 56 for the $\pi^{+}\pi^{-}D^{0}\bar{D}^{0}$ final state with selection efficiency of 90.1\% and background rejection rate of 45.5\%, and less than 40 for the $\pi^{+}\pi^{-}D^{+}D^{-}$ final state with selection efficiency of 90.3\% and background rejection rate of 29.5\%.

In Fig.~\ref{D0vsDbar} the invariant mass of $D$ meson candidates is plotted versus that
of  the $\bar{D}$ meson candidates  at $\sqrt{s}=4.42$~GeV after the selection described above.
The signal region indicated by the red line in Fig.~\ref{D0vsDbar} is defined
as $-6 < \Delta\hat{M} <10$~MeV/$c^{2}$ and $|\Delta{M}| <
35$~MeV/$c^2$ for $D^{0}\bar{D}^{0}$ pairs, and $-5 < \Delta\hat{M} <
10$~MeV/$c^{2}$ and $|\Delta{M}| < 25$~MeV/$c^2$ for $D^{+}D^{-}$ pairs,
where the $\Delta\hat{M} = \hat{M} - m_D$ and $\Delta{M} =
M(D)-M(\bar{D})$ with $m_D$ being the nominal $D$ meson
mass~\cite{pdg}.  The sideband regions (indicated by the pink rectangles in Fig.~\ref{D0vsDbar}) are defined
as $-6 < \Delta\hat{M} <10$~MeV/$c^{2}$ and $50 < |\Delta{M}| <
120$~MeV/$c^2$ for $D^{0}\bar{D}^{0}$ pairs, and $-5 < \Delta\hat{M} <
10$~MeV/$c^{2}$ and $40 <|\Delta{M}| < 90$~MeV/$c^2$ for $D^{+}D^{-}$ pairs.

\begin{figure*}[htbp]
  \centering
   \begin{overpic}[width=0.42\textwidth]{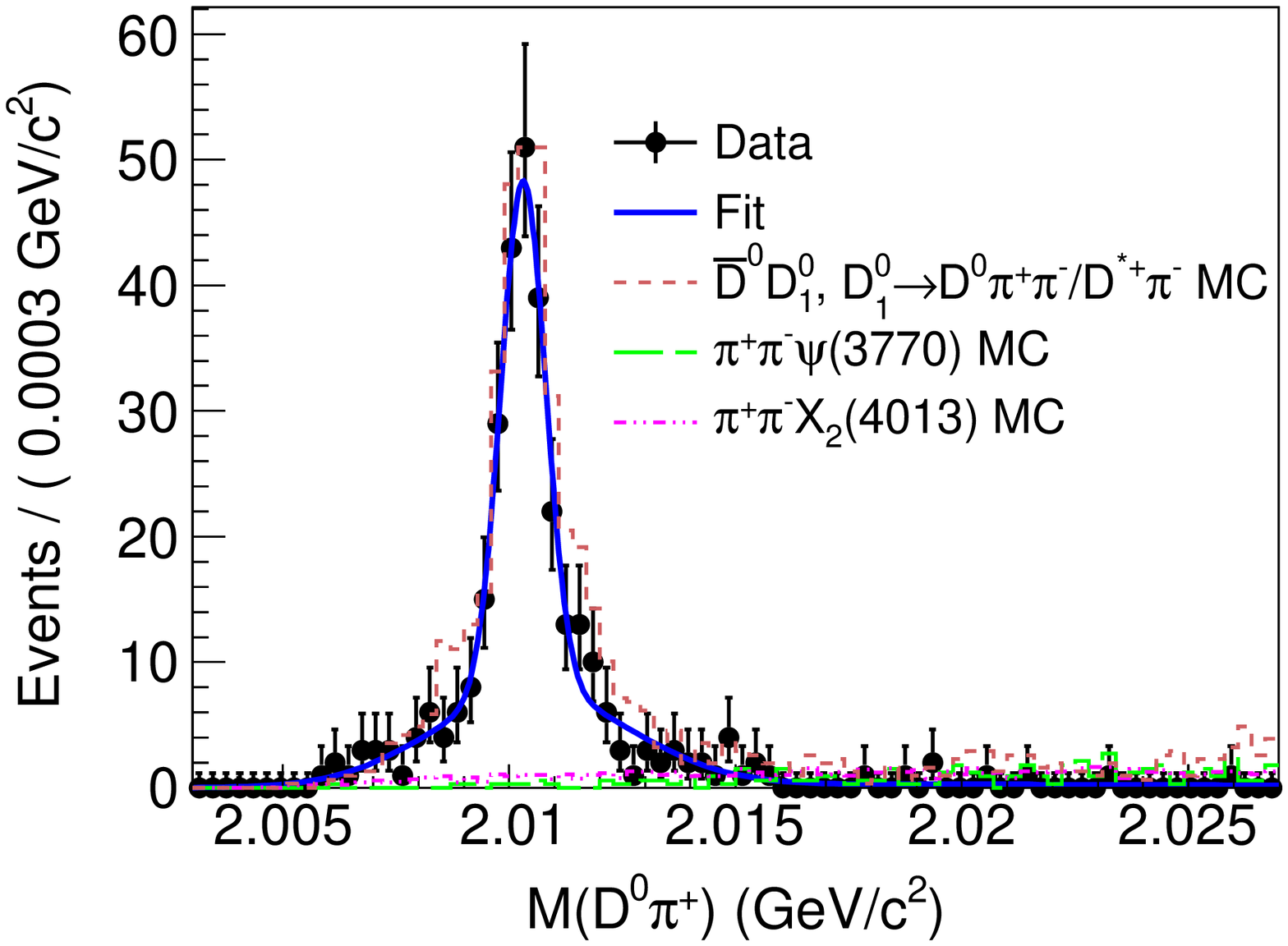}
   \put(80,60){(a)}
   \end{overpic}
   \begin{overpic}[width=0.42\textwidth]{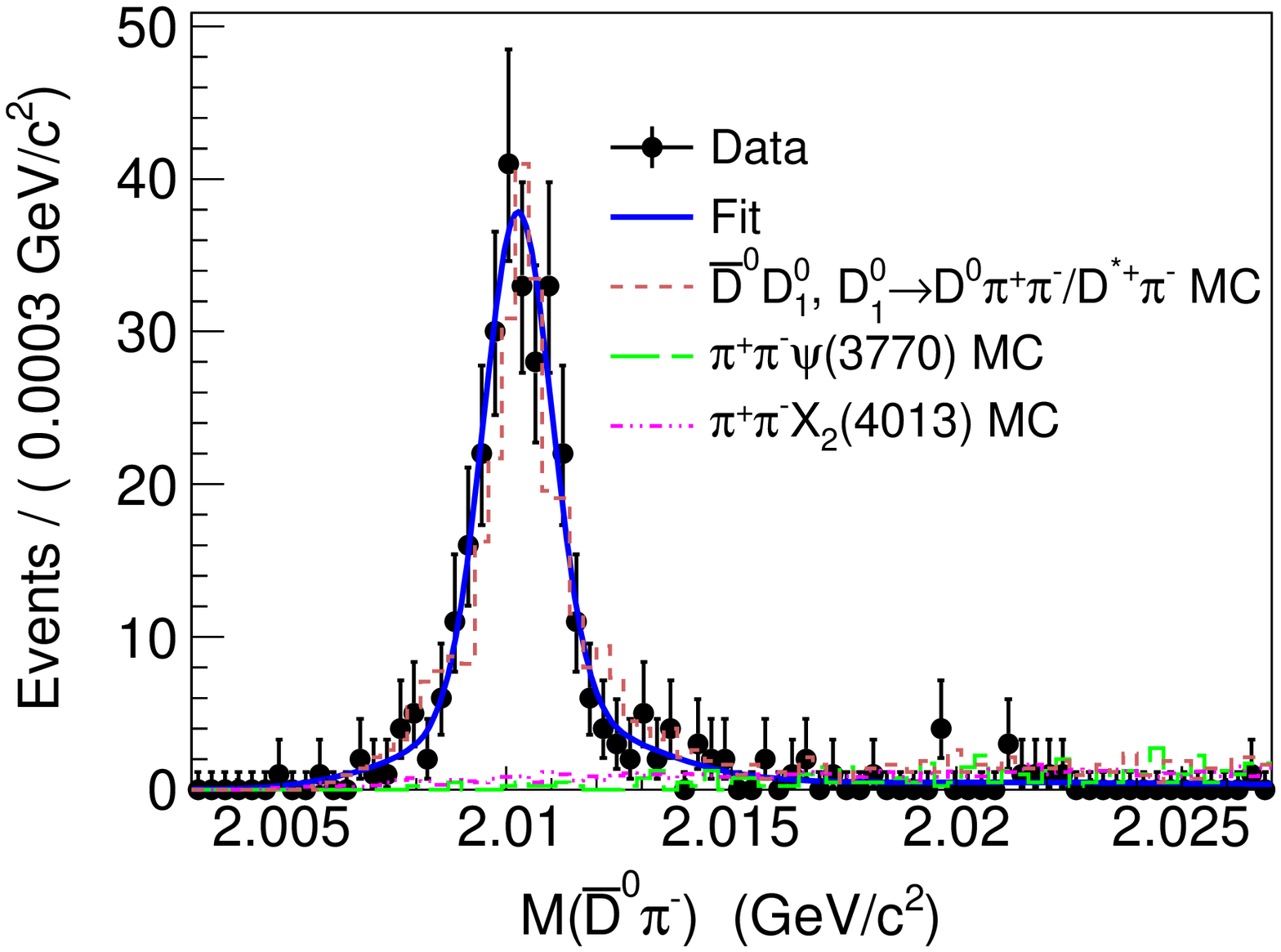}
   \put(80,60){(b)}
   \end{overpic}
\caption{Distributions of the invariant masses $M(D^{0}\pi^{+})$ (a) and
  $M(\bar{D}^{0}\pi^{-})$ (b) at $\sqrt{s}= 4.42$~GeV.
  The black dots with error bars are data and the blue solid lines are the fit results. The brown dashed lines are the distributions from $\bar{D}^{0}D^{0}_{1}$ and the green long-dashed lines the distributions from $\pi^{+}\pi^{-}\psi(3770)$. The pink dotted-dotted-dashed lines show the distribution from $\pi^{+}\pi^{-} X_{2}(4013)$. The distributions from $\bar{D}^{0}D^{0}_{1}$ are normalized to the maximum bin content of data, and the distributions from $\pi^{+}\pi^{-}\psi(3770)$ and $\pi^{+}\pi^{-} X_{2}(4013)$ are normalized arbitrarily.}
  \label{fitofDstar}
\end{figure*}

In order to suppress the background contribution of $e^{+}e^{-}\to
D^{(*)}\bar{D}^{(*)}(\pi)$, we examine if there is a $D^{*}$
($\bar{D}^{*}$) signal in the $D^0\pi^+$ ($\bar{D}^0\pi^-$)
combination. The
distributions of $M(D^{0}\pi^{+})$ and
$M(\bar{D}^{0}\pi^{-})$ are shown in
Fig.~\ref{fitofDstar}. To improve the mas resolution, $M(D^{0}\pi^{+})$ is calculated as $M(D^{0}\pi^{+})-M(D^{0})+m_{D^0}$  and $M(\bar{D}^{0}\pi^{-})$ as
$M(\bar{D}^{0}\pi^{-})-M(\bar{D}^{0})+m_{\bar{D}^0}$, thereby eliminating the effect of the mass resolution
from the reconstruction of the $D^{0}$ ($\bar{D}^0$) meson.
The criteria
$M(D^{0}\pi^{+})>2.017$~GeV/$c^{2}$ and
$M(\bar{D}^{0}\pi^{-})> 2.017$~GeV/$c^{2}$
are applied to the processes $e^{+}e^{-}\to
\pi^{+}\pi^{-}\psi(3770)$, $\psi(3770)\to D^{0}\bar{D}^{0}$,
$e^{+}e^{-}\to \rho^{0} X_{2}(4013)$, $X_{2}(4013)\to
D^{0}\bar{D}^{0}$, and $e^{+}e^{-}\to
D_{1}(2420)^{0}\bar{D}^{0}, D_{1}(2420)^{0}\to D^{0}\pi^{+}\pi^{-}$. The criteria $M(D^{0}\pi^{+})<2.017$~GeV/$c^{2}$ and
$M(\bar{D}^{0}\pi^{-})>2.017$~GeV/$c^{2}$
are applied to the process $e^{+}e^{-}\to D_{1}(2420)^{0}\bar{D}^{0}, D_{1}(2420)^{0}\to D^{*+}\pi^{-}$, and the criteria $M(D^{0}\pi^{+})>2.017$~GeV/$c^{2}$ and $M(\bar{D}^{0}\pi^{-})<2.017$~GeV/$c^{2}$ are applied to the charged conjugate process.

The inclusive MC sample is used to investigate possible background contributions. There is neither a
peaking background contribution found near 3.773~GeV/$c^{2}$ and
4.013~GeV/$c^{2}$ in the $D\bar{D}$ invariant  mass distribution
nor near 2.42~GeV/$c^{2}$ in the $D\pi\pi$ invariant
mass distribution. From the study of the MC samples with highly
excited charmed mesons, we find that only the process $e^{+}e^{-}\to
D_{2}^{*}(2460)^{0}\bar{D}^{0}$, $D_{2}^{*}(2460)^{0}\to D^{*+}\pi^{-}$
produces a peak near 2.46~GeV/$c^{2}$ in the signal region of the  $D_{1}(2420)^{0}$ in the invariant mass distribution of
$D^{*+}\pi^{-}$. Therefore,  $e^{+}e^{-}\to
D_{2}^{*}(2460)^{0}\bar{D}^{0}$, $D_{2}^{*}(2460)^{0}\to D^{*+}\pi^{-}$ is considered
as a component of the background contribution in the study of $e^{+}e^{-}\to
D_{1}(2420)^{0}\bar{D}^{0}$, $D_{1}(2420)^{0}\to D^{*+}\pi^{-}$.

There will be some non-$D\bar{D}$ backgrounds remaining in the signal region. According to
 the study of the inclusive MC, in the $D\bar{D}$ and $D\pi\pi$ invariant
mass distribution, non-$D\bar{D}$ backgrounds and sidebands events are consistent with each other.
Therefore, the events from the sidebands are used to describe non-$D\bar{D}$ backgrounds in this analysis.

\section{SIGNAL yield determination}

\subsection{\boldmath $e^{+}e^{-}\to \pi^{+}\pi^{-}\psi(3770)$}
\begin{figure*}[htbp]
  \centering
  \begin{overpic}[width=0.329\textwidth]{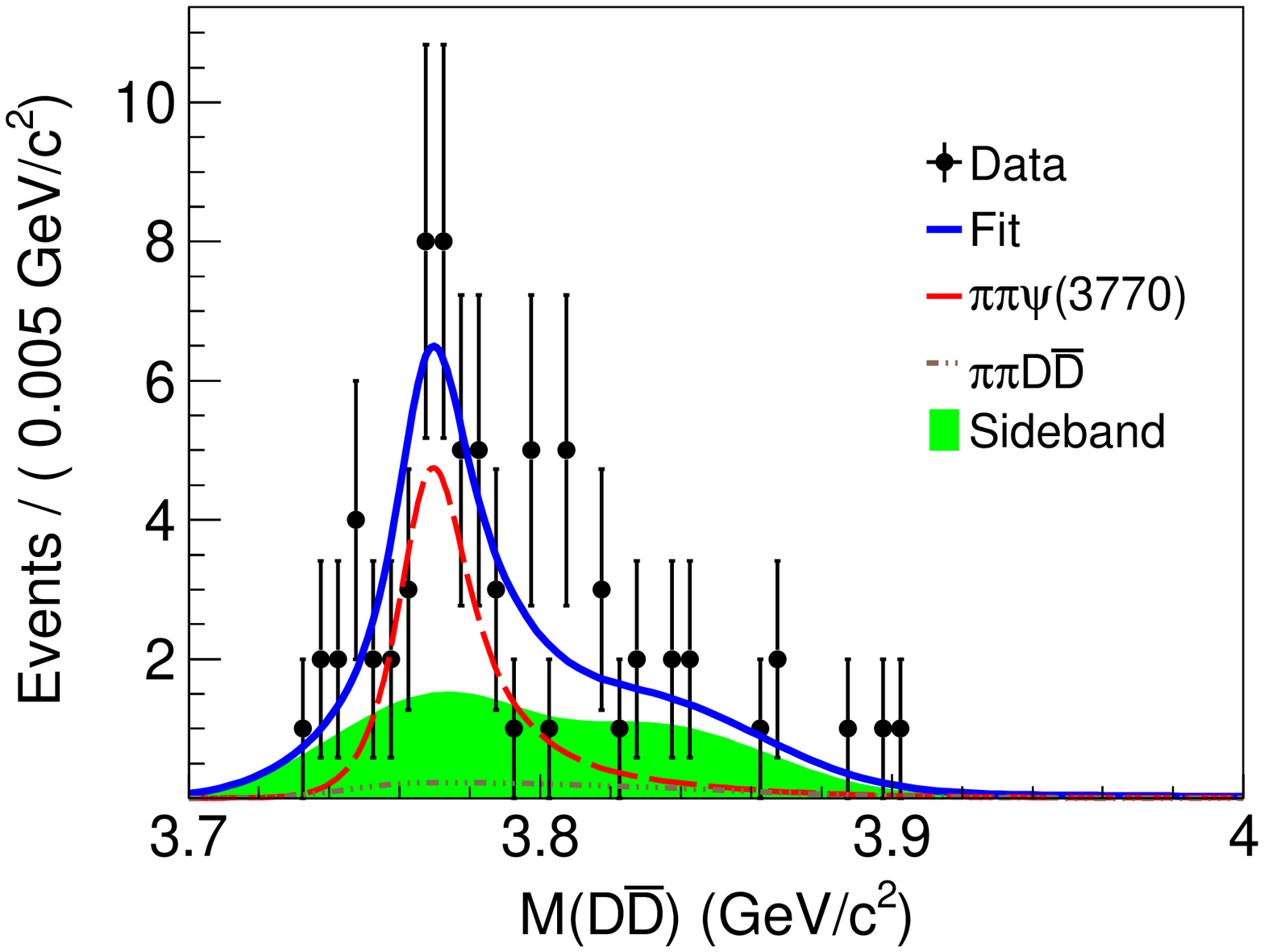}
   \put(75,65){(a)}
   \end{overpic}
   \begin{overpic}[width=0.329\textwidth]{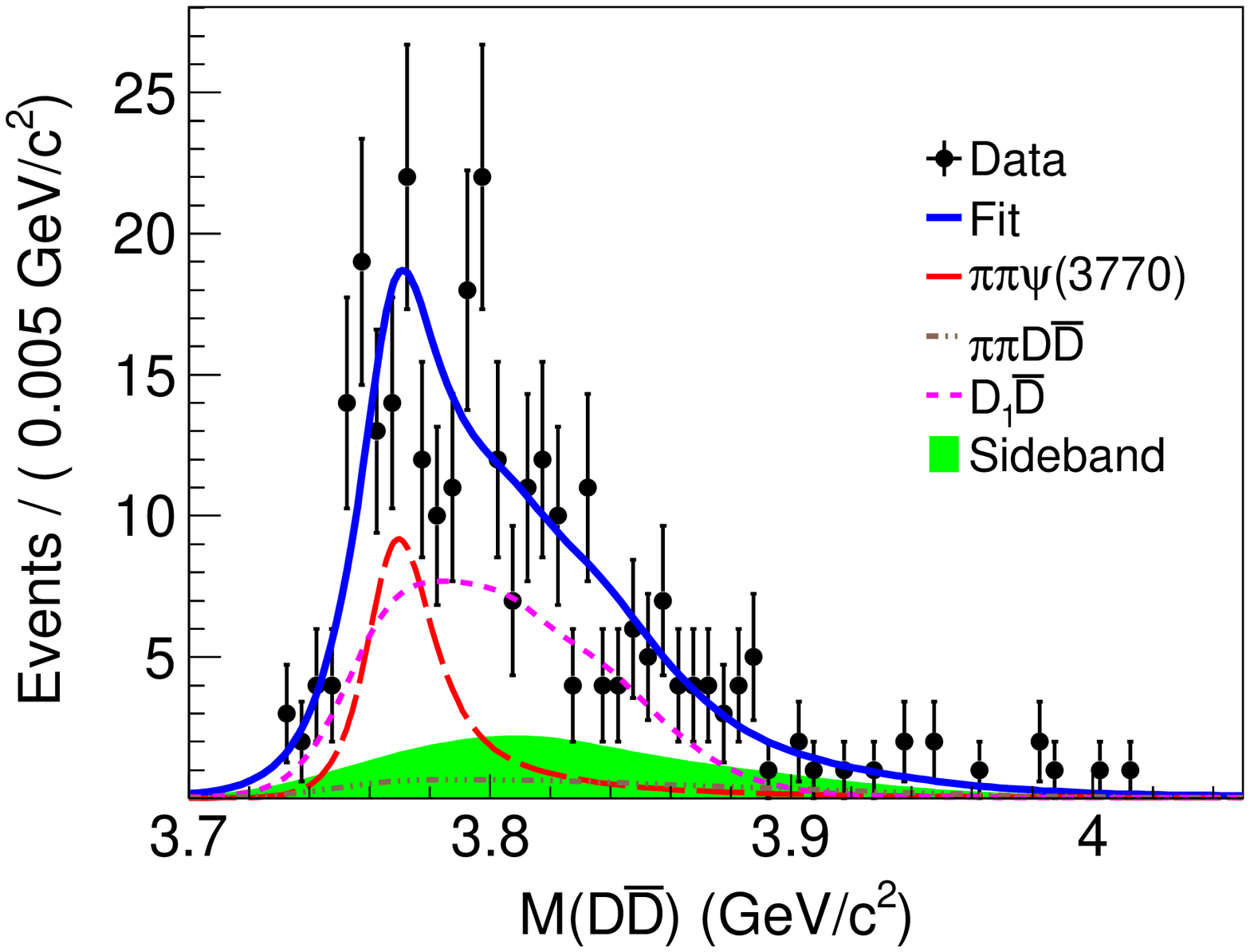}
   \put(75,65){(b)}
   \end{overpic}
   \begin{overpic}[width=0.329\textwidth]{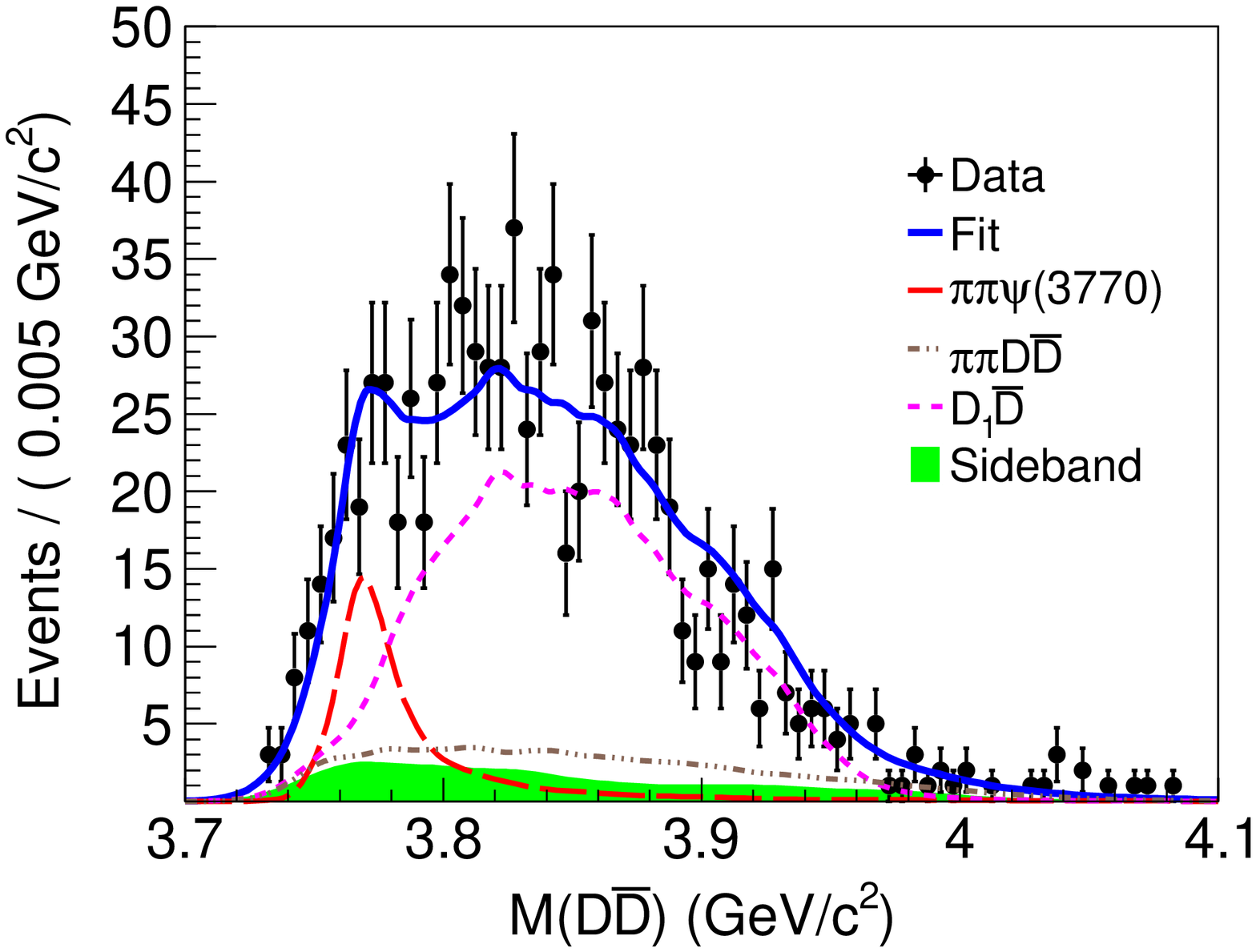}
   \put(75,65){(c)}
   \end{overpic}
   \caption{Fit to the $D\bar{D}$ invariant
mass distribution at $\sqrt{s} = $ 4.26 (a), 4.36 (b), and 4.42 (c)~GeV. The black dots with error bars
are data and the blue solid lines are the fit results. The red long-dashed lines indicate the contribution of the $\pspp$ and the pink
dashed lines the contribution of the $D_{1}(2420)\bar{D}$ final state. The brown dotted-dashed lines show the
$\pi^{+}\pi^{-} D\bar{D}$ background contributions and the green shaded histograms are the distributions from the sideband regions. }
  \label{FitDDbar}
\end{figure*}

After imposing all the requirements mentioned above, the $D\bar{D}$ invariant  mass distributions are shown in
Fig.~\ref{FitDDbar}. $M(D\bar{D})$ is used for the expression $M(D\bar{D})-M(D)-M(\bar{D})+2m_{D}$ to obtain
a better mass resolution by eliminating the mass resolution effect
coming from the reconstruction of the $D$ and $\bar{D}$ mesons. A peak at around 3.77~GeV/$c^{2}$ can be seen,
but there is no evidence for an intermediate state $\psi(1^{3}D_{3})$.

\begin{figure*}[t]
  \centering
   \begin{overpic}[width=0.329\textwidth]{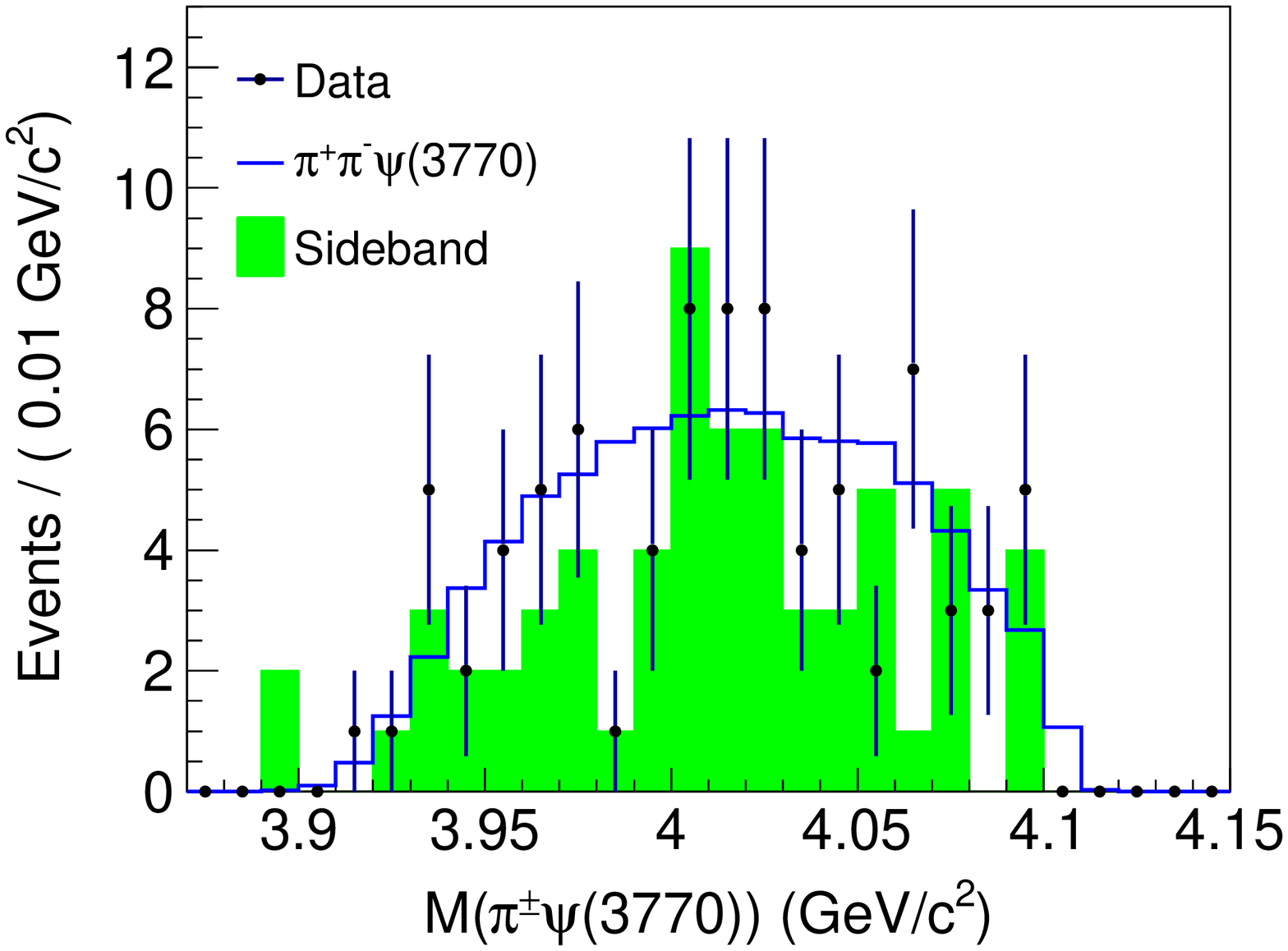}
   \put(75,65){(a)}
   \end{overpic}
   \begin{overpic}[width=0.329\textwidth]{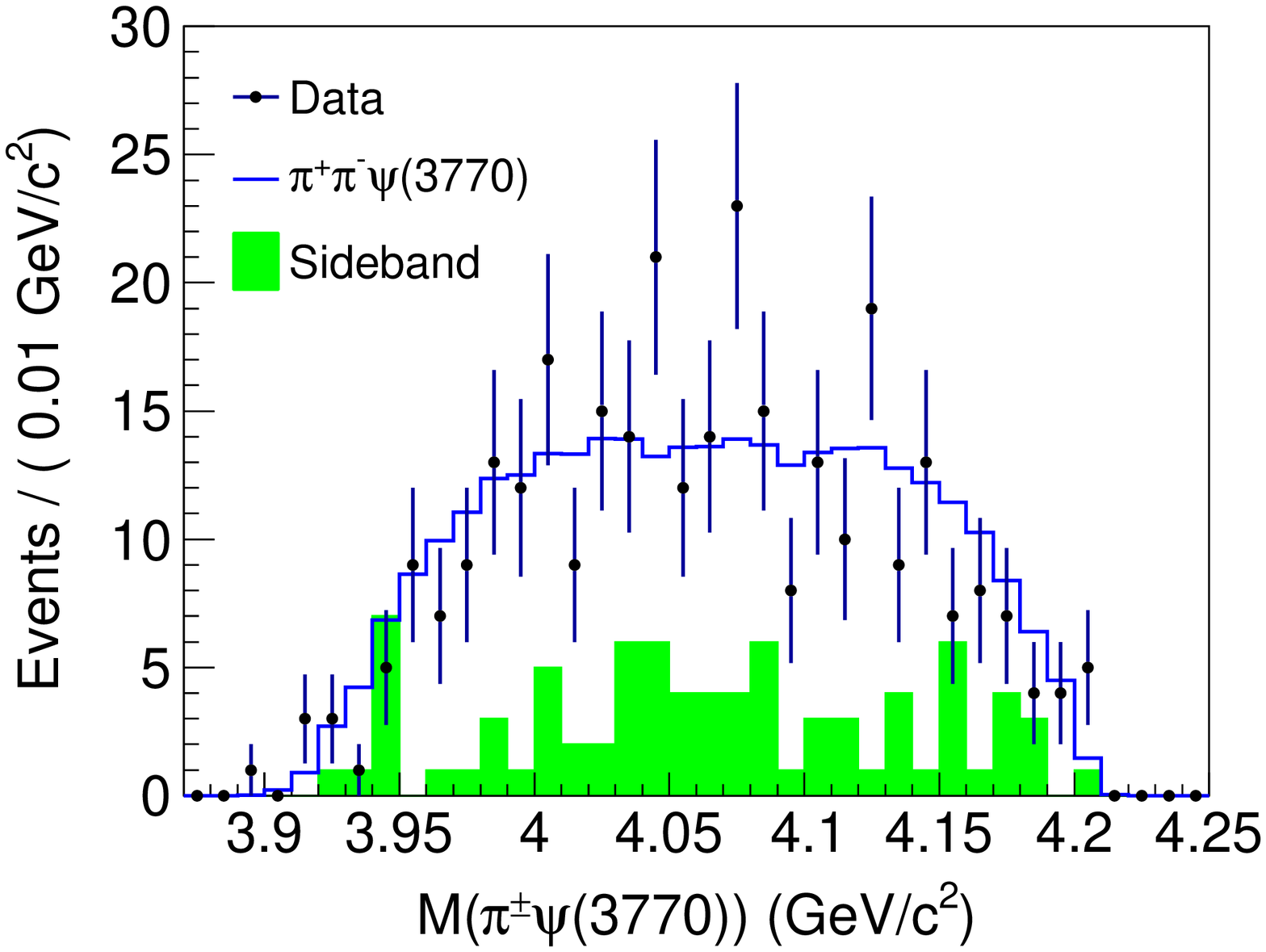}
   \put(75,65){(b)}
   \end{overpic}
   \begin{overpic}[width=0.329\textwidth]{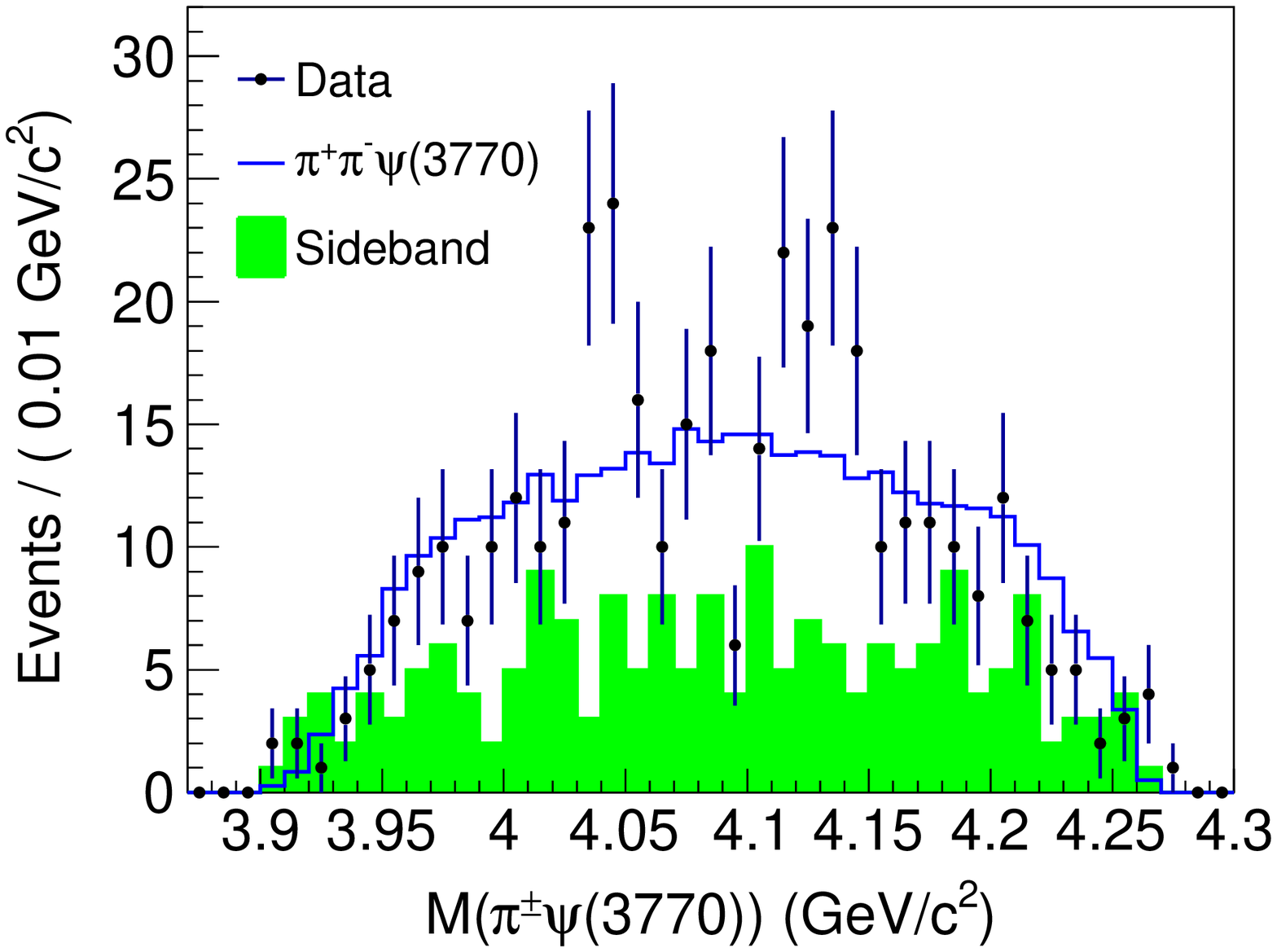}
   \put(75,65){(c)}
   \end{overpic}
   \caption{Distribution of the $M(\pi^{\pm}\psi(3770))$ invariant mass at $\sqrt{s}$ = 4.26 (a), 4.36 (b), 4.42 (c)~GeV. The black dots with error bars are data. The blue histograms show the distributions of the MC simulation of the process $e^{+}e^{-}\rightarrow\pi^{+}\pi^{-}\psi(3770)$ (phase space distributed), while the green
shaded histograms are the distributions from the sideband regions.}
  \label{Mpipmpsipp}
\end{figure*}

To determine the signal yield of $e^{+}e^{-}\to \pi^{+}\pi^{-}\psi(3770)$,
an unbinned maximum likelihood fit is performed to the $M(D\bar{D})$ spectra as shown in Fig.~\ref{FitDDbar}. The
signal contribution is described by the MC simulated shape which is modeled using non-parametric kernel-estimation~\cite{keyspdf}. The
background component includes the channel $D_{1}(2420)\bar{D}$, the four-body decay $\pi^{+}\pi^{-} D\bar{D}$ (both described
with the shape taken from the MC simulation which are also modeled with non-parametric kernel-estimation) and the non-$D\bar{D}$ background distribution (described by the $D\bar{D}$ sideband events). In the fit, the signal yields are free parameters with lower limits set to 0.
The yields of the $D_{1}(2420)\bar{D}$ and the $\pi^{+}\pi^{-} D\bar{D}$ background contributions are free
parameters. The number of non-$D\bar{D}$ background events is
fixed to the number of events from the sidebands. The signal yields at $\sqrt{s}$
= 4.26, 4.36, and 4.42~GeV returned by the fit are  $30.7 \pm 9.9$, $68.7 \pm 21.8$, and $99.2\pm 21.0$ events, respectively. The statistical significance of the signal yield
is determined to be $3.2\sigma$, $3.3\sigma$, and $5.2\sigma$,
respectively, by comparing the log-likelihood values with and
without the signal hypothesis and taking the change in the number of
degrees of freedom into account. With the same method,  the data samples taken at
other c.m.\ energies are also studied as shown in Fig.~\ref{FitDDbar_all} of Appendix A.
The signal yields are listed in Table~\ref{RCFDDbar}. In this analysis, if the statistical significance of the signal is less than 1$\sigma$,  we will scan the likelihood distribution as a function of the signal yield greater than 0 and use the difference between the most probable values and the thresholds of 68.3\% total integral area as the errors.

We also search for structures in the  $\pi^{\pm}\psi(3770)$ invariant
mass distribution at the energy points where the $\pi\pi\psi(3770)$
signal is most prominent. The $\pi^{\pm}\psi(3770)$ distribution after requiring $M(D\bar{D})$ $\in$ [3.73, 3.80] GeV/$c^{2}$ around the $\psi(3770)$ mass is shown in Fig.~\ref{Mpipmpsipp}. There are hints for peaks at 4.04 and 4.13~GeV/$c^2$ in $\sqrt{s}$ = 4.42~GeV data, but the statistical significance is low.

\subsection{\boldmath  $e^{+}e^{-}\to \rho^{0} X_{2}(4013)$}

\begin{figure*}[htbp]
   \centering
   \begin{overpic}[width=0.329\textwidth]{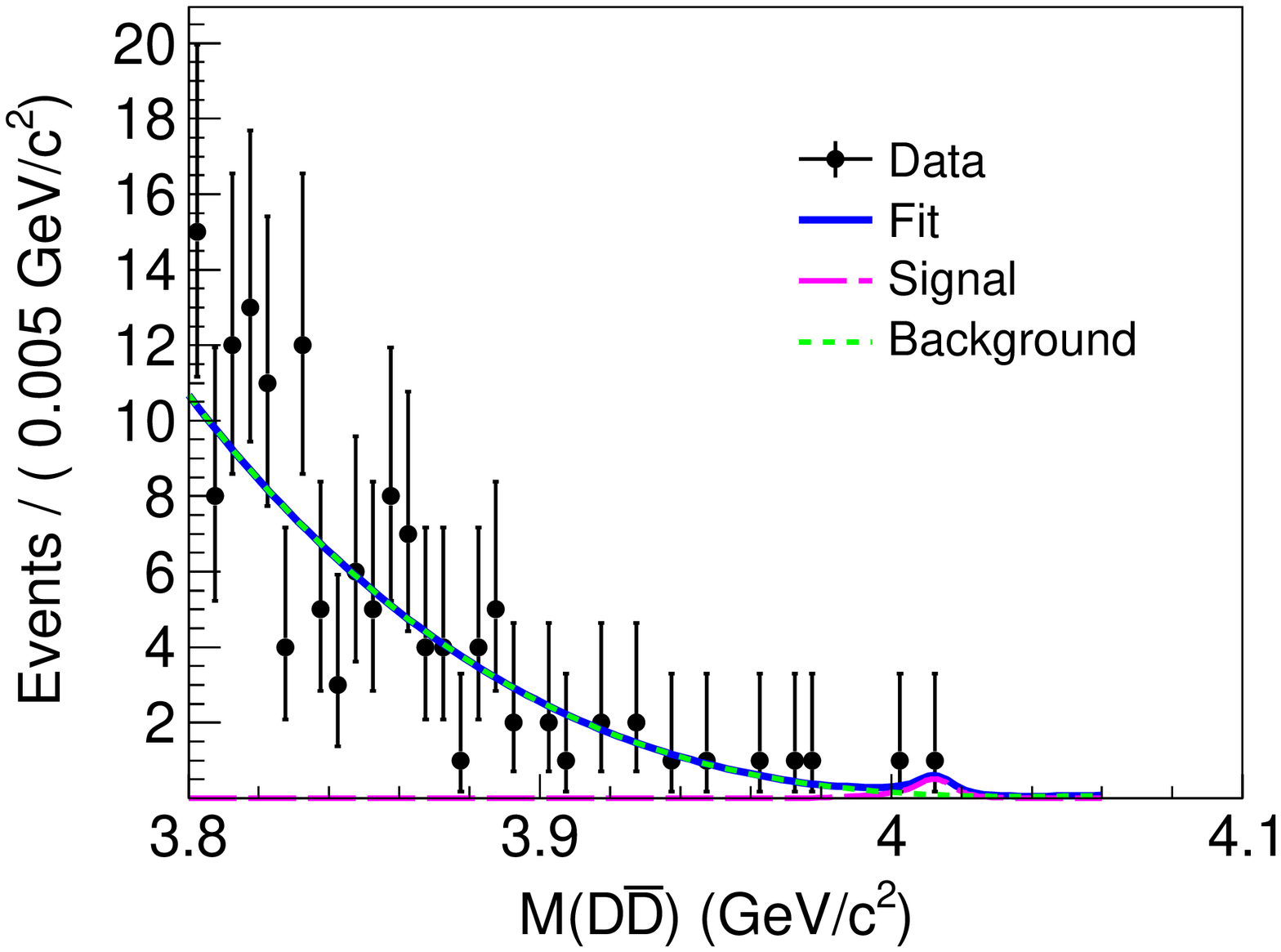}
   \put(75,65){(a)}
   \end{overpic}
   \begin{overpic}[width=0.329\textwidth]{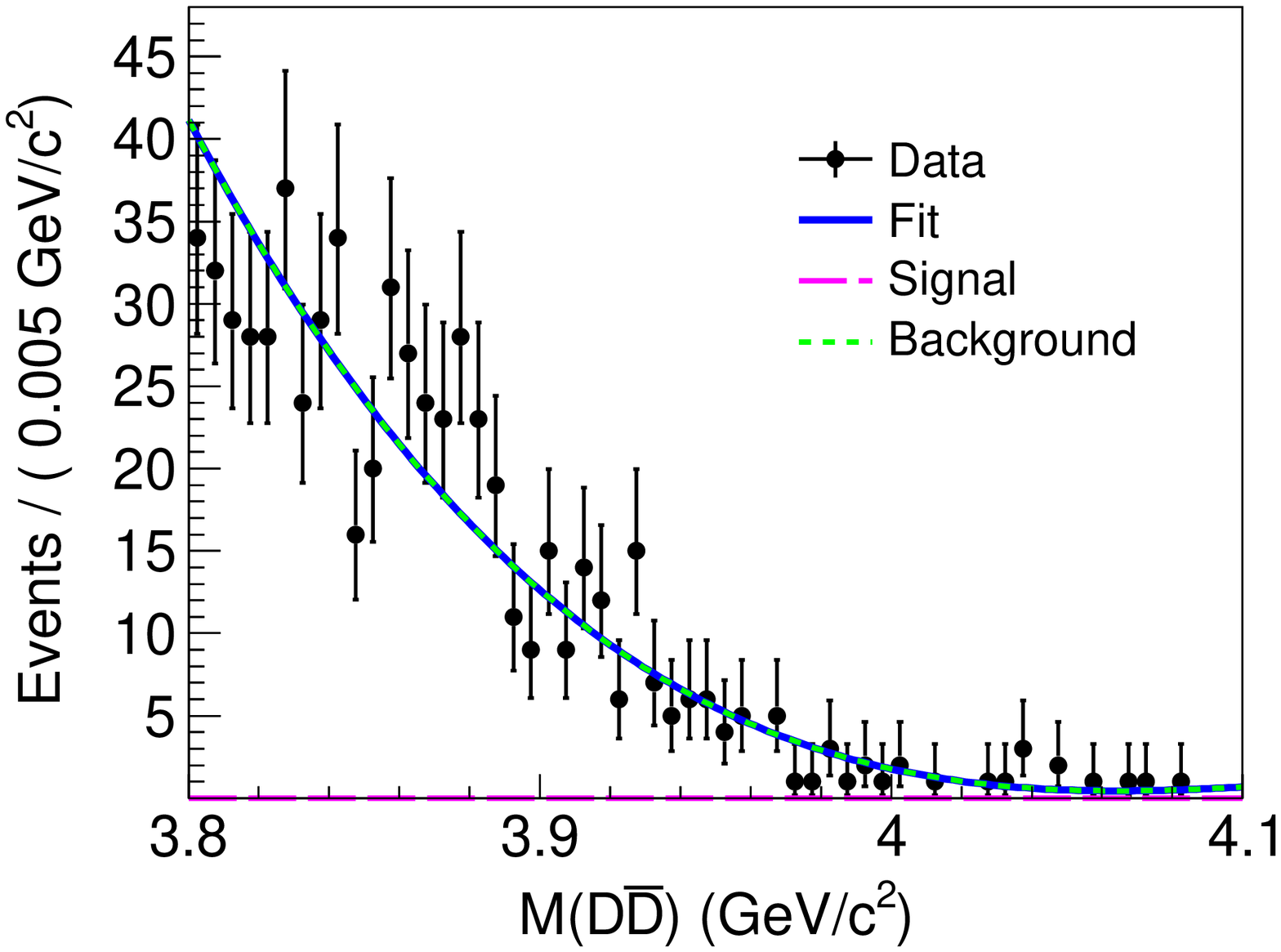}
   \put(75,65){(b)}
   \end{overpic}
   \begin{overpic}[width=0.329\textwidth]{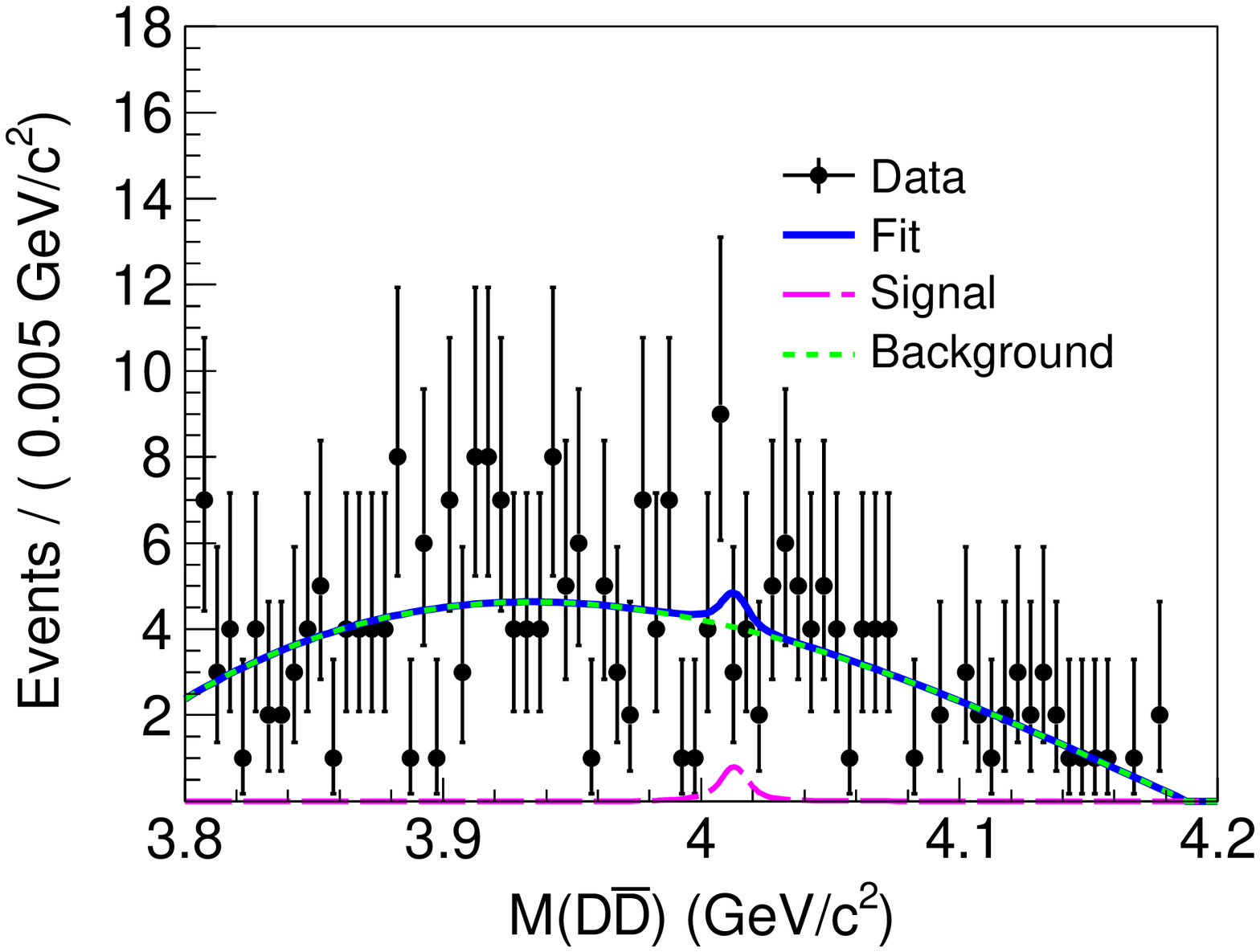}
   \put(75,65){(c)}
   \end{overpic}
   \caption{Fit to the region of large $D\bar{D}$ invariant  masses at $\sqrt{s} = $ 4.36 (a), 4.42 (b), and 4.60 (c)~GeV.  The black dots with error bars are data, the blue solid curves are the fit results, the pink long-dashed lines show the X(4013) signal contribution and the green dashed lines  describe the background distribution.}
  \label{FitDDbar_X}
\end{figure*}

For the search for the $X_{2}(4013)$ resonance, the region of large $\ddb$ invariant masses is investigated.  Figure~\ref{FitDDbar_X} shows the distributions after imposing all the requirements above. There is no
obvious signal visible around 4.013~GeV/$c^{2}$. We try
to fit these distributions with the signal shape of the process $\rho^{0} X_{2}(4013)$ taken from the MC simulation and a third order polynomial as background distribution as shown in Fig.~\ref{FitDDbar_X}. The signal yields are $1.1_{-1.1}^{+1.5}$, $0.0_{-0.0}^{+1.8}$, and
$2.7_{-2.7}^{+5.3}$  events with the statistical significance of $1.5\sigma$,
$0\sigma$, and $0.5\sigma$ for the data sets at $\sqrt{s}$ = 4.36, 4.42, and
4.60~GeV, respectively. Results are listed in Table~\ref{RCFX4013}.

\begin{figure*}[htbp]
  \centering
   \begin{overpic}[width=0.329\textwidth]{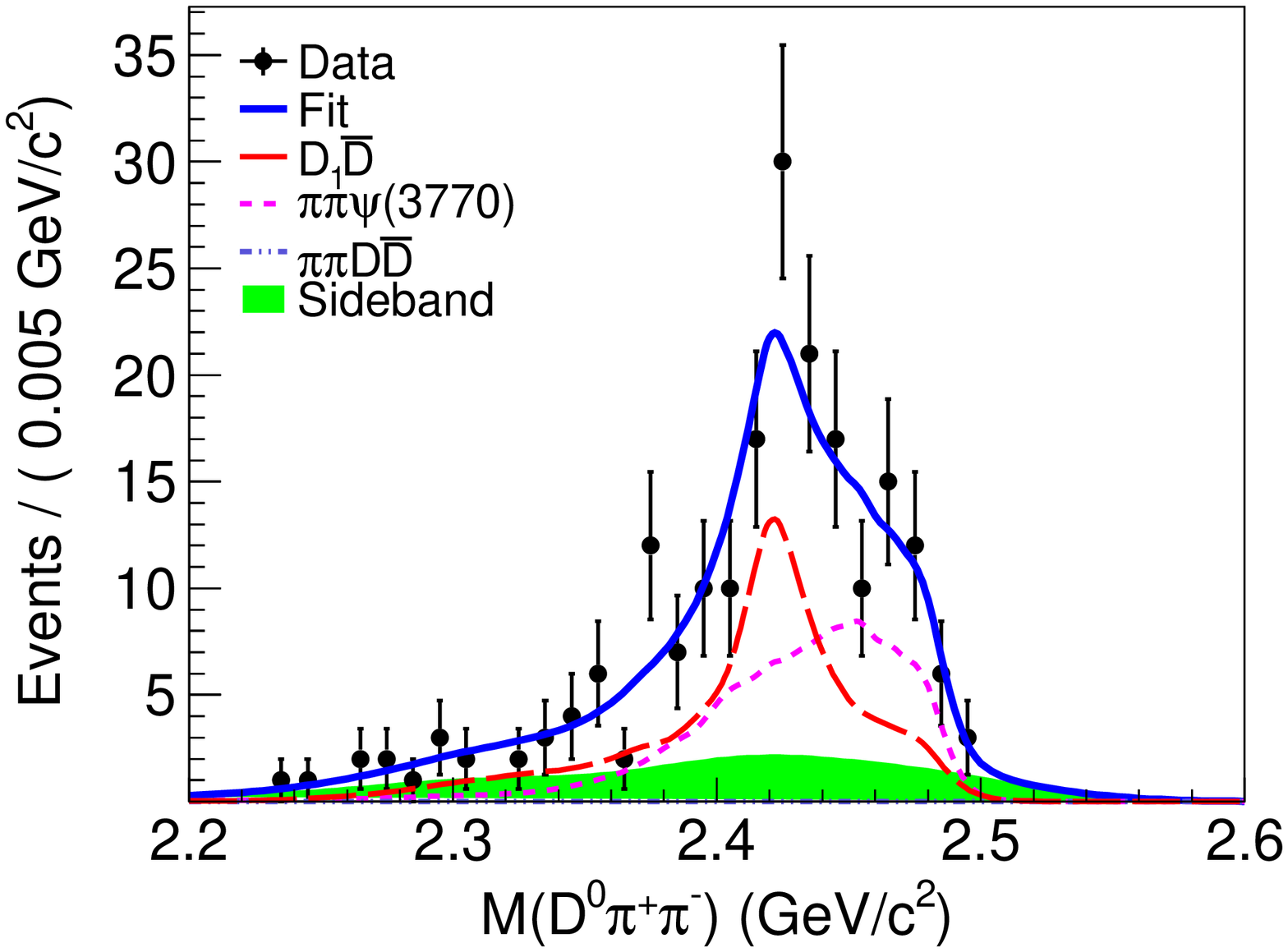}
   \put(75,65){(a)}
   \end{overpic}
   \begin{overpic}[width=0.329\textwidth]{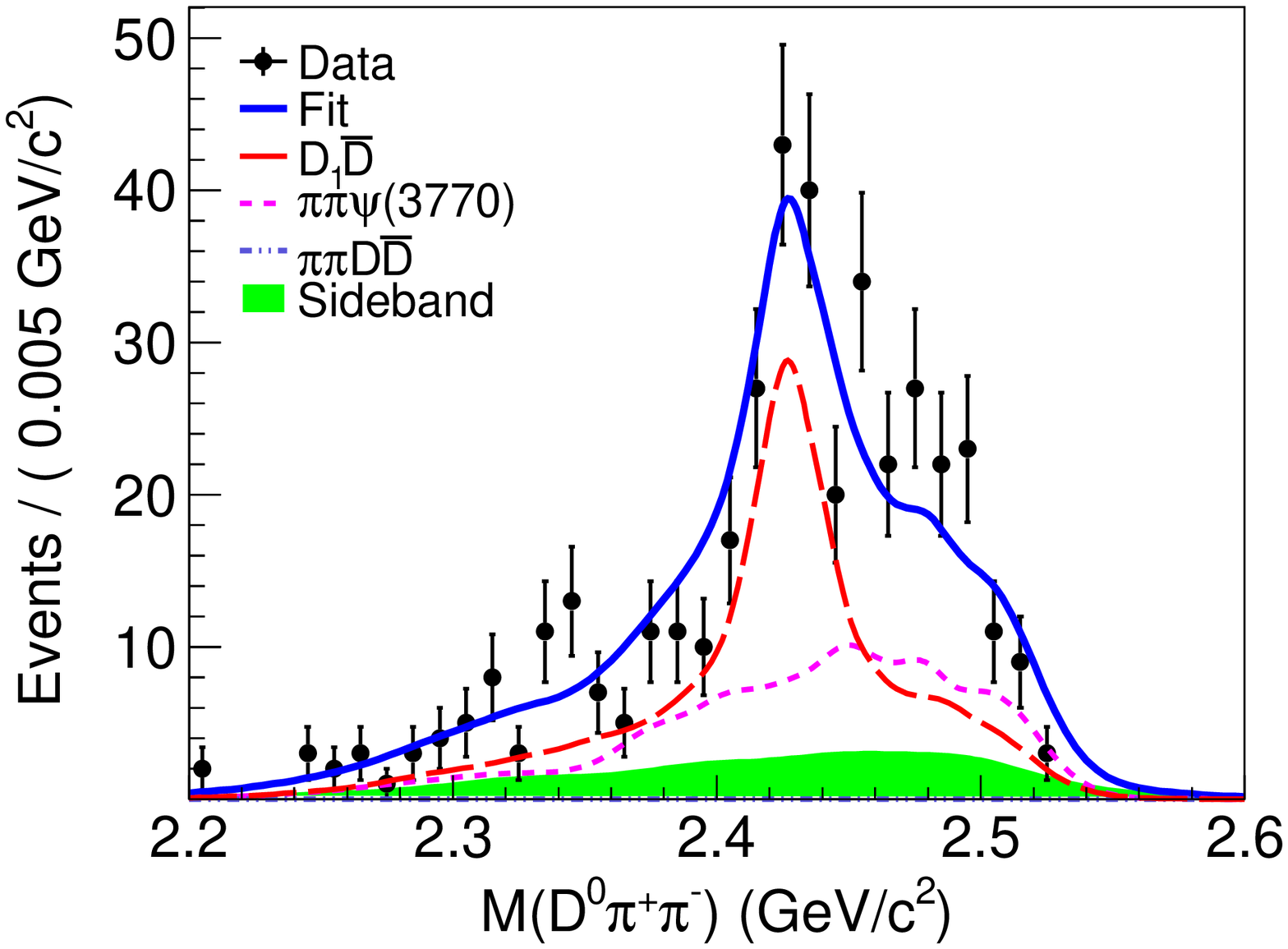}
   \put(75,65){(b)}
   \end{overpic}
   \begin{overpic}[width=0.329\textwidth]{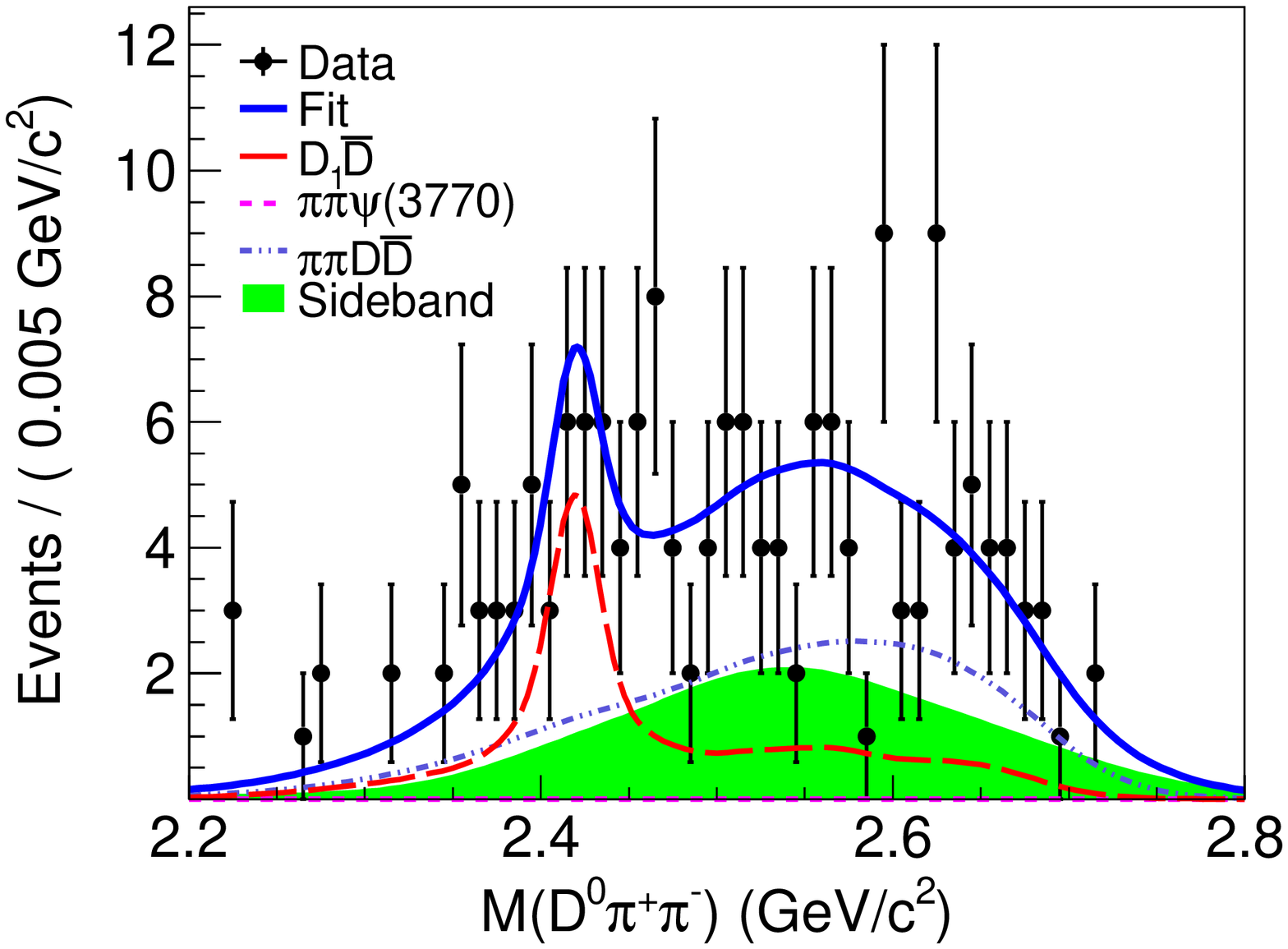}
   \put(75,65){(c)}
   \end{overpic}
   \begin{overpic}[width=0.329\textwidth]{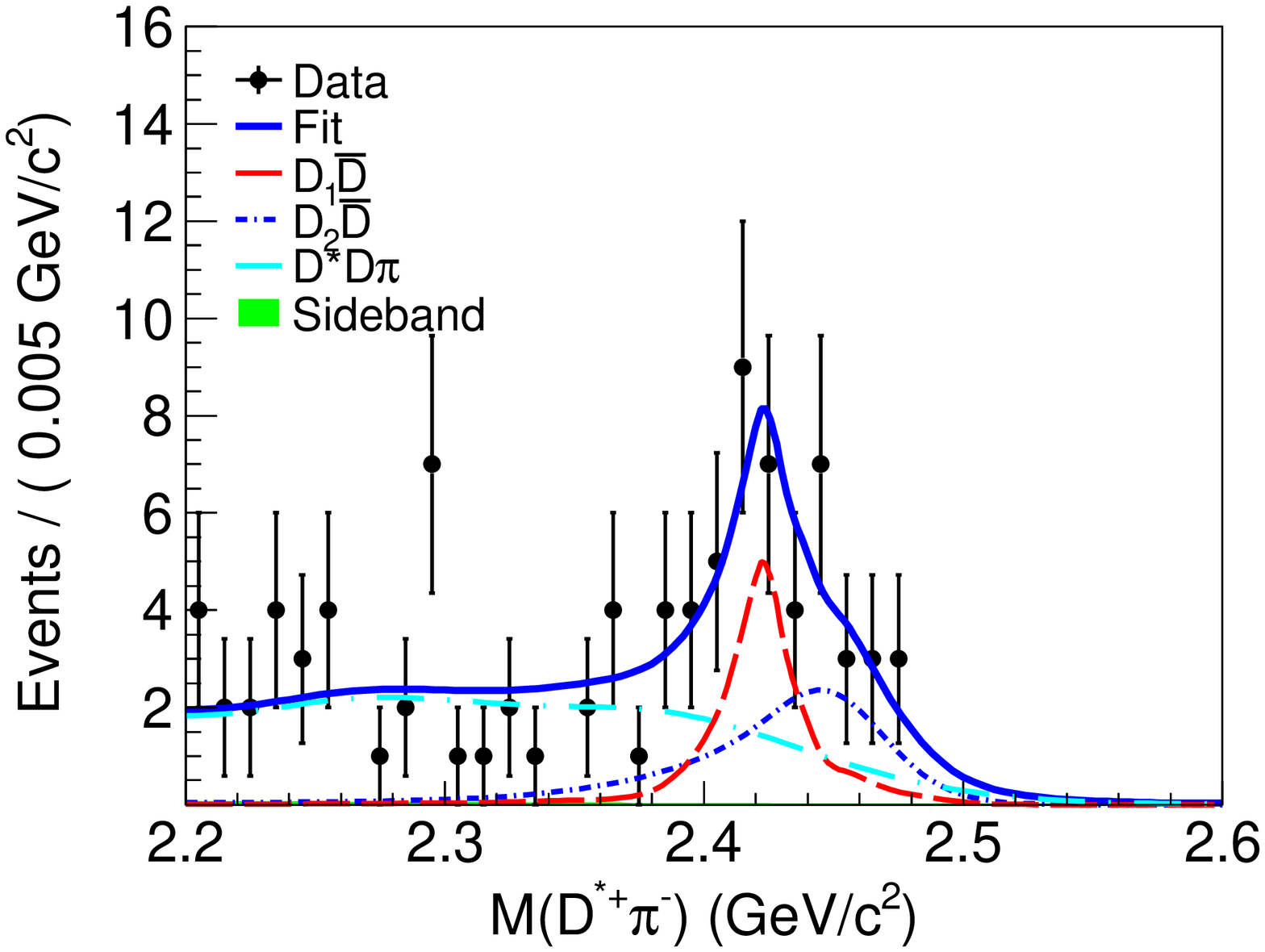}
   \put(75,65){(d)}
   \end{overpic}
   \begin{overpic}[width=0.329\textwidth]{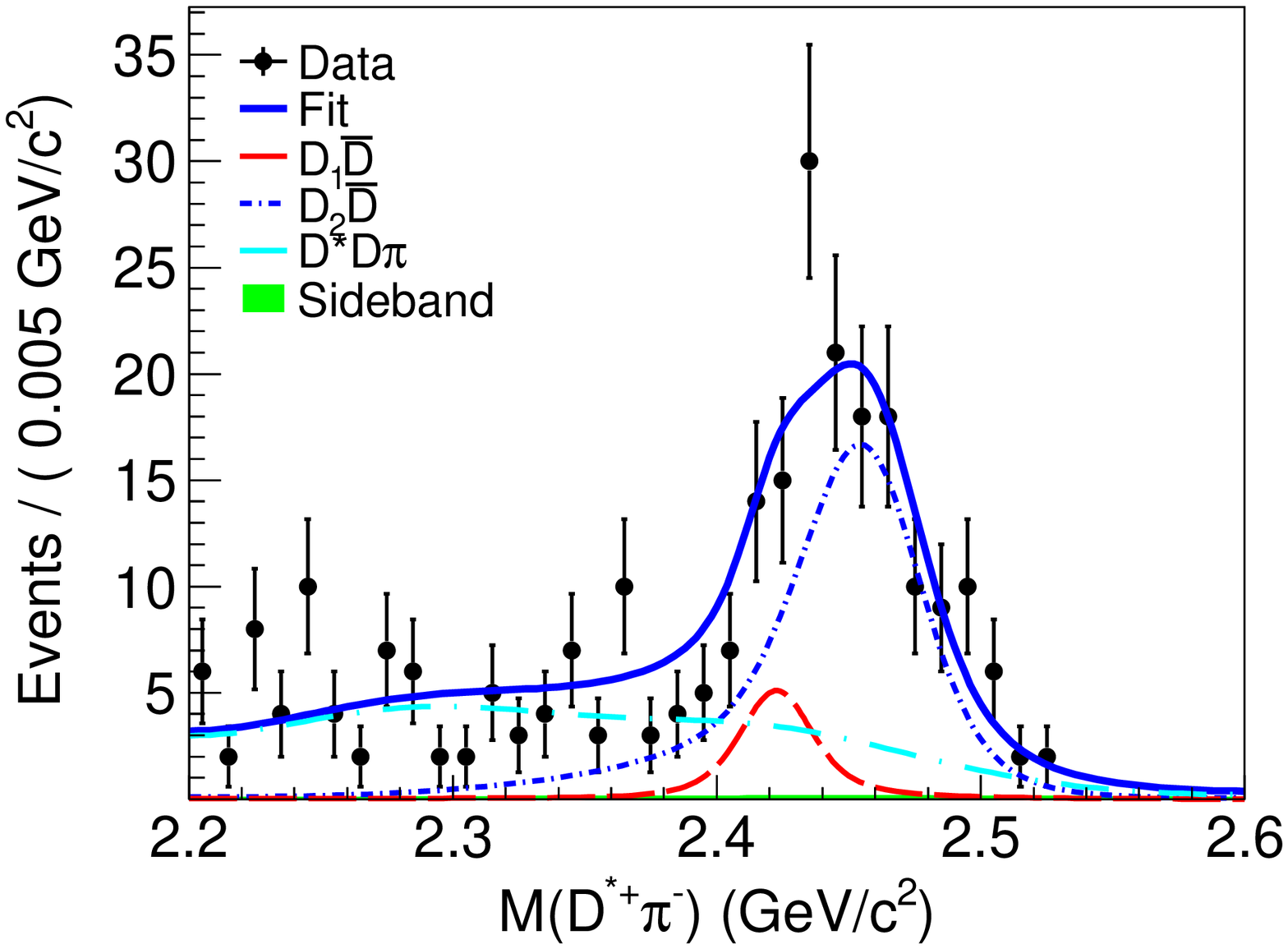}
   \put(75,65){(e)}
   \end{overpic}
   \begin{overpic}[width=0.329\textwidth]{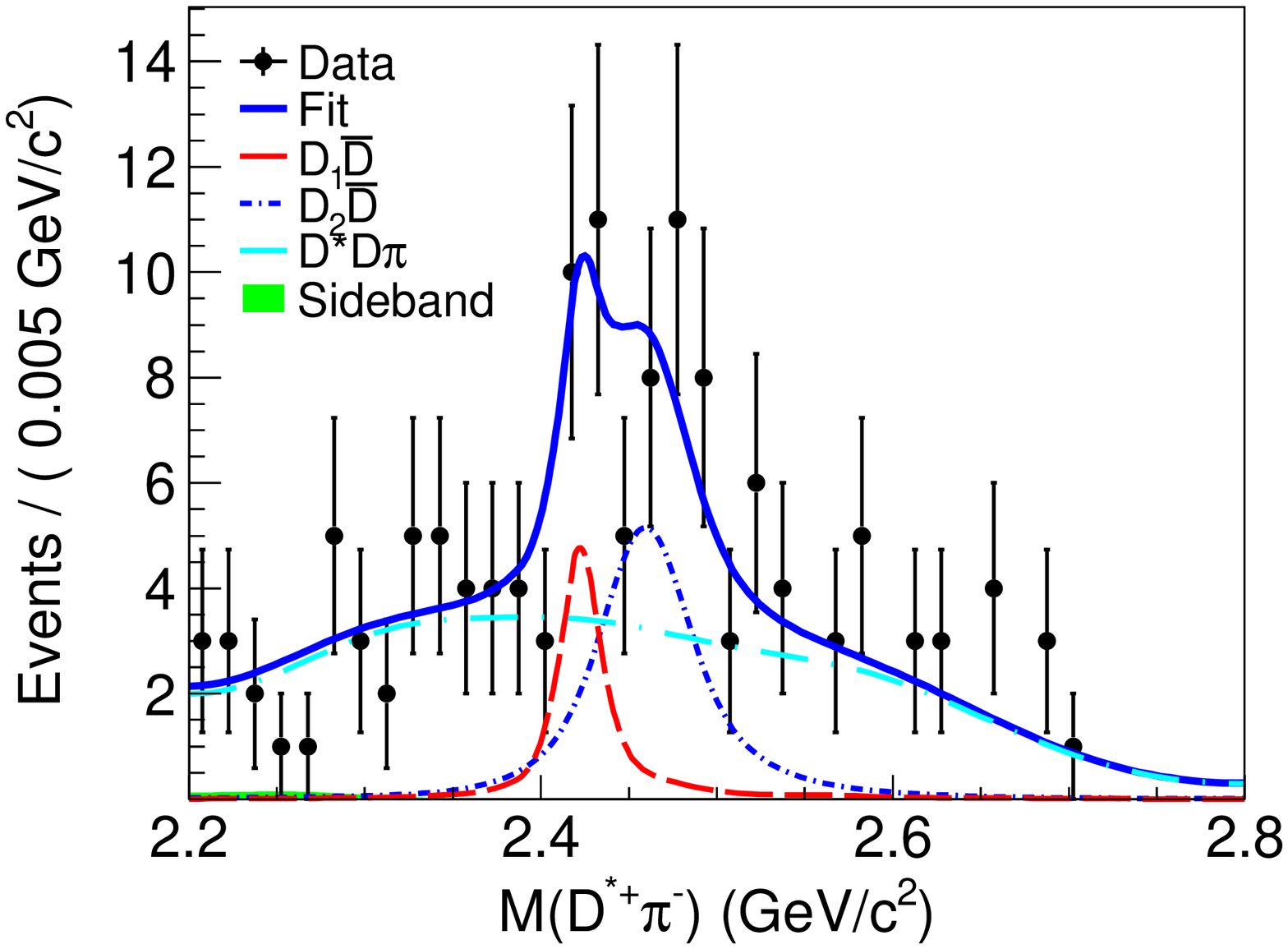}
   \put(75,65){(f)}
   \end{overpic}
   \begin{overpic}[width=0.329\textwidth]{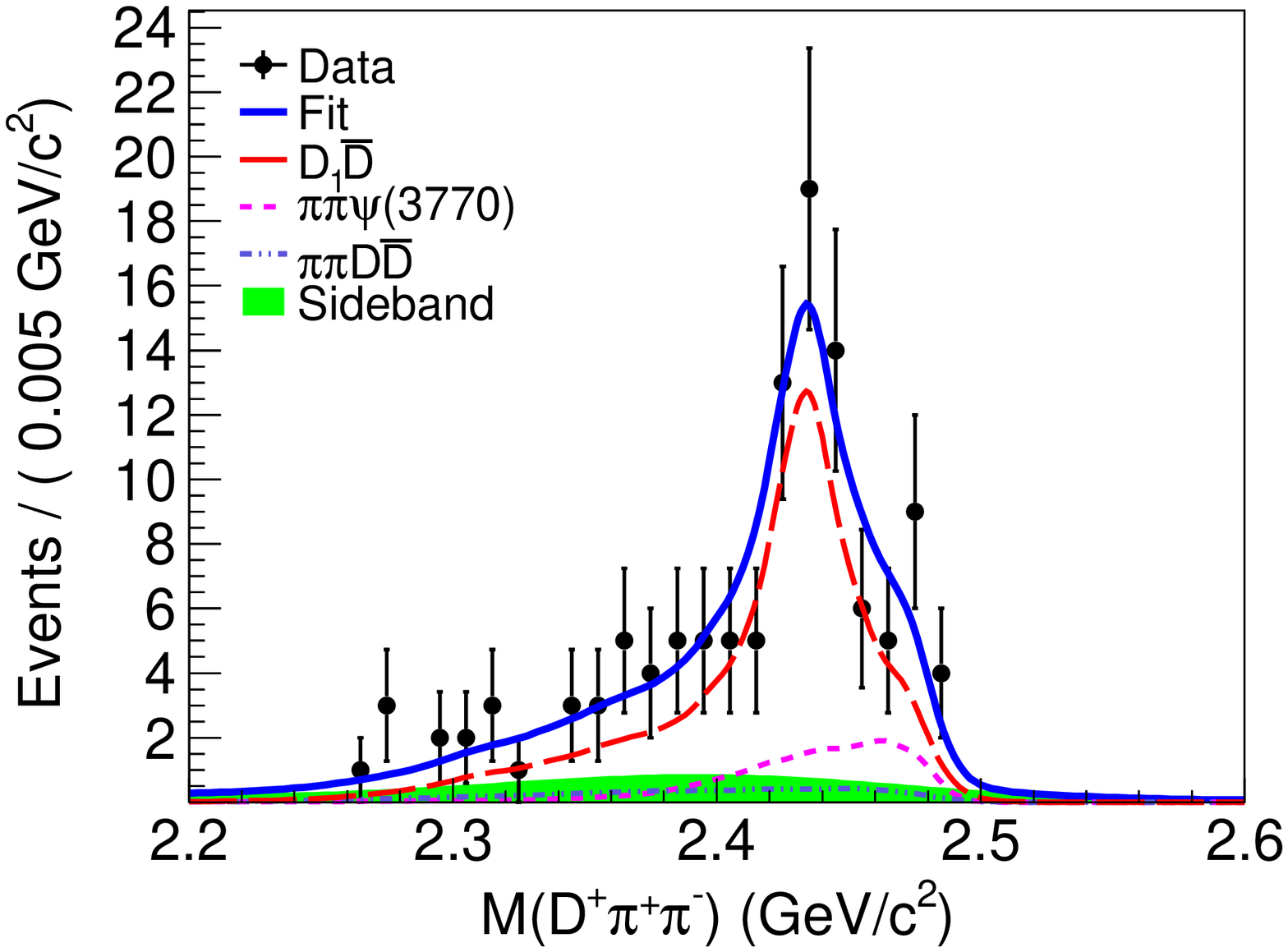}
   \put(75,65){(g)}
   \end{overpic}
   \begin{overpic}[width=0.329\textwidth]{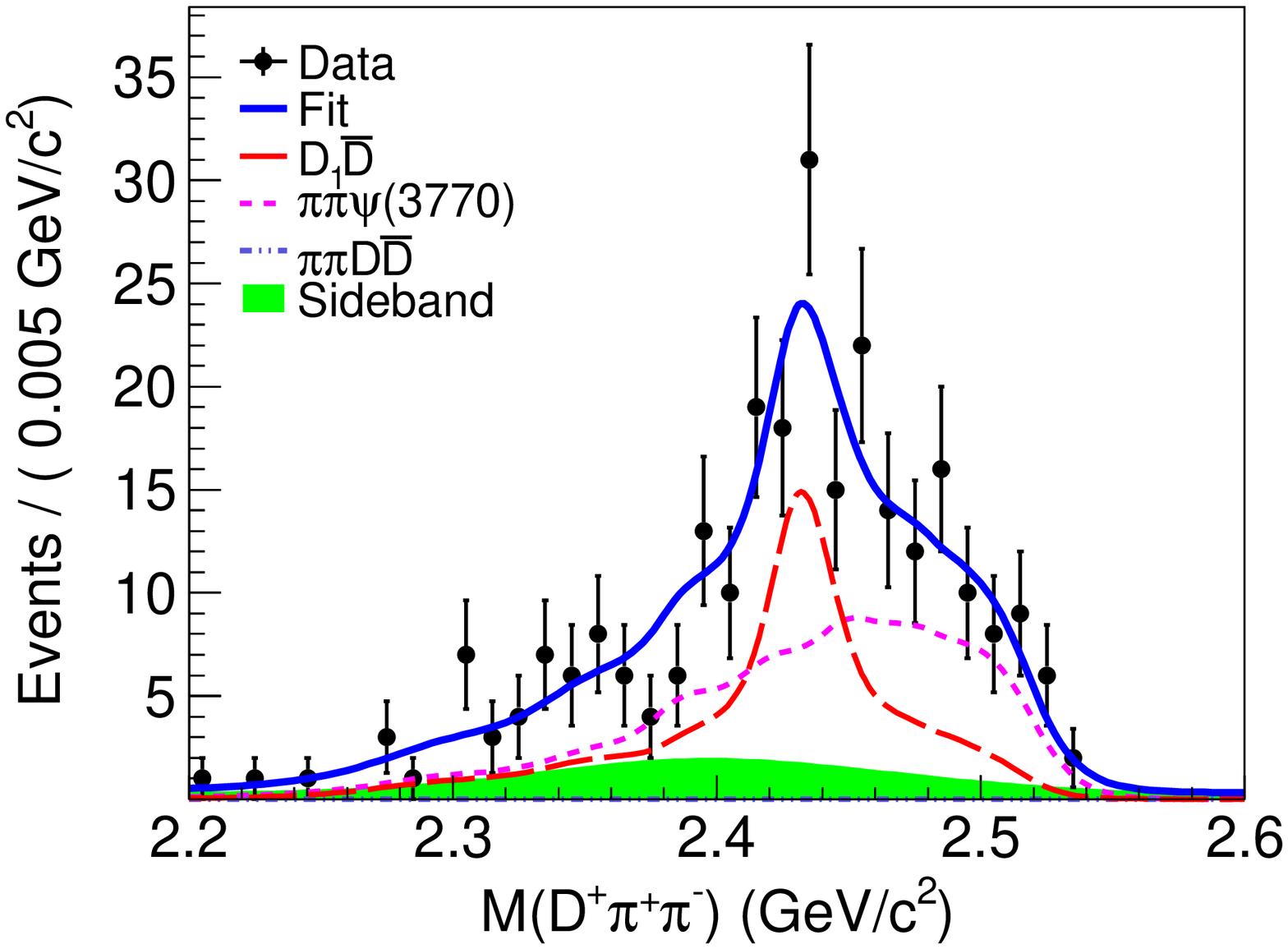}
   \put(75,65){(h)}
   \end{overpic}
   \begin{overpic}[width=0.329\textwidth]{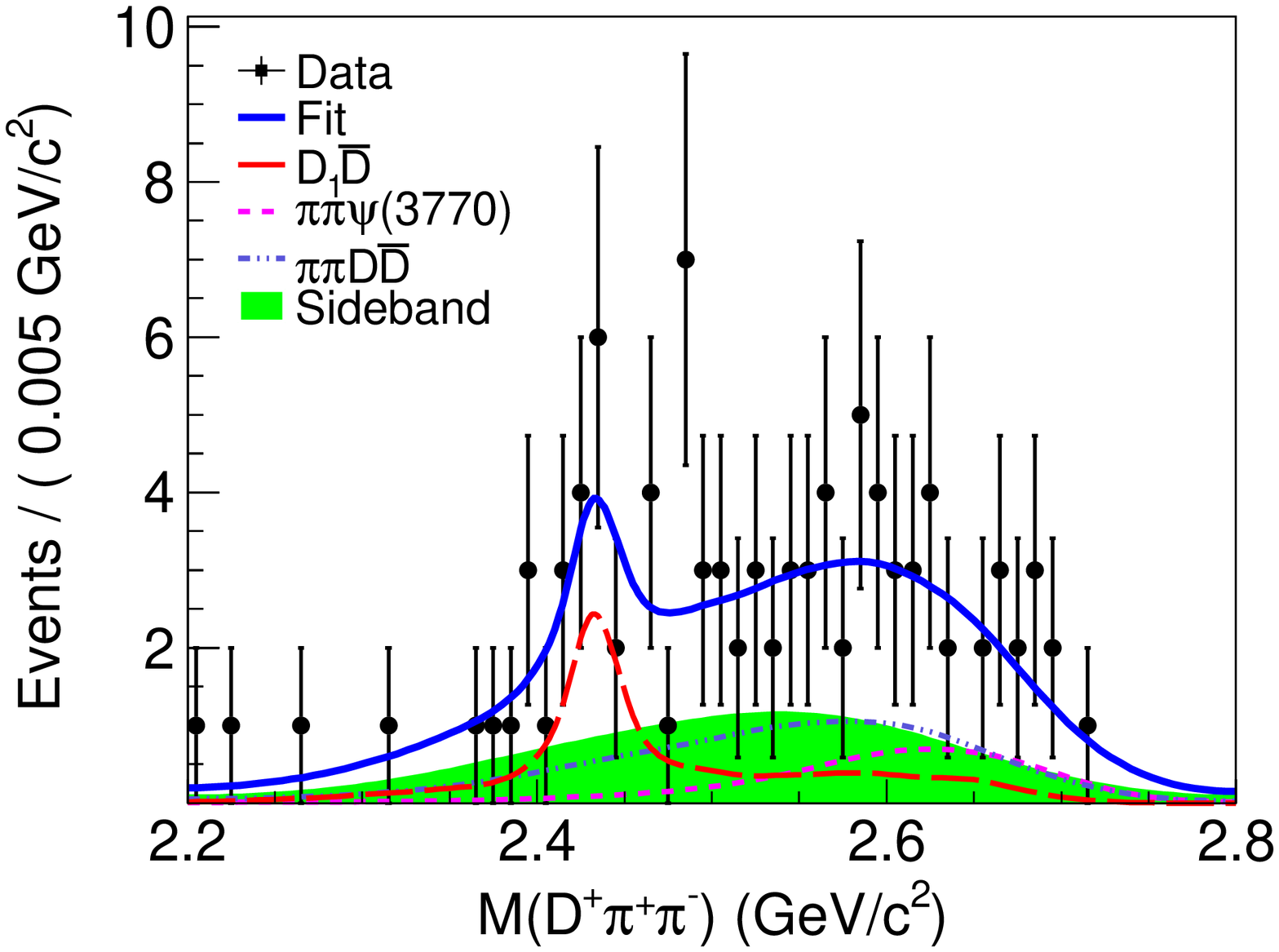}
   \put(75,65){(i)}
   \end{overpic}
\caption{Fit to the $D^{0}\pi^{+}\pi^{-}$
(first row), $D^{*+}\pi^{-}$ (second row) and $D^{+}\pi^{+}\pi^{-}$ (third row) invariant  mass distribution at $\sqrt{s} = $ 4.36 (a, d, g), 4.42 (b, e, h), and 4.60 (c, f, i)~GeV. The black dots with error bars are data, the blue solid curves the fit results, and the red long-dashed lines the $D_{1}(2420)$ signal contributions. The pink dashed lines are the contributions of the final state
$\pp\psi(3770)$, the blue dotted-dashed lines these of the $\pp D\bar{D}$ or $D_{2}^{*}(2460)^{0}\bar{D}$ final state, and the light blue dotted-long-dashed lines the $D^{*}\bar{D}\pi$ background contributions, while the green shaded histograms are the distributions from the sideband regions.}
  \label{FitD1}
\end{figure*}

\begin{table*}[htbp]
\renewcommand\arraystretch{1.18}
 \setlength{\tabcolsep}{3.23mm}{
\caption{Results for the process $e^{+}e^{-}\to \pi^{+}\pi^{-}\psi(3770)$.
Shown in the table are the number of observed events
$N^{\rm obs}$, the integrated luminosity $L_{\rm int}$, the
radiation correction factor $1+\delta^{r}$, the vacuum polarization
correction factor $\frac{1}{|1-\Pi|^{2}}$, the summation over the products of
branching fraction and efficiency $\sum_{i,j}\epsilon_{i,j}
\mathcal{B}_{i}\mathcal{B}_{j}$ (left) for the $\ddn$ and (right) for the $\ddc$ mode, the Born cross section $\sigma^{B}$ and the statistical significance S.
The upper limits are at 90\% C.L.} \label{RCFDDbar}
\begin{tabular}{l c c c c c c c}
\hline \hline
    $\sqrt{s}$ (GeV)      &$N^{\rm obs}$
    &$L_{\rm int}({\rm pb}^{-1})$  &$1+\delta^{r}$
    &$\frac{1}{|1-\Pi|^{2}}$  &$\sum_{i,j}\epsilon_{i,j} \mathcal{B}_{i}\mathcal{B}_{j}$
    &$\sigma^{B}$ (pb) &S\\
    \hline
     4.0854&     $0.0_{-0.0}^{+1.1}$     &~~~~52.6    &0.78 &1.052  &0.0005 $\mid$ 0.0005   &$0.0_{-0.0}^{+57.2}\pm0.0$ ($<$120)   &- \\
     4.1886&     $0.0_{-0.0}^{+1.6}$     &~~~~43.1    &0.84 &1.056  &0.0043 $\mid$ 0.0032   &$0.0_{-0.0}^{+11.9}\pm0.0$ ($<$25)   &- \\
     4.2077&     $0.0_{-0.0}^{+2.2}$     &~~~~54.6    &0.84 &1.057  &0.0047 $\mid$ 0.0036   &$0.0_{-0.0}^{+11.6}\pm0.0$ ($<$24)   &- \\
     4.2171&     $0.0_{-0.0}^{+1.8}$     &~~~~54.1    &0.86 &1.057  &0.0048 $\mid$ 0.0036   &$0.0_{-0.0}^{+9.2}\pm0.0$  ($<$20)   &- \\
     4.2263&     14.3$\pm$8.6    &1047.3      &0.81 &1.056  &0.0049 $\mid$ 0.0037           &3.9$\pm$2.4$\pm$0.6  ($<$7.9)   &1.4$\sigma$ \\
     4.2417&     $0.0_{-0.0}^{+2.0}$     &~~~~55.6    &0.81 &1.056  &0.0050 $\mid$ 0.0038   &$0.0_{-0.0}^{+10.1}\pm0.0$ ($<$21)   &- \\
     4.2580&     30.7$\pm$9.9    &~~825.7     &0.77 &1.054  &0.0050 $\mid$ 0.0038           &11.1$\pm$3.6$\pm$1.7  ($<$16)    &3.2$\sigma$\\
     4.3079&     $0.0_{-0.0}^{+2.7}$     &~~~~44.9    &0.90 &1.052  &0.0047 $\mid$ 0.0036   &$0.0_{-0.0}^{+16.5}\pm0.0$ ($<$36)  &- \\
     4.3583&     68.7$\pm$21.8   &~~539.8     &0.95 &1.051  &0.0039 $\mid$ 0.0029           &39.5$\pm$12.5$\pm$5.9 ($<$59)    &3.3$\sigma$ \\
     4.3874&     14.7$\pm$6.6    &~~~~55.2    &0.93 &1.051  &0.0035 $\mid$ 0.0027           &93.3$\pm$41.9$\pm$14.0 ($<$166)  &2.7$\sigma$\\
     4.4156&     99.2$\pm$21.0   &1028.9      &0.95 &1.053  &0.0025 $\mid$ 0.0020           &46.0$\pm$9.7$\pm$5.9             &5.2$\sigma$\\
     4.4671&     $1.5_{-1.5}^{+3.6}$     &~~109.9     &0.93 &1.055  &0.0025 $\mid$ 0.0019   &6.6$\pm$16.8$\pm$0.8 ($<$39)  &0.1$\sigma$\\
     4.5271&     $0.0_{-0.0}^{+1.4}$     &~~110.0     &0.95 &1.055  &0.0022 $\mid$ 0.0017   &$0.0_{-0.0}^{+6.8}\pm0.0$ ($<$18)  &-\\
     4.5745&     4.4$\pm$2.7     &~~~~47.7    &0.94 &1.055  &0.0020 $\mid$ 0.0016           &54.0$\pm$33.1$\pm$6.9 ($<$123)  &1.4$\sigma$\\
     4.5995&     $3.7_{-3.7}^{+7.9}$     &~~566.9     &0.96 &1.055  &0.0019 $\mid$ 0.0015   &$3.9_{-3.9}^{+8.4}\pm0.5$ ($<$20)  &0.5$\sigma$\\
    \hline \hline
\end{tabular}}
\end{table*}

\begin{table*}[htbp]
\renewcommand\arraystretch{1.18}
\setlength{\tabcolsep}{3.5mm}{
\caption{Results for the reaction channel $e^{+}e^{-}\to \rho^{0}X_{2}(4013)$. For the symbols
see Table~\ref{RCFDDbar}, the penultimate column
is the Born cross section $\sigma^{B}$ times the branching fraction of $X_{2}(4013)\to D\bar{D}$.
The upper limits are at 90\% C.L.} \label{RCFX4013}
\begin{tabular}{c c c c c c c c }
\hline \hline    $\sqrt{s}$ (GeV)  &$N^{\rm obs}$  &$L_{\rm int}({\rm pb}^{-1})$
&$1+\delta^{r}$  &$\frac{1}{|1-\Pi|^{2}}$  &$\sum_{i,j}\epsilon_{i,j}\mathcal{B}_{i}\mathcal{B}_{j}$
&$\sigma^{B}\cdot\mathcal{B}_{X_{2}(4013)\to D\bar{D}}$ (pb)  &S\\
    \hline
     4.3583     & $1.1_{-1.1}^{+1.5}$  &~~539.8        &0.92  &1.051  &0.0018$\mid$0.0017  &$1.2_{-1.2}^{+1.6}\pm0.2$ ($<$4.8) &1.5$\sigma$\\
     4.4156     & $0.0_{-0.0}^{+1.8}$ &1028.9         &0.93  &1.053  &0.0045$\mid$0.0035  &$0.0_{-0.0}^{+0.4}\pm0.0$ ($<$1.0) &-\\
     4.5995     & $2.7_{-2.7}^{+5.3}$ &~~566.9        &0.95  &1.055  &0.0054$\mid$0.0042  &$1.0_{-1.0}^{+1.9}\pm0.2$ ($<$5.5) &0.5$\sigma$\\
    \hline \hline
\end{tabular}}
\end{table*}

\subsection{\boldmath $e^{+}e^{-} \rightarrow D_{1}(2420)\bar{D}$}

After imposing all the requirements above, the  $D\pi^{+}\pi^{-}$  invariant mass distributions are shown in
Fig.~\ref{FitD1}. A peak at around 2.42~GeV/$c^{2}$ can
be seen.

To determine the signal yield for the reaction $e^{+}e^{-}\to
D_{1}(2420)^{0}\bar{D}^{0}$, the $D^0\pp$ invariant  mass distribution is
fitted with the signal shape taken from the MC simulation convolved with
a Gaussian function to take into account the shift of the reconstructed mass to the generated one and the difference in the mass
resolution between data and MC simulation as shown in Fig.~\ref{FitD1}. As background components, the channels
 $\pi^{+}\pi^{-}\psi(3770)$ and
$\pi^{+}\pi^{-} D^{0}\bar{D}^{0}$ are included in the fit as well as a non-$D\bar{D}$ component, which is fixed to the shape and number of events expected from the sideband regions. The numbers of $\pi^{+}\pi^{-}\psi(3770)$ events are fixed to the values calculated using the cross sections we measured (see Table~\ref{RCFX4013}). The yields of the signal events and of the $\pi^{+}\pi^{-} D^{0}\bar{D}^{0}$ events are allowed to vary in the fit. The signal yields at $\sqrt{s}$ = 4.36, 4.42, and 4.60~GeV
are $114.7 \pm 13.8$, $252.3\pm 39.4$, and $43.8\pm 15.1$  events with a
statistical significance of $3.2\sigma$, $7.4\sigma$, and
$3.3\sigma$, respectively.

\begin{table*}[!htbp]
\renewcommand\arraystretch{1.18}
\setlength{\tabcolsep}{3.1mm}{
\caption{Results for the process $e^{+}e^{-}\to
D_{1}(2420)^{0}\bar{D}^{0}$ with $D_{1}(2420)^{0}\rightarrow D^{0}\pi^{+}\pi^{-} + c.c.$ For the symbols see Table~\ref{RCFDDbar}, the penultimate column is the Born cross
section
$\sigma^{B}$ times the branching fraction of $D_{1}(2420)^{0}\to
D^{0}\pi^{+}\pi^{-}$. The upper limits are at 90\% C.L.}
\label{RCFD0D1}
\begin{tabular}{l c c c c c c c}
\hline \hline
    $\sqrt{s}$ (GeV)      &$N^{\rm obs}$  &$L_{\rm int}({\rm pb}^{-1})$  &$1+\delta^{r}$  &$\frac{1}{|1-\Pi|^{2}}$  &$\sum_{i,j}\epsilon_{i,j} \mathcal{B}_{i}\mathcal{B}_{j}$     &$\sigma^{B}\cdot\mathcal{B}_{D_{1}^{0}\to D^{0}\pi^{+}\pi^{-}}$(pb) &S \\
    \hline
     4.3079&     2.4$\pm$1.7      &~~~~44.9   &0.87 &1.052  &0.0049  &11.8$\pm$8.4$\pm$2.4 ($<$31)                           &1.4$\sigma$ \\
     4.3583&     114.7$\pm$13.8    &~~539.8  &0.94 &1.051  &0.0041   &52.2$\pm$6.3$\pm$9.8  ($<$66)                                 &3.2$\sigma$ \\
     4.3874&     15.5$\pm$10.8     &~~~~55.2   &0.95 &1.051  &0.0037 &76.6$\pm$53.4$\pm$14.4 ($<$142)                          &1.3$\sigma$ \\
     4.4156&     252.3$\pm$39.4   &1028.9 &0.94 &1.053  &0.0028      &89.5$\pm$14.0$\pm$12.8                                 &7.4$\sigma$ \\
     4.4671&     6.9$\pm$5.6      &~~109.9  &0.92 &1.055  &0.0028    &23.3$\pm$18.9$\pm$3.3 ($<$52)                           &1.5$\sigma$ \\
     4.5271&     5.6$\pm$2.9      &~~110.0  &0.94 &1.055  &0.0024    &21.3$\pm$11.0$\pm$3.6 ($<$44)                           &1.3$\sigma$ \\
     4.5745&     $2.8_{-2.8}^{+2.3}$     &~~~~47.7   &0.94 &1.055  &0.0023  &$25.7_{-25.7}^{+21.1}\pm4.4$ ($<$80)             &0.7$\sigma$ \\
     4.5995&     43.8$\pm$15.1    &~~566.9  &0.95 &1.055  &0.0022    &35.0$\pm$12.1$\pm$5.9 ($<$56)                          &3.3$\sigma$ \\
    \hline \hline
\end{tabular}}
\end{table*}

A similar fit is performed to the  $D^{*+}\pi^{-}$ invariant mass distribution as shown in Fig.~\ref{FitD1}. The signal shape is taken from the MC simulation in the same way as described above. As
background components, the channels
$D^{*}\bar{D}\pi$ and $D_{2}^{*}(2460)^{0}\bar{D}$ are included in the fit as well as the non-$D\bar{D}$ component taken from the sideband regions. The signal yields at $\sqrt{s}$ = 4.36, 4.42, and 4.60~GeV
are $17.8\pm 9.3$, $22.3\pm 13.2$, and $12.6\pm 7.3$ events with the
statistical significance of $1.6\sigma$, $2.4\sigma$, and
$1.5\sigma$, respectively.

\begin{table*}[htbp]
\renewcommand\arraystretch{1.18}
\setlength{\tabcolsep}{3.3mm}{
\caption{ Results for the reaction channel $e^{+}e^{-}\to
D_{1}(2420)^{0}\bar{D}^{0}$ with $D_{1}(2420)^{0}\rightarrow D^{*+}\pi^{-} + c.c.$ (For symbols see Table~\ref{RCFD0D1}).}
\label{RCFD0starD1}
\begin{tabular}{c c c c c c c c}
\hline \hline
    $\sqrt{s}$(GeV)      &$N^{obs}$ &$L_{\rm int}({\rm pb}^{-1})$  &$1+\delta^{r}$  &$\frac{1}{|1-\Pi|^{2}}$  &$\sum_{i,j}\epsilon_{i,j} \mathcal{B}_{i}\mathcal{B}_{j}$     &$\sigma^{B}\cdot\mathcal{B}_{D_{1}^{0}\to D^{*+}\pi^{-}}$(pb) &S\\
    \hline
     4.3079&    $1.8_{-1.8}^{+1.6}$     &~~~~44.9    &0.92  &1.052  &0.00052  &$80.1_{-80.1}^{+71.2}\pm13.8$ ($<$231)    & 0.8$\sigma$ \\
     4.3583&    17.8$\pm$9.3    &~~539.8     &0.92  &1.051  &0.00059          &57.7$\pm$30.2$\pm$9.3   ($<$107)    & 1.6$\sigma$ \\
     4.3874&    $0.4_{-0.4}^{+3.0}$     &~~~~55.2    &0.94  &1.051  &0.00061  &$12.0_{-12.0}^{+90.2}\pm1.9$  ($<$210)    & 0.5$\sigma$ \\
     4.4156&    22.3$\pm$13.2   &1028.9      &0.93  &1.053  &0.00055          &40.0$\pm$23.7$\pm$6.0   ($<$78)    & 2.4$\sigma$ \\
     4.4671&    $5.1_{-5.1}^{+3.0}$     &~~109.9     &0.92  &1.055  &0.00059  &$80.5_{-80.5}^{+47.4}\pm12.1$  ($<$194)    & 0.9$\sigma$\\
     4.5271&    $0.0_{-0.0}^{+2.2}$     &~~110.0     &0.94  &1.055  &0.00062  &$0.0_{-0.0}^{+32.6}\pm0.0$    ($<$71)    & - \\
     4.5745&    $0.5_{-0.5}^{+2.0}$     &~~~~47.7    &0.94  &1.055  &0.00064  &$16.4_{-16.4}^{+70.0}\pm2.9$   ($<$174)    & 0.1$\sigma$ \\
     4.5995&    12.6$\pm$7.3    &~~566.9     &0.95  &1.055  &0.00065          &34.4$\pm$19.9$\pm$6.0   ($<$70)    & 1.5$\sigma$ \\
    \hline \hline
\end{tabular}}
\end{table*}

To determine the signal yield of $e^{+}e^{-}\to
D_{1}(2420)^{+}D^{-}$, the $D^+\pp$ invariant mass distribution is fitted with a procedure similar to the neutral mode as shown in Fig.~\ref{FitD1}. The
signal yields at $\sqrt{s}$ = 4.36, 4.42, and 4.60~GeV are $68.4\pm
17.3$, $132.8\pm 31.4$, and $17.7\pm 10.2$  events with the statistical
significance of $3.1\sigma$, $3.0\sigma$, and $2.1\sigma$,
respectively.

  The data samples taken at other c.m.\ energies are also studied with the same method. The fits are shown in Figs.~\ref{FitD1_D0}-\ref{FitD1_Dp} in Appendix B. Signal yields are listed in Tables~\ref{RCFD0D1}-\ref{RCFDpD1}.

\begin{table*}[htbp]
\renewcommand\arraystretch{1.18}
\setlength{\tabcolsep}{3.0mm}{
\caption{Results for the reaction channel
$e^{+}e^{-}\to D_{1}(2420)^{+}D^{-}$ with $D_{1}(2420)^{+}\rightarrow D^{+}\pi^{+}\pi^{-} + c.c.$ (For symbols see Table~\ref{RCFD0D1}).}
 \label{RCFDpD1}
\begin{tabular}{c c c c c c c c}
\hline \hline
 $\sqrt{s}$ (GeV)      &$N^{\rm obs}$ &$L_{\rm int}({\rm pb}^{-1})$
 &$1+\delta^{r}$  &$\frac{1}{|1-\Pi|^{2}}$  &$\sum_{i,j}\epsilon_{i,j} \mathcal{B}_{i}\mathcal{B}_{j}$
 &$\sigma^{B}\cdot\mathcal{B}_{D_{1}^{+}\to D^{+}\pi^{+}\pi^{-}}$ (pb) &S \\
    \hline
     4.3079&     $0.0_{-0.0}^{+2.2}$      &~~~~44.9    &0.88  &1.052  &0.0041   &$0.0_{-0.0}^{+13.0}\pm0.0$  ($<$26)         & -     \\
     4.3583&     68.4$\pm$17.3            &~~539.8     &0.95  &1.051  &0.0032   &39.7$\pm$10.0$\pm$7.6 ($<$54)   & 3.1$\sigma$   \\
     4.3874&     $1.4_{-1.4}^{+3.5}$      &~~~~55.2    &0.94  &1.051  &0.0031   &$8.3_{-8.3}^{+20.2}\pm1.6$ ($<$49)       & -     \\
     4.4156&     132.8$\pm$31.4           &1028.9      &0.94  &1.053  &0.0024   &54.0$\pm$12.8$\pm$7.6 ($<$76)   & 3.0$\sigma$   \\
     4.4671&     9.5$\pm$6.8              &~~109.9     &0.92  &1.055  &0.0023   &38.8$\pm$27.8$\pm$5.4 ($<$72)   & 1.7$\sigma$   \\
     4.5271&     2.3$\pm$1.9              &~~110.0     &0.94  &1.055  &0.0020   &10.3$\pm$8.5$\pm$1.6  ($<$29)   & 1.0$\sigma$   \\
     4.5745&     4.8$\pm$2.7              &~~~~47.7    &0.94  &1.055  &0.0020   &51.4$\pm$28.9$\pm$8.0 ($<$110)  & 2.0$\sigma$   \\
     4.5995&     17.7$\pm$10.2            &~~566.9     &0.94  &1.055  &0.0017   &18.9$\pm$10.9$\pm$2.9 ($<$36)   & 2.1$\sigma$   \\
    \hline \hline
\end{tabular}}
\end{table*}

\section{CROSS SECTION RESULTS}

\begin{figure*}[htbp]
  \centering
  \begin{overpic}[width=0.45\textwidth]{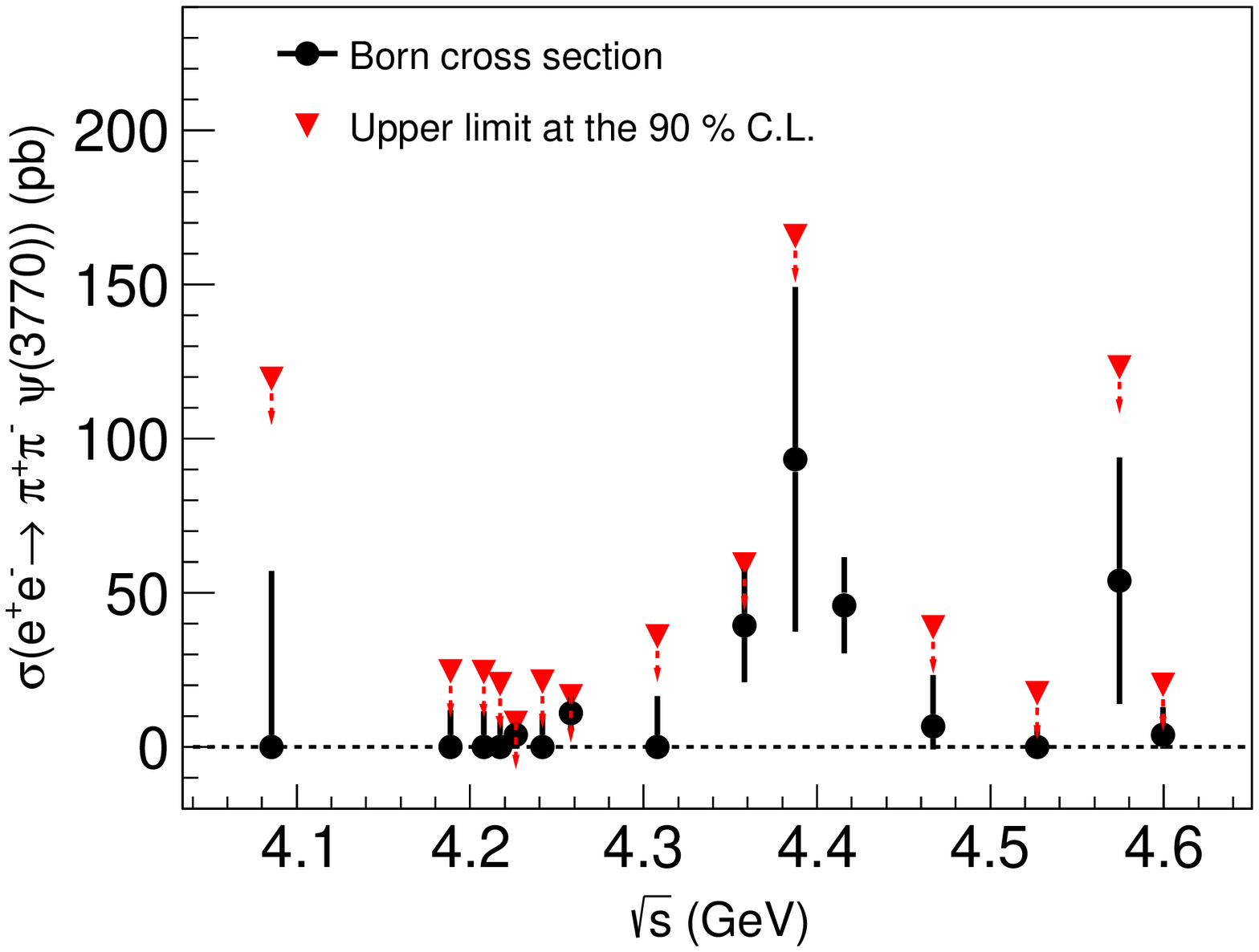}
   \put(83,65){(a)}
   \end{overpic}
   \begin{overpic}[width=0.45\textwidth]{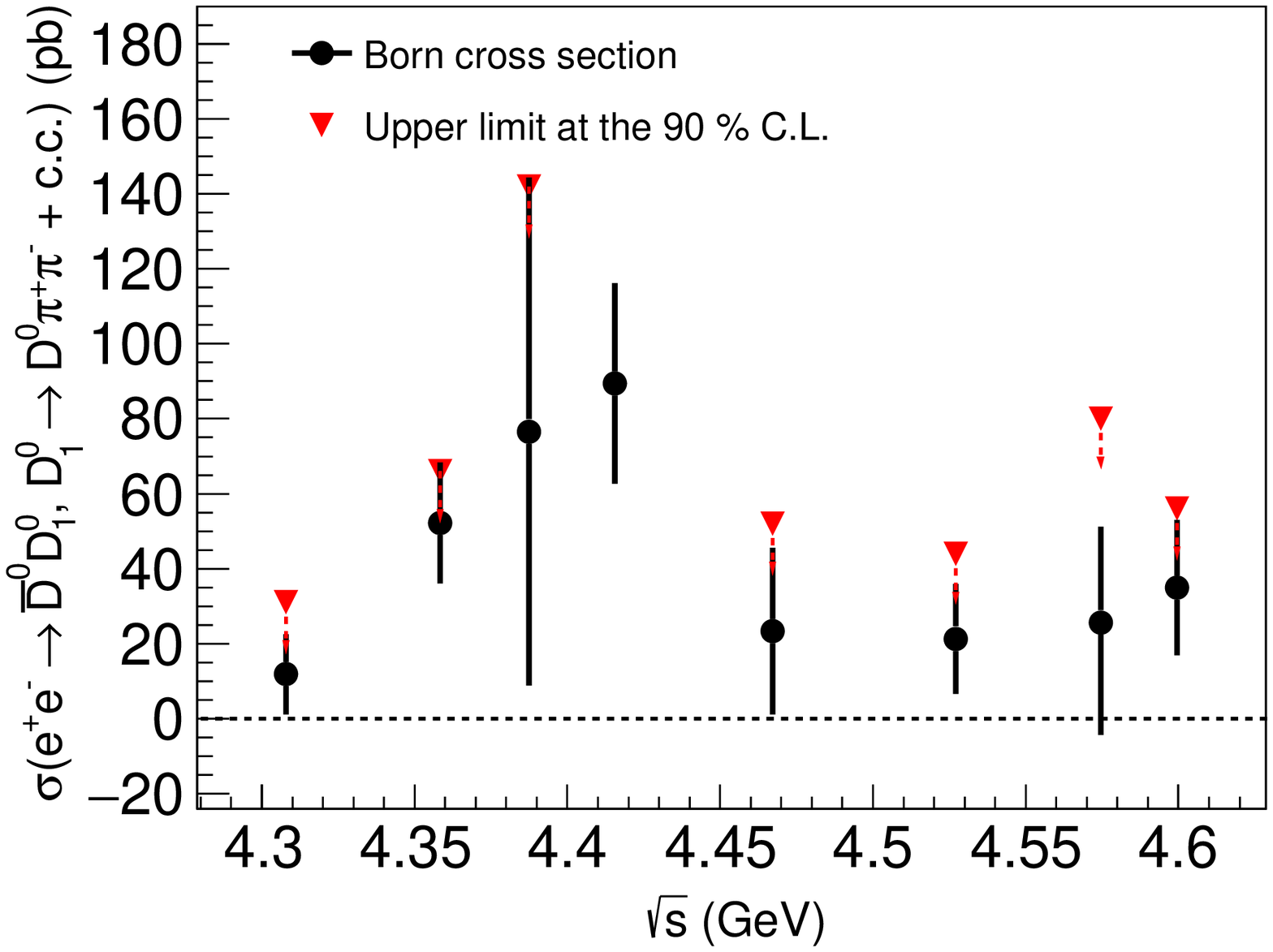}
   \put(83,65){(b)}
   \end{overpic}
   \begin{overpic}[width=0.45\textwidth]{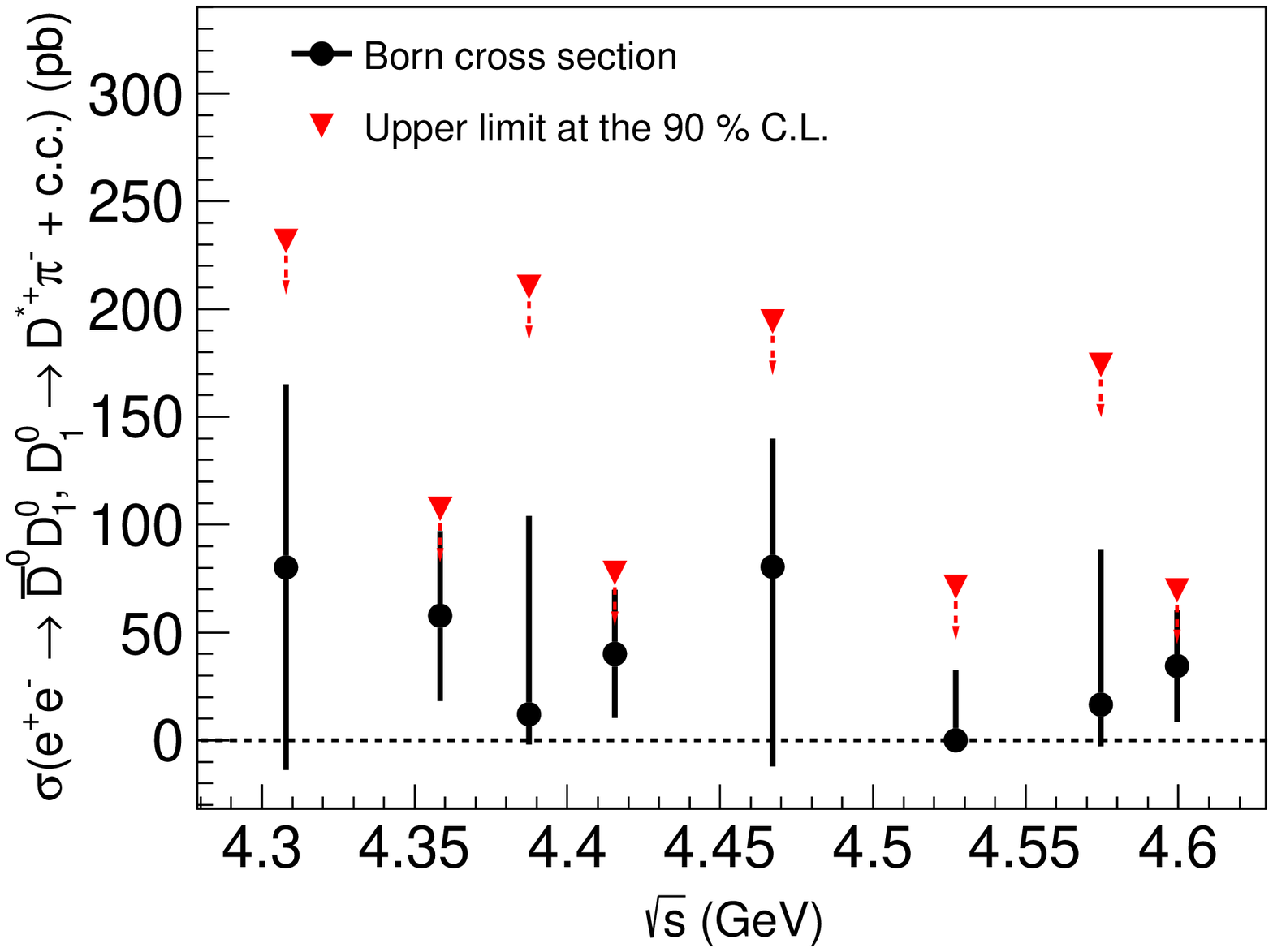}
   \put(83,65){(c)}
   \end{overpic}
   \begin{overpic}[width=0.45\textwidth]{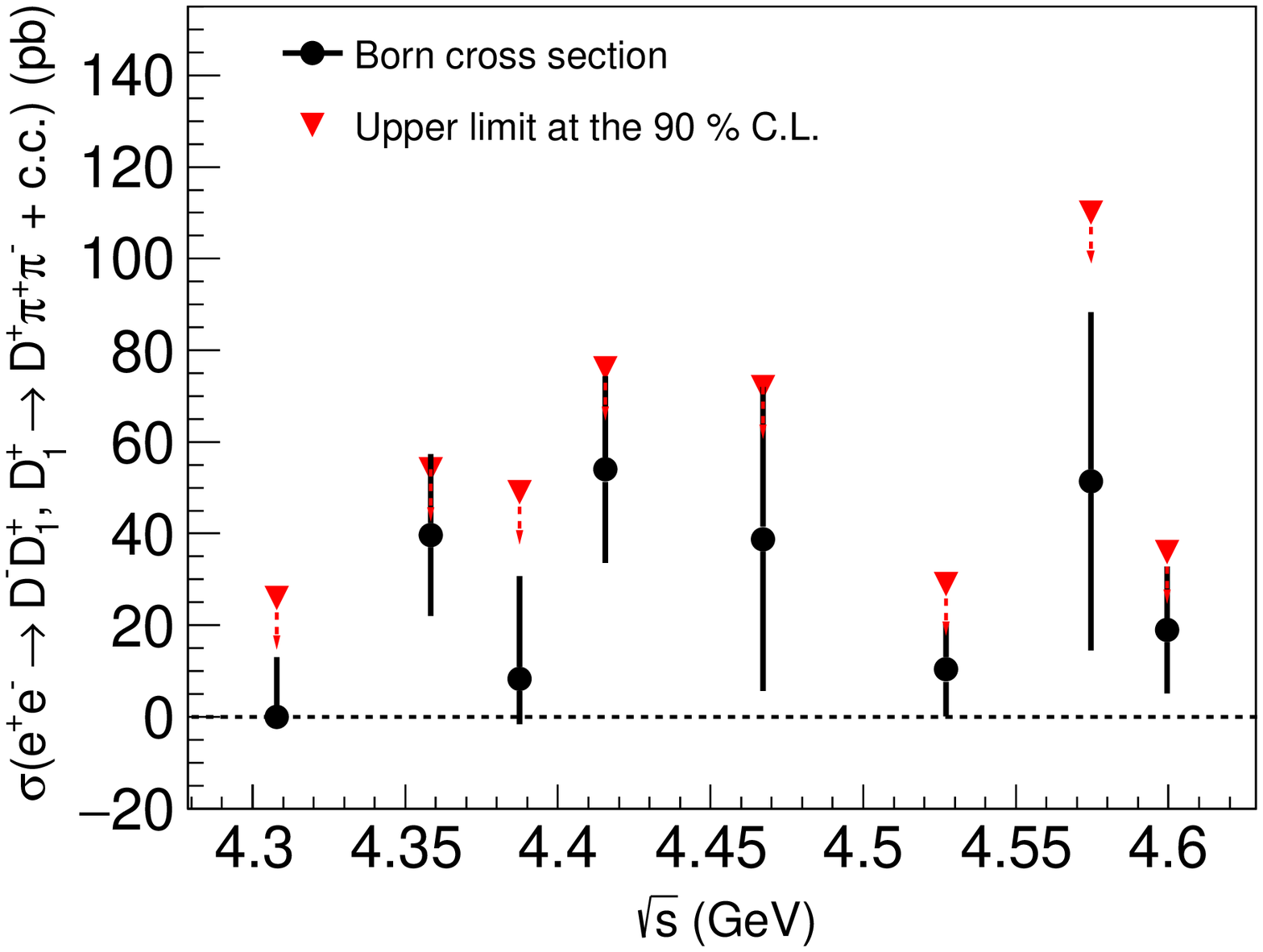}
   \put(83,65){(d)}
   \end{overpic}
\caption{Born cross sections of the processes (black dots) $e^{+}e^{-}\to
\pi^{+}\pi^{-}\psi(3770)$ (a), $e^{+}e^{-}\to
D_{1}(2420)^{0}\bar{D}^{0}\to \pp\ddn$ (b), $e^{+}e^{-}\to
D_{1}(2420)^{0}\bar{D}^{0}\to D^{*+}\bar{D}^{0}\pi^{-} \to \pp\ddn$ (c), and
 $e^{+}e^{-}\to D_{1}(2420)^{+}D^{-}\to \pp\ddc$ (d). The error bars include the statistical and systematic uncertainties. The red triangles are the upper limits on the Born cross sections at the 90 \% C.L.}
\label{Cpipipsi3770}
\end{figure*}

The Born cross section of  $e^{+}e^{-}\to \pi^{+}\pi^{-}\psi(3770)$
is calculated with
\begin{linenomath*}
\begin{equation}
  \sigma^{B} = \frac{N^{\rm obs}}{\mathcal{L}_{\rm int}
 f_{r} f_{v} (\mathcal{B}_{N}
 \sum_{i,j}\epsilon_{i,j} \mathcal{B}_{i}\mathcal{B}_{j}+
 \mathcal{B}_{C} \sum_{k,l}\epsilon_{k,l}\mathcal{B}_{k}\mathcal{B}_{l})},
 \label{xspspp}
\end{equation}
\end{linenomath*}
\noindent where $N^{\rm obs}$ is the number of observed events,
$\mathcal{L}_{\rm int}$ the integrated luminosity and
$\epsilon_{i,j}$ the selection efficiency for $e^{+}e^{-} \to
\pi^{+}\pi^{-}\psi(3770)$, $\psi(3770)\to D^{0}\bar{D}^{0}$,
$D^{0}\to i$, $\bar{D}^{0}\to j$. $\mathcal{B}_{N}$
and $\mathcal{B}_{C}$ are the branching
fractions for $\pspp\to \ddn$ and $\pspp\to \ddc$ and $\mathcal{B}_{i}$ ($\mathcal{B}_{j}$) is the
branching fraction for $D^{0}\to i$ ($\bar{D}^{0}\to j$) taken from PDG~\cite{pdg}. The same
applies to $\epsilon_{k,l}$ and $\mathcal{B}_{k}$
($\mathcal{B}_{l}$) for charged mode.
$f_{v}=\frac{1}{|1-\Pi|^{2}}$ is the vacuum polarization factor~\cite{vpcalculation}
and $f_{r}=(1+\delta^{r})$ is the radiative correction factor which is defined as
\begin{linenomath*}
\begin{equation}
    (1+\delta^{r})= \frac{\sigma^{\rm obs}}{\sigma^{\rm dressed}} = \frac{\int \,\sigma^{\rm dressed}(s(1-x))F(x,s)d x}{\sigma^{\rm dressed}(s)}.
\end{equation}
\end{linenomath*}
 \noindent $F(x,s)$ is the radiator function, which is calculated from QED~\cite{QEDcalculation} with an accuracy of 0.1\%, $x\equiv 2E^{*}_{\gamma}/\sqrt{s}=1-m^{2}/s$, where $E^{*}_{\gamma}$ is the ISR photon energy and m is the invariant mass of the final state after radiating the photon. $\sigma^{\rm dressed}(s)$ is the energy dependent dressed cross section of $e^{+}e^{-}\to \pi^{+}\pi^{-}\psi(3770)$.
 Here the observed signal events are assumed to originate from the $Y(4260)$ resonance to obtain
the efficiency and the ISR correction factor. Then the measured line shape is used as input to calculate the efficiency
and ISR correction factor again. This procedure is repeated until the difference between two subsequent iterations is
comparable with the statistical uncertainties.
 The Born cross sections are listed in Table~\ref{RCFDDbar} and shown in Fig.~\ref{Cpipipsi3770}.
 At the energy points where no significant $\psi(3770)$ signal yields are observed (significance $<$ 5$\sigma$) the upper limits on the cross sections  at the $90\%$ confidence level (C.L.) are calculated using the Bayesian method~\cite{pdg} with a flat prior. By fitting the  $D\bar{D}$ invariant mass distribution with fixed values for the signal yield, we obtain a scan of the likelihood distribution as a function of the cross section. To take all systematic uncertainties into consideration we convolve the likelihood distribution with a Gaussian function with a width corresponding to the total systematic uncertainty.  The upper limit on $\sigma$ at the 90\% C.L.\ is obtained from
\begin{linenomath*}
\begin{equation}
\int^{\sigma}_{0} L(x)dx/\int^{\infty}_{0} L(x)dx = 0.9  .
\end{equation}
\end{linenomath*}
\noindent  All upper limits on the cross sections are also listed in Table~\ref{RCFDDbar}.

For the reaction channel $e^{+}e^{-}\to \rho^{0}X_{2}(4013)$ the upper limit of
the product of the Born cross section and the branching fraction
to $\ddb$  is measured, assuming $\mathcal{B}_{X_{2}(4013)\to D^{0}\bar{D}^{0}} =
\mathcal{B}_{X_{2}(4013)\to D^{+}D^{-}} = 0.5 \times \,\mathcal{B}_{X_{2}(4013)\to
D\bar{D}}$. The efficiency and ISR correction factor in Eq.~(\ref{xspspp}) is
taken from the MC sample with the $X_{2}(4013)$ resonance as intermediate state and the cross-section following the $Y(4260)$ line shape.
The upper limits on the cross sections at $90\%$ C.L. are estimated using the same method as described above.
All results and upper limits are listed in Table~\ref{RCFX4013}.

For the reaction channel $e^{+}e^{-}\to D_{1}(2420)\bar{D}$ with $D_{1}(2420)\to
X(D\pi^{+}\pi^{-}$ or $D^{*}\pi)$, the product of the Born cross section times the
branching fraction of $D_{1}(2420)\to X$ is calculated using
\begin{linenomath*}
\begin{equation}
  \sigma^{B}\times \mathcal{B}_{D_{1}(2420)\to X} =
  \frac{N^{\rm obs}}{\mathcal{L}_{\rm int} f_{v} f_{r}
 \sum_{i,j}\epsilon_{i,j} \mathcal{B}_{i} \mathcal{B}_{j}},
\end{equation}
\end{linenomath*}
\noindent where $\epsilon_{i,j}$ is the selection efficiency for each process $e^{+}e^{-} \to
D_{1}(2420)\bar{D}$ ($D_{1}(2420)\to D\pi\pi/D^{*}\pi$, $D\to i$,
$\bar{D}\to j$). The low momentum of the $\pi$ meson from $D^{*}\rightarrow D \pi$  decay reduces the efficiency for the decay channel $D_{1}(2420)\to D^{*}\pi$ in comparison to $D_{1}(2420)\to D\pi\pi$.  The other variables are the same as defined in
Eq.~(\ref{xspspp}). For the $D_{1}(2420)\to D^{*}\pi$ channel, the branching fraction $\mathcal{B}_{D^{*} \to D \pi}$ is taken into account.   The cross sections as a function of c.m.\ energy are shown in Fig.~\ref{Cpipipsi3770}.
At the energy points where no significant $D_{1}$ signals are observed (significance $< 5\sigma$), the upper limits on the cross sections at the $90\%$ C.L.\ are estimated using the same method as described above.  All numbers are shown in the Tables~\ref{RCFD0D1},  ~\ref{RCFD0starD1}, and~\ref{RCFDpD1} for the neutral
and the charged modes, respectively.

\section{Systematic uncertainty estimation}

The systematic uncertainties in the cross section measurements
mainly stem from the integrated luminosity, the tracking and photon
detection efficiency, various intermediate branching
fractions, the ISR correction factor, the signal and
background shapes, the fit range, the signal region of double tag of the $D$ mesons, and the cross section of the $\pi^{+}\pi^{-}\psi(3770)$ final state. The estimation of the systematic uncertainties is described in the following and the results at $\sqrt{s}$ = 4.36, 4.42,
and 4.60~GeV are listed in
Tables~\ref{sys_err_psipp}-\ref{sys_err_Dp}. The results for all other energy points are listed in Appendix C.

\begin{table}[htbp]
\caption{Relative systematic uncertainties (in \%)
on the $\sigma(e^{+}e^{-}\to \pi^{+}\pi^{-}\psi(3770))$ measurement.}
\label{sys_err_psipp}
\begin{tabular}{l r r r}
\hline \hline
     Sources / $\sqrt{s}$ (GeV)              &4.3583     &4.4156     &4.5995\\
    \hline
    Integrated luminosity                       & 1.0     & 1.0    & 1.0  \\
    Efficiency related               &8.9      &8.8     &8.6   \\
    Radiative correction             &0.6      &2.1     &0.2  \\
    Signal shape                      &3.4      &3.8     &3.8   \\
    Background shape                 &9.9      &7.6     &7.6   \\
    Fit range                        &2.0      &2.4     &2.4   \\
    Signal region of double tag      &5.3      &2.0     &2.0   \\
    \hline
    Total                            &14.9     &12.8     &12.5      \\
    \hline
    \hline
\end{tabular}
\end{table}

\begin{table}[htbp]
\caption{Relative systematic uncertainties (in \%) on the
$\sigma(e^{+}e^{-}\to \rho^{0}X_{2}(4013)\to \pp\ddb$) measurement.}
\label{sys_err_X4013}
\begin{tabular}{l r r r}
\hline \hline
     Sources / $\sqrt{s}$ (GeV)                &4.3583   &4.4156   &4.5995\\
     \hline
    Integrated luminosity                       & 1.0   & 1.0    & 1.0 \\
    Efficiency related               &9.0    &9.0     &9.0	\\
    Radiative correction             &5.6    &14.5    &5.9  \\
    Signal shape                      &9.9    &0.0       &11.5 \\
    Background shape                 &3.2    &0.0       &4.6  \\
    Fit range                        &4.5    &0.0       &4.4  \\
    Signal region of double tag      &3.3    &0.0       &12.1  \\
    \hline
    Total                            &15.9   &17.1  &20.9  \\
    \hline
    \hline
\end{tabular}
\end{table}

\begin{table}[htbp]
\caption{Relative systematic uncertainties (in \%) on the $\sigma(e^{+}e^{-}\to D_{1}(2420)^{0}\bar{D}^{0}, D_{1}(2420)^{0}\to D^{0}\pi^{+}\pi^{-} + c.c.$) measurement.}
\label{sys_err_D0}
\begin{tabular}{l  r  r  r}
\hline \hline
   Sources / $\sqrt{s}$ (GeV)             &4.3583     &4.4156     &4.5995\\
    \hline
    Integrated luminosity                    & 1.0   & 1.0   & 1.0 \\
    Efficiency related            &4.9    &5.0    &4.9   \\
    Radiative correction          &0.6    &2.1    &0.2   \\
    Signal shape                  &7.2    &5.5    &8.2   \\
    Background shape              &3.2    &1.2    &7.4   \\
    Fit range                     &0.9    &1.0    &0.9   \\
    Signal region of double tag   &2.6    &4.2    &4.8   \\
    $\sigma(\pi\pi\psi(3770))$    &11.4   &0.5    &1.1   \\
    \hline
    Total                         &18.8   &14.3   &17.0   \\
    \hline
    \hline
\end{tabular}
\end{table}

\begin{table}[htbp]
\caption{Relative systematic uncertainties (in \%) on the $\sigma(
e^{+}e^{-}\to D_{1}(2420)^{0}\bar{D}^{0}, D_{1}(2420)^{0}\to D^{*+}\pi^{-} + c.c.$) measurement.}
\label{sys_err_D0star}
\begin{tabular}{l  r  r  r}
\hline \hline
   Sources / $\sqrt{s}$ (GeV)       &4.3583     &4.4156    &4.5995\\
    \hline
    Integrated luminosity                     & 1.0   & 1.0  & 1.0 \\
    Efficiency related            &4.9    &5.0   &4.9 \\
    Radiative correction          &0.6    &2.1   &0.2   \\
    Signal shape                   &5.3    &2.8   &4.9  \\
    Background shape              &0.2    &1.8   &0.1  \\
    Fit range                     &2.7    &3.9   &0.8   \\
    Signal region of double tag   &4.9    &2.6   &9.8  \\
    \hline
    Total                         &16.0   &15.1   &17.5    \\
    \hline
    \hline
\end{tabular}
\end{table}

\begin{table}[htbp]
\caption{Relative systematic uncertainties (in $\%$) on the $\sigma(
e^{+}e^{-}\to D_{1}(2420)^{+}D^{-}, D_{1}(2420)^{+}\to D^{+}\pi^{+}\pi^{-} + c.c.$) measurement.}
\label{sys_err_Dp}
\begin{tabular}{l  r r r}
\hline \hline
    Sources / $\sqrt{s}$ (GeV)   &4.3583     &4.4156     &4.5995 \\
    \hline
    Integrated luminosity          &1.0     &1.0     &1.0 \\
    Efficiency related             &13.4    &13.2    &13.0\\
    Radiative correction           &0.2     &1.0     &1.5\\
    Signal shape                   &7.4     &4.2     &4.1 \\
    Background shape               &1.6     &0.8     &3.8  \\
    Fit range                      &1.7     &0.9     &1.7  \\
    Signal region of double tag    &1.1     &1.3     &5.2\\
    $\sigma(\pi\pi\psi(3770))$     &11.1    &0.3     &1.7   \\
    \hline
    Total                          &19.1     &14.0     &15.4       \\
    \hline
    \hline
\end{tabular}
\end{table}

(a) The uncertainty from the integrated luminosity measurement
using Bhabha ($e^{+}e^{-}\to e^{+}e^{-}$) scattering events is
estimated to be 1.0\%~\cite{luminosity}.

(b) The systematic uncertainty from the efficiency includes the
uncertainties from MC statistics, particle identification,
charged track, photon, $\pi^{0}$, and $K_{S}^{0}$ reconstruction,
as well as the branching fractions of the various $D$ decays. The reconstruction
uncertainty for each charged track is 1\%~\cite{Tracking}.
The uncertainty from the photon reconstruction is
1\% per photon~\cite{Photon}, and the uncertainty from the $\pi^{0}$
reconstruction is 1\% per $\pi^{0}$~\cite{Photon}. The uncertainty from
the $K^{0}_{S}$ reconstruction is 4\% per $K^{0}_{S}$~\cite{K_S0}. The uncertainty from the particle identification is 1\% per track~\cite{Tracking}.
The systematic uncertainty for the branching fraction $\mathcal{B}(D^{*+}\to D^{0}\pi^{+})$ is 0.74\%. Those for $\mathcal{B}(D^{0}\to K^{-}\pi^{+})$,
$\mathcal{B}(D^{0}\to K^{-}\pi^{+}\pi^{0})$, $\mathcal{B}(D^{0}\to
K^{-}\pi^{+}\pi^{+}\pi^{-})$, and $\mathcal{B}(D^{0}\to
K^{-}\pi^{+}\pi^{+}\pi^{-}\pi^{0})$ are 1.02\%, 5.60\%,
2.85\%, and 9.52\%, respectively, and those for
$\mathcal{B}(D^{+}\to K^{-}\pi^{+}\pi^{+})$, $\mathcal{B}(D^{+}\to
K^{-}\pi^{+}\pi^{+}\pi^{0})$, $\mathcal{B}(D^{+}\to
K^{0}_{S}\pi^{+})$, $\mathcal{B}(D^{+}\to
K^{0}_{S}\pi^{+}\pi^{0})$, $\mathcal{B}(D^{+}\to
K^{0}_{S}\pi^{+}\pi^{-}\pi^{+})$ are 2.54\%, 2.61\%,
3.92\%, 2.35\%, and 2.95\%, respectively~\cite{pdg}. The total efficiency related systematic uncertainty is the
 sum of all these individual uncertainties in quadrature.

(c) ISR photons are simulated by using the {\sc kkmc} package. The shape of the  c.m.\ energy dependent cross section
affects the radiative correction factor
and the reconstruction efficiency. For the reactions $e^{+}e^{-}\to \pi^{+}\pi^{-}\psi(3770)$
and $e^{+}e^{-}\to D_{1}(2420)\bar{D}$,  the difference between the last two iterations  is
taken as the systematic uncertainty. Since we have no knowledge on the production cross section for the reaction $e^{+}e^{-}\to
\rho^{0}X_{2}(4013)$, we assume that the cross section of $\rho^{0}X_{2}(4013)$ follows the
$Y(4260)$ or the $\psi(4415)$ line shape. The difference between
these two assumptions is taken as the systematic uncertainty.

(d) For the determination of the systematic uncertainty caused by the signal shape, additional MC samples are produced by varying
the width of the signal resonance by one standard
deviation of its world average value~\cite{pdg}. The largest difference of the cross section compared
with the nominal value is taken as the systematic uncertainty of
the signal shape.

(e) The systematic uncertainty caused by the background shape, which is taken from
MC simulation of the final states $\pi^{+}\pi^{-}\psi(3770)$, $D_{1}(2420)\bar{D}$ and $D_{1}(2460)\bar{D}$, is estimated by generating alternative MC samples where the width of the $\psi(3770)$, $D_{1}(2420)$ and $D_{1}(2460)$ resonances is changed by one
standard deviation of the world average value~\cite{pdg}. The largest difference of the cross section compared with the nominal fit value
is taken as the systematic uncertainty of the background shape. The systematic uncertainty originating from the sideband selection is estimated by
changing the sideband windows by 10~MeV/$c^2$. The largest difference of the cross section compared with
the nominal mass window is taken as systematic
uncertainty. For $e^{+}e^{-}\to \rho^{0}X_{2}(4013)$,  the
background shape is changed from a third order polynomial to a fourth order polynomial, and the
difference is taken as the systematic uncertainty of background shape.

(f) The systematic uncertainty caused by the choice of the fit range is estimated
by varying the limits of the fit range by 20~MeV/$c^{2}$. The largest difference of the cross section from the nominal value is taken as systematic uncertainty.

(g) In order to estimate the systematic uncertainty due to the
selection of the signal window for the double $D$-tag method,  the whole
analysis is repeated by changing the signal region from
$|\Delta{M}|<35$~MeV/$c^2$, $-6<\Delta\hat{M}<10$~MeV$/c^{2}$ to
$|\Delta{M}|<39$~MeV$/c^2$, $-8<\Delta\hat{M}<12$~MeV/$c^{2}$ for
the $D^{0}\bar{D}^{0}$ mode and from $|\Delta{M}|<25$~MeV/$c^2$,
$-5<\Delta\hat{M}<10$~MeV/$c^{2}$ to $|\Delta{M}|<29$~MeV/$c^2$,
$-7<\Delta\hat{M}<12$~MeV$/c^{2}$ for the $D^{+}D^{-}$ mode. The
difference of the cross section from the nominal value is taken as systematic uncertainty.

(h) The systematic uncertainty caused by the fixed number of $\pi^{+}\pi^{-}\psi(3770)$ events
in the fit of $M(D\pi^{+}\pi^{-})$ is estimated by varying the fixed number by one standard deviation.
The largest deviation from the nominal cross section is taken as systematic uncertainty.

Tables~\ref{sys_err_psipp} to \ref{sys_err_Dp} summarize all the systematic uncertainties.
The overall systematic uncertainty for each process and c.m.\ energy is obtained by summing up all
sources of systematic uncertainties in quadrature, assuming
they are uncorrelated.

\section{Summary and discussion}

In this analysis, the processes $\EE\to \pp\ddb$ are studied by using the data samples collected at $\sqrt{s}$ = 4.09, 4.19, 4.21, 4.22,
4.23, 4.245, 4.26, 4.31, 4.36, 4.39, 4.42, 4.47, 4.53, 4.575, and
4.60~GeV.

We observe the process $e^{+}e^{-}\to \pi^{+}\pi^{-}\psi(3770)$  for the first time with a statistical significance of 5.2$\sigma$ at $\sqrt{s}$ = 4.42 GeV and see evidence for this process with a statistical significance of 3.2$\sigma$ and 3.3$\sigma$ at $\sqrt{s}$ = 4.26 and 4.36 GeV, respectively. However,  no
evidence for the $\psi(1^{3}D_{3})$ state is found. The Born cross section of
$e^{+}e^{-}\to \pi^{+}\pi^{-}\psi(3770)$ is measured as shown in
Fig.~\ref{Cpipipsi3770}. It can be compared with the
cross section of the process $\EE\to \pp\psi(1^3D_2)$~\cite{X3823}. If we take
 $\BR(\psi(1^3D_2)\to \gamma\chi_{c1})\approx
 \frac{250~{\rm keV}}{390~{\rm keV}}\approx 0.64$~\cite{BranchX3823},
the Born cross section of $\EE\to \pp\psi(1^3D_1)$ is an order of
magnitude larger than that of $\EE\to \pp\psi(1^3D_2)$ at the
same c.m.\ energies~\cite{X3823}. The $e^{+}e^{-}\to \pp\psi(3770)$
line shape looks similar to that of $\EE\to
\pp\psi(1^3D_2)$~\cite{X3823}. Whether the events are from the
$Y(4390)$ or the $\psi(4415)$ resonance or from any other resonance cannot be distinguished
based on the current statistics. For the data points with enough statistics for the $\EE\to \pp\pspp$ final state, no
significant structure (\emph{i.e.} a possible $Z_c$ state) is observed in the $\pi^{\pm}\pspp$ system.

We also search for the state $X_{2}(4013)$, the proposed heavy-quark-spin-symmetry partner of the
$X(3872)$, by analyzing the process $e^{+}e^{-} \to \rho^{0} X_{2}(4013)$  with $X_{2}(4013) \to D\bar{D}$. No significant
signal for the $X_{2}(4013)$  is observed in any data sample. The upper
limit (at the
90\% C.L.) of $\sigma(\EE \to \rho^{0} X_{2}(4013))\cdot
\mathcal{B}(X_{2}(4013) \to \ddb)$ is estimated as 5.0, 1.0, and
5.1~pb at $\sqrt{s}$ = 4.36, 4.42, and 4.60~GeV, respectively.

The process $e^{+}e^{-} \to D_{1}(2420)^{0} \bar{D}^{0}$, $D_{1}(2420)^{0} \to D^{0} \pi^{+} \pi^{-}$ is observed for the first time with a statistical significance  of 7.4$\sigma$ at $\sqrt{s}$ = 4.42 GeV, and we see evidence for this process with  statistical significance of 3.2$\sigma$ and 3.3$\sigma$ at $\sqrt{s}$ = 4.36 and 4.60 GeV, respectively. There is also evidence for the process $e^{+}e^{-} \to D_{1}(2420)^{+} D^{-}, D_{1}(2420)^{+} \to D^{+} \pi^{+} \pi^{-}$  with  statistical significance of 3.1$\sigma$ and 3.0$\sigma$ at $\sqrt{s}$ = 4.36 and 4.42 GeV, respectively.  There is no evidence for the process $\EE\to D_{1}(2420)^{0}\bar{D}^{0}, D_{1}(2420)^{0} \to D^{*+} \pi^{-}$.  The Born cross sections of  $e^{+}e^{-}\to D_{1}(2420)^{0}\bar{D}^{0}, D_{1}(2420)^{0}\to D^{0} \pi^{+} \pi^{-} / D_{1}(2420)^{0} \to D^{*+}\pi^{-}$  and $e^{+}e^{-} \to D_{1}(2420)^{+} D^{-}, D_{1}(2420)^{+} \to D^{+} \pi^{+} \pi^{-}$ are measured at
$\sqrt{s}$ = 4.31, 4.36, 4.39, 4.42, 4.47, 4.53, 4.575, and
4.60~GeV as shown in Fig.~\ref{Cpipipsi3770}. No fast rise of the cross section above the $D_{1}(2420)\bar{D}$
threshold is visible as indicated in Ref.~\cite{Y4260asDD1}, whose point is also disfavored by Ref.~\cite{hanqing}. There is no other  obvious structure visible either.

\acknowledgments

The BESIII collaboration thanks the staff of BEPCII and the IHEP computing center for their strong support. This work is supported in part by National Key Basic Research Program of China under Contract No. 2015CB856700; National Natural Science Foundation of China (NSFC) under Contracts Nos. 11625523, 11635010, 11735014; National Natural Science Foundation of China (NSFC) under Contract No. 11835012; the Chinese Academy of Sciences (CAS) Large-Scale Scientific Facility Program; Joint Large-Scale Scientific Facility Funds of the NSFC and CAS under Contracts Nos. U1532257, U1532258, U1732263, U1832207; CAS Key Research Program of Frontier Sciences under Contracts Nos. QYZDJ-SSW-SLH003, QYZDJ-SSW-SLH040; 100 Talents Program of CAS; INPAC and Shanghai Key Laboratory for Particle Physics and Cosmology; German Research Foundation DFG under Contract No. Collaborative Research Center CRC 1044; Istituto Nazionale di Fisica Nucleare, Italy; Koninklijke Nederlandse Akademie van Wetenschappen (KNAW) under Contract No. 530-4CDP03; Ministry of Development of Turkey under Contract No. DPT2006K-120470; National Science and Technology fund; The Knut and Alice Wallenberg Foundation (Sweden) under Contract No. 2016.0157; The Swedish Research Council; U. S. Department of Energy under Contracts Nos. DE-FG02-05ER41374, DE-SC-0010118, DE-SC-0012069; University of Groningen (RuG) and the Helmholtzzentrum fuer Schwerionenforschung GmbH (GSI), Darmstadt.

\onecolumngrid

\begin{appendix}

\section{Fits to the  $D\bar{D}$ invariant mass distributions }

\begin{figure}[H]
  \centering
   \begin{overpic}[width=0.329\textwidth]{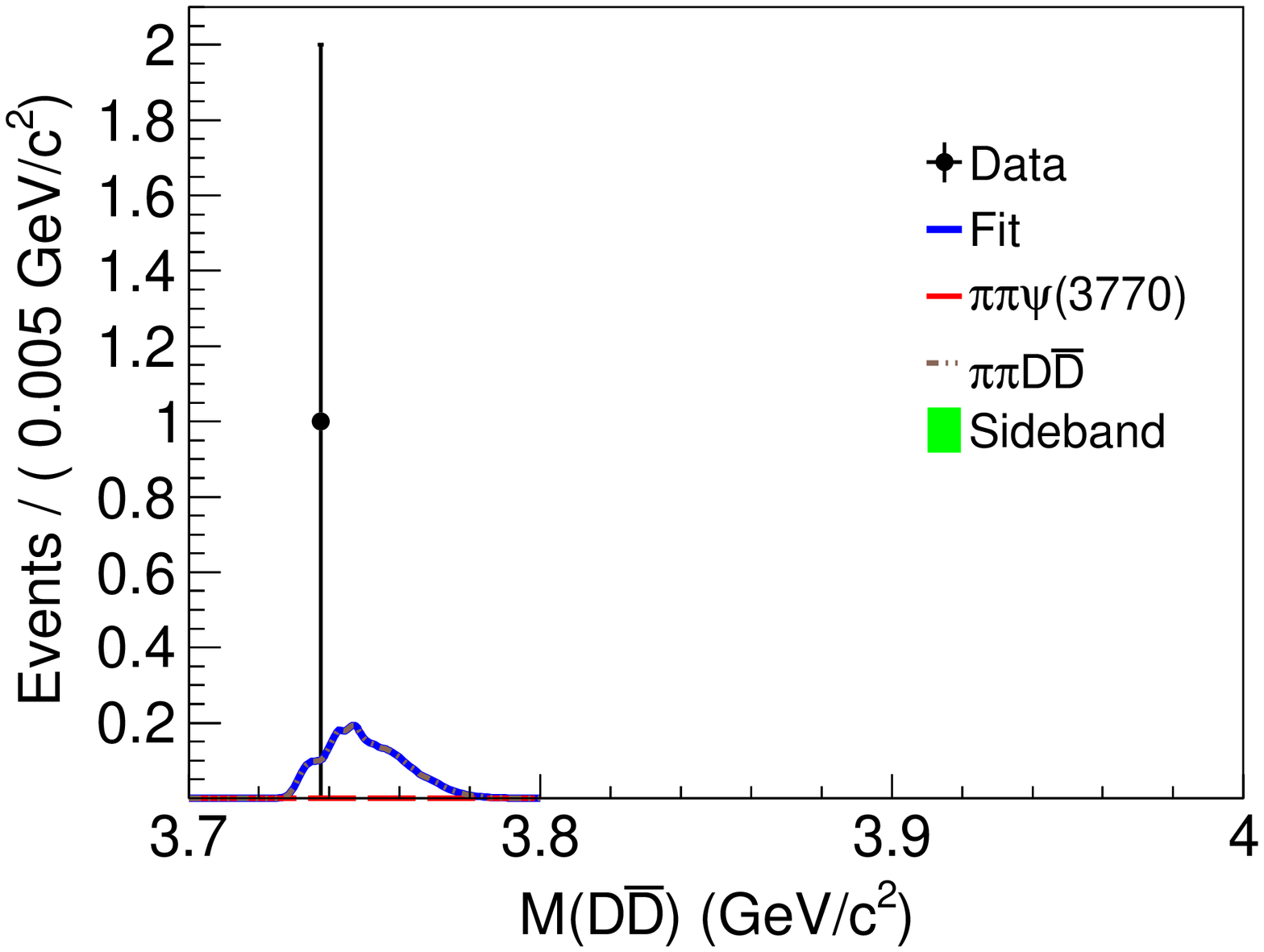}
   \put(80,64){(a)}
   \end{overpic}
   \begin{overpic}[width=0.329\textwidth]{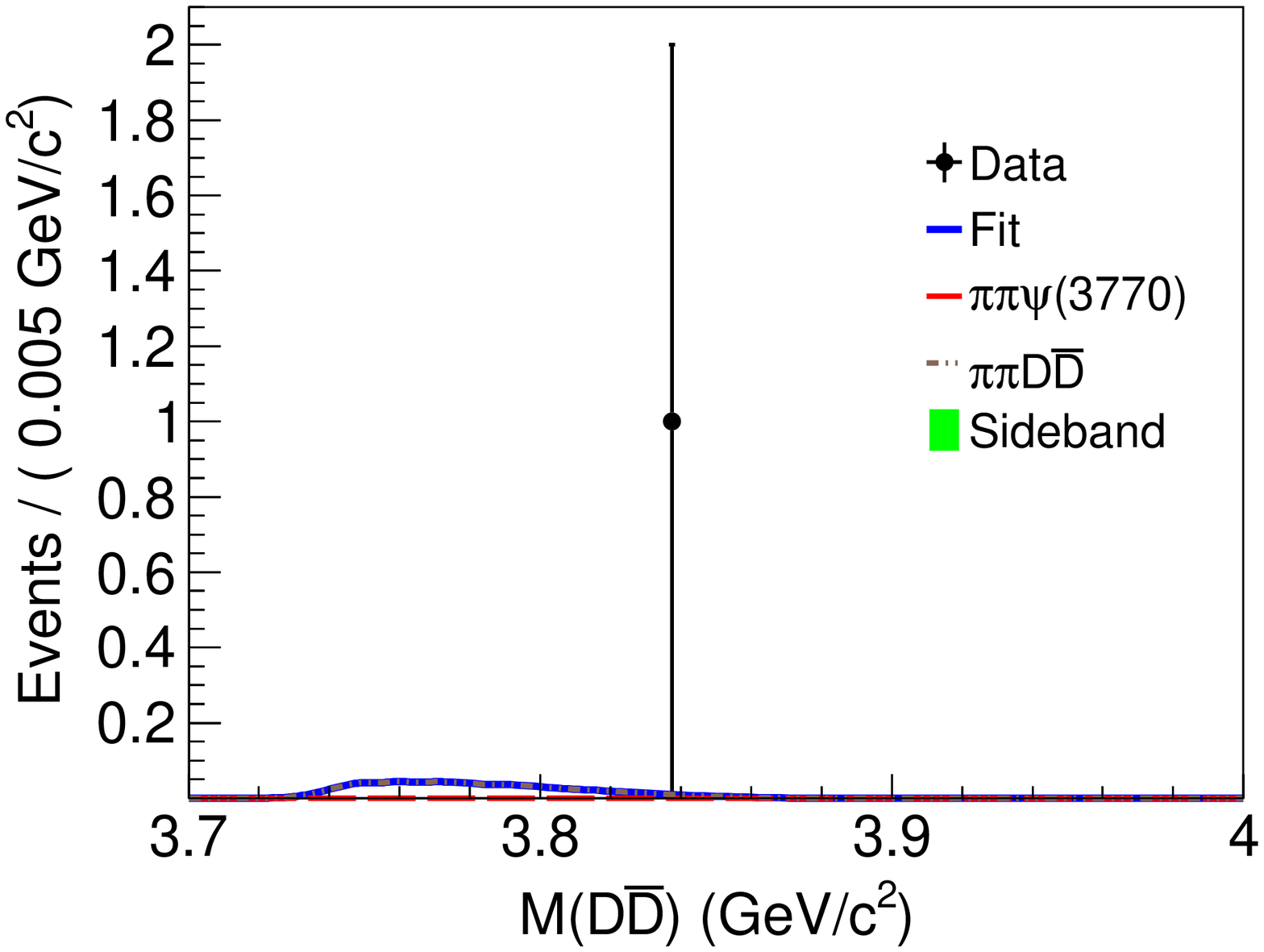}
   \put(80,64){(b)}
   \end{overpic}
   \begin{overpic}[width=0.329\textwidth]{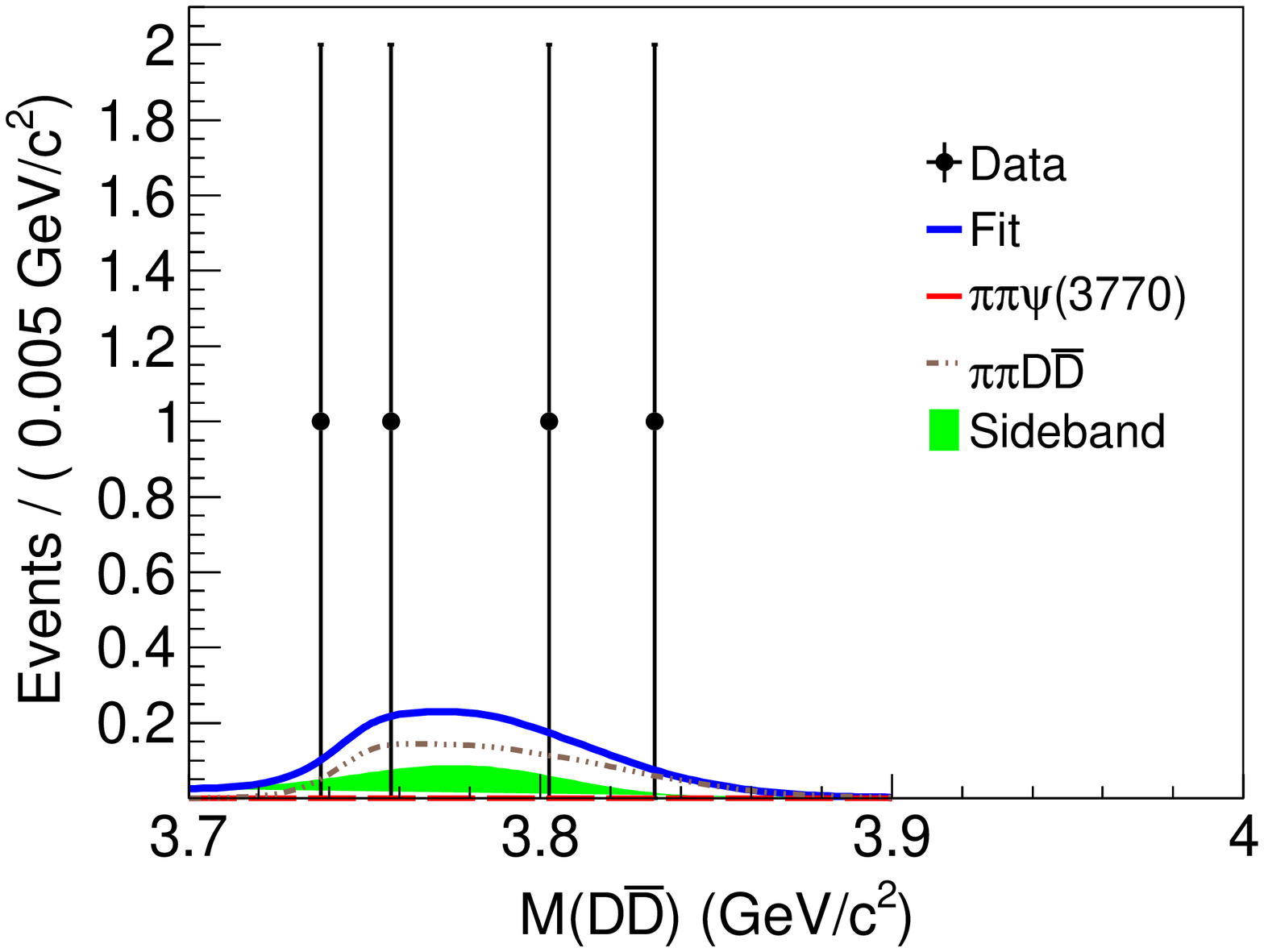}
   \put(80,64){(c)}
   \end{overpic}
   \begin{overpic}[width=0.329\textwidth]{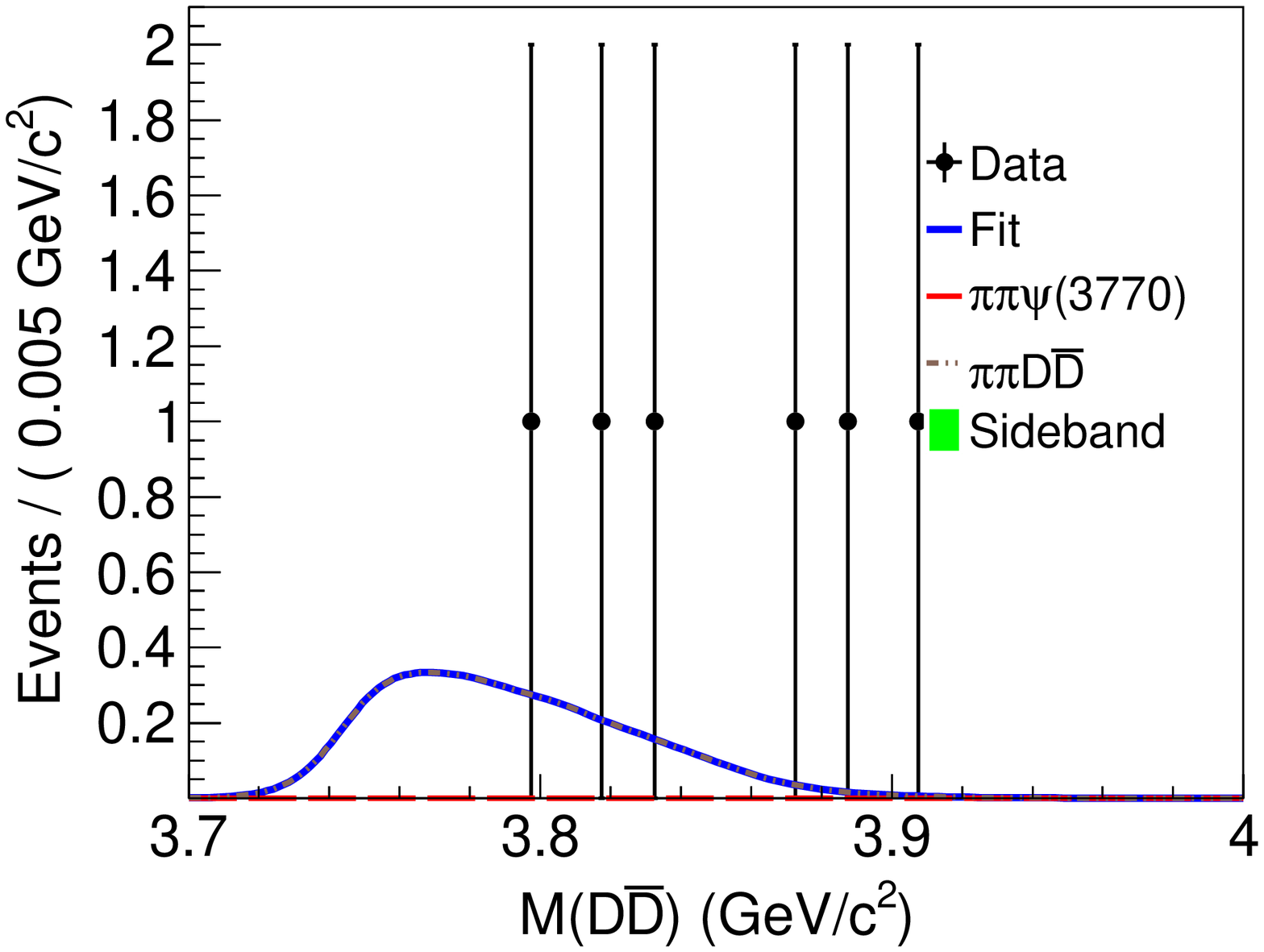}
   \put(80,64){(d)}
   \end{overpic}
   \begin{overpic}[width=0.329\textwidth]{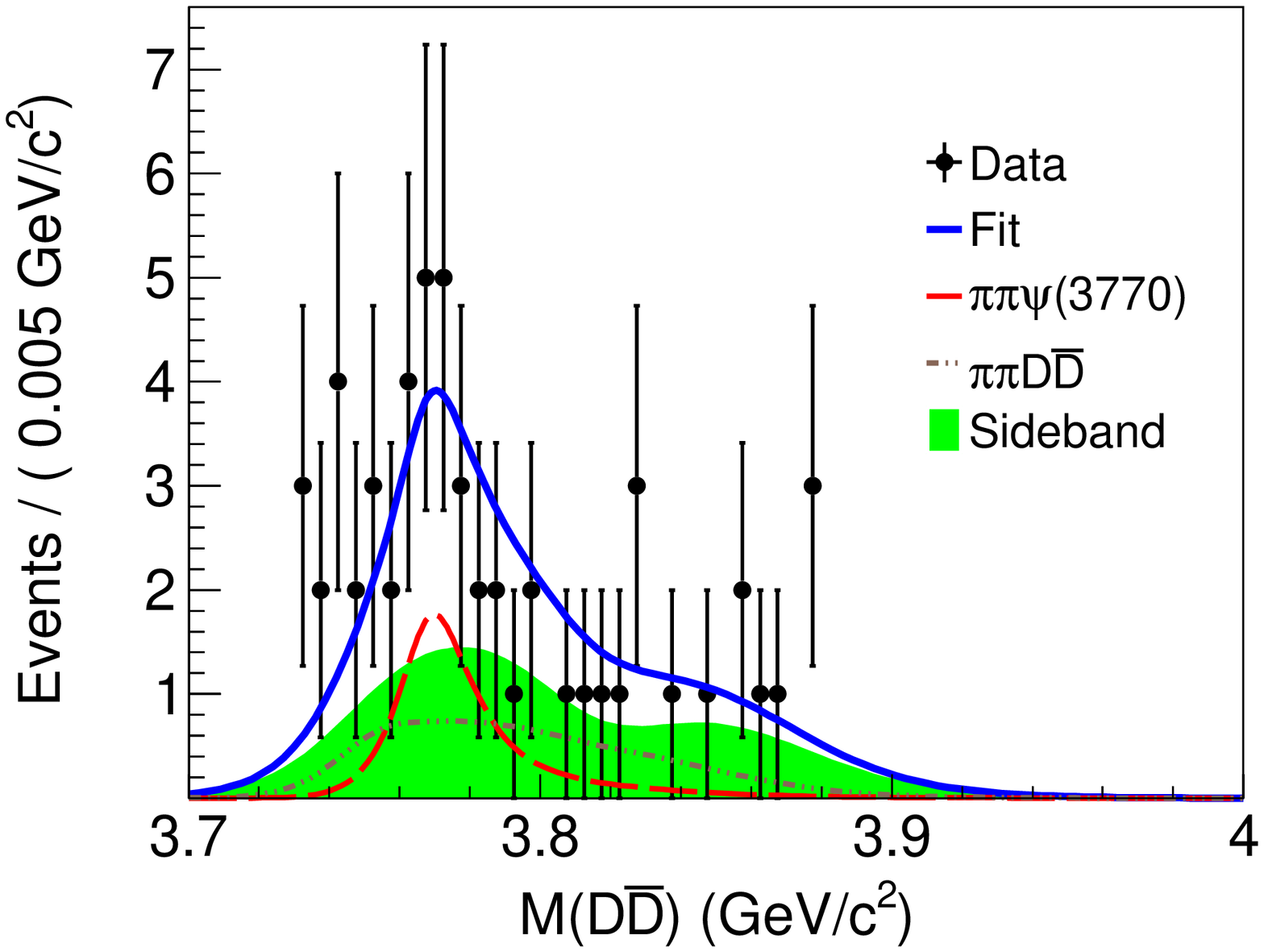}
   \put(80,64){(e)}
   \end{overpic}
   \begin{overpic}[width=0.329\textwidth]{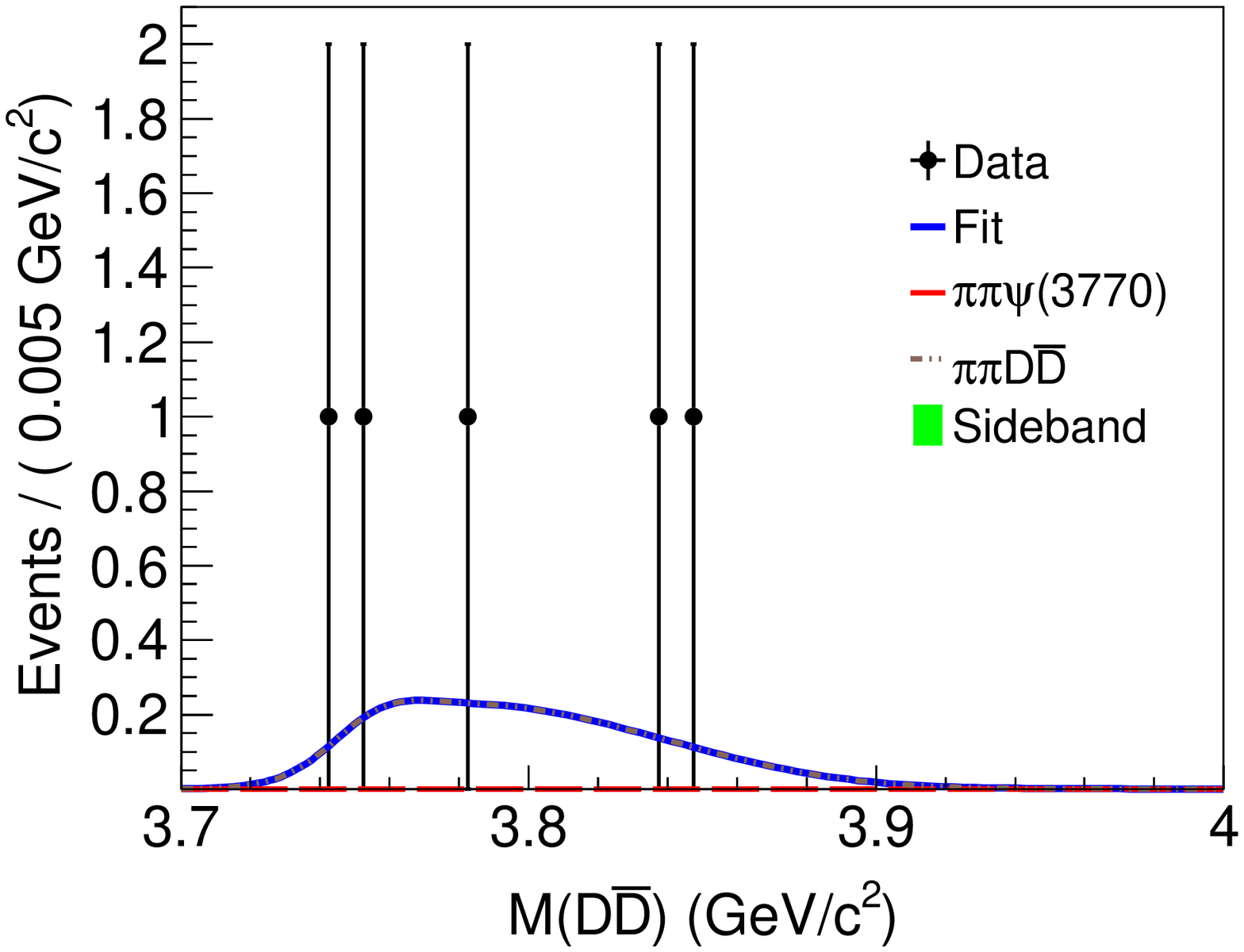}
   \put(80,64){(f)}
   \end{overpic}
   \begin{overpic}[width=0.329\textwidth]{DDbarfit_D0andDpbin_4260.eps}
   \put(80,64){(g)}
   \end{overpic}
   \begin{overpic}[width=0.329\textwidth]{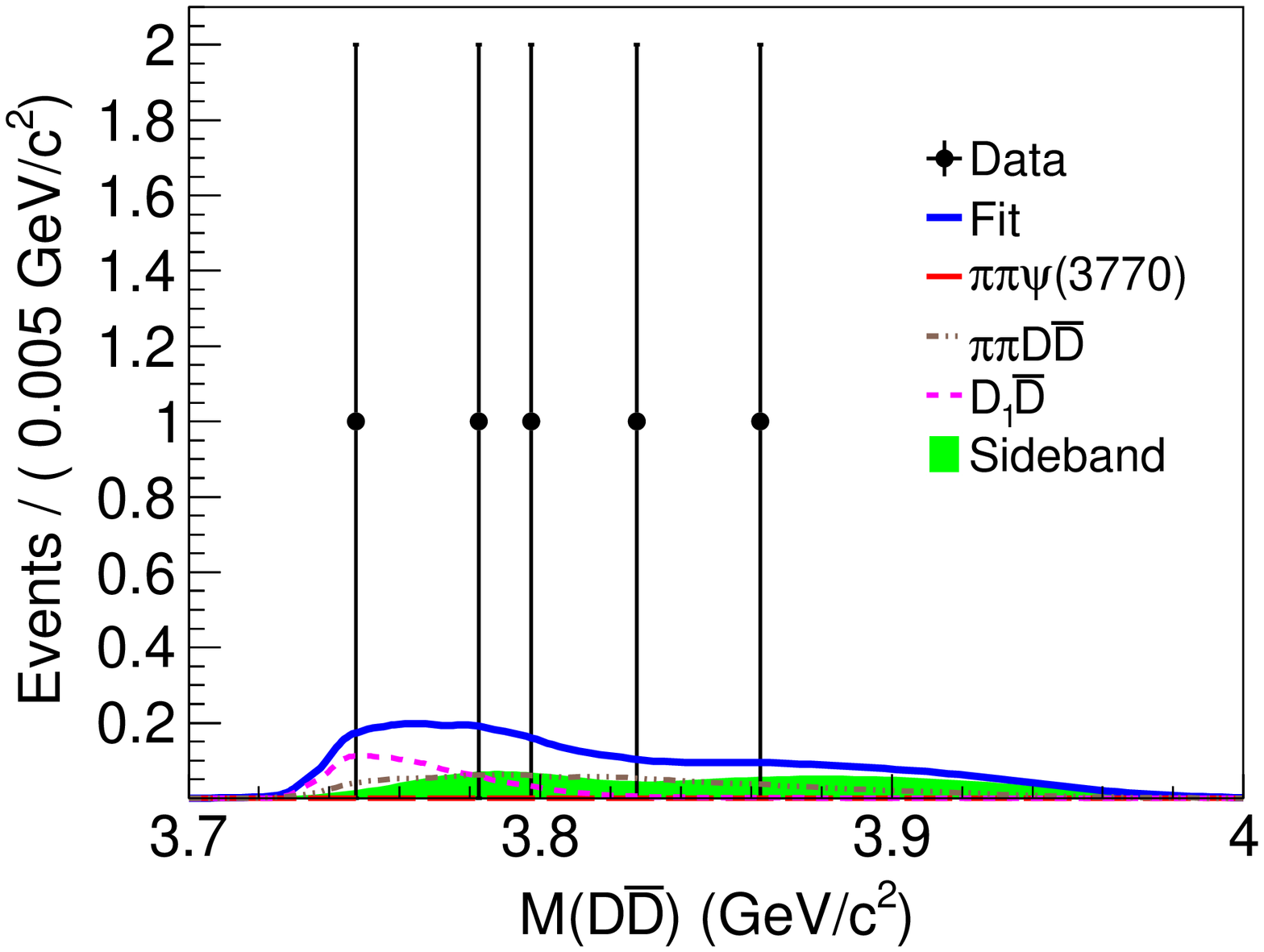}
   \put(80,64){(h)}
   \end{overpic}
   \begin{overpic}[width=0.329\textwidth]{DDbarfit_D0andDpbin_4360.eps}
   \put(80,64){(i)}
   \end{overpic}
   \begin{overpic}[width=0.329\textwidth]{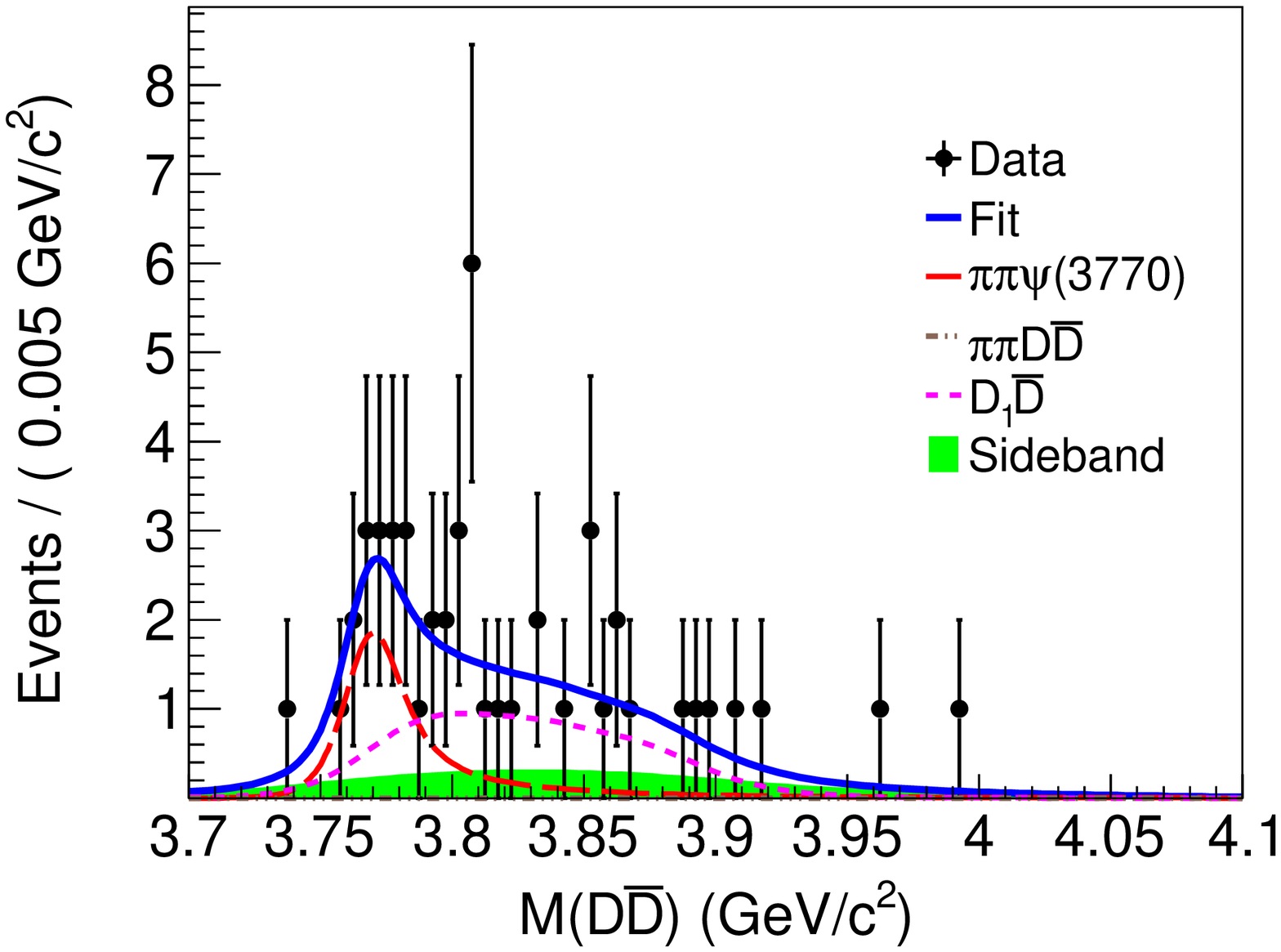}
   \put(80,64){(j)}
   \end{overpic}
   \begin{overpic}[width=0.329\textwidth]{DDbarfit_D0andDpbin_4420.eps}
   \put(80,64){(k)}
   \end{overpic}
   \begin{overpic}[width=0.329\textwidth]{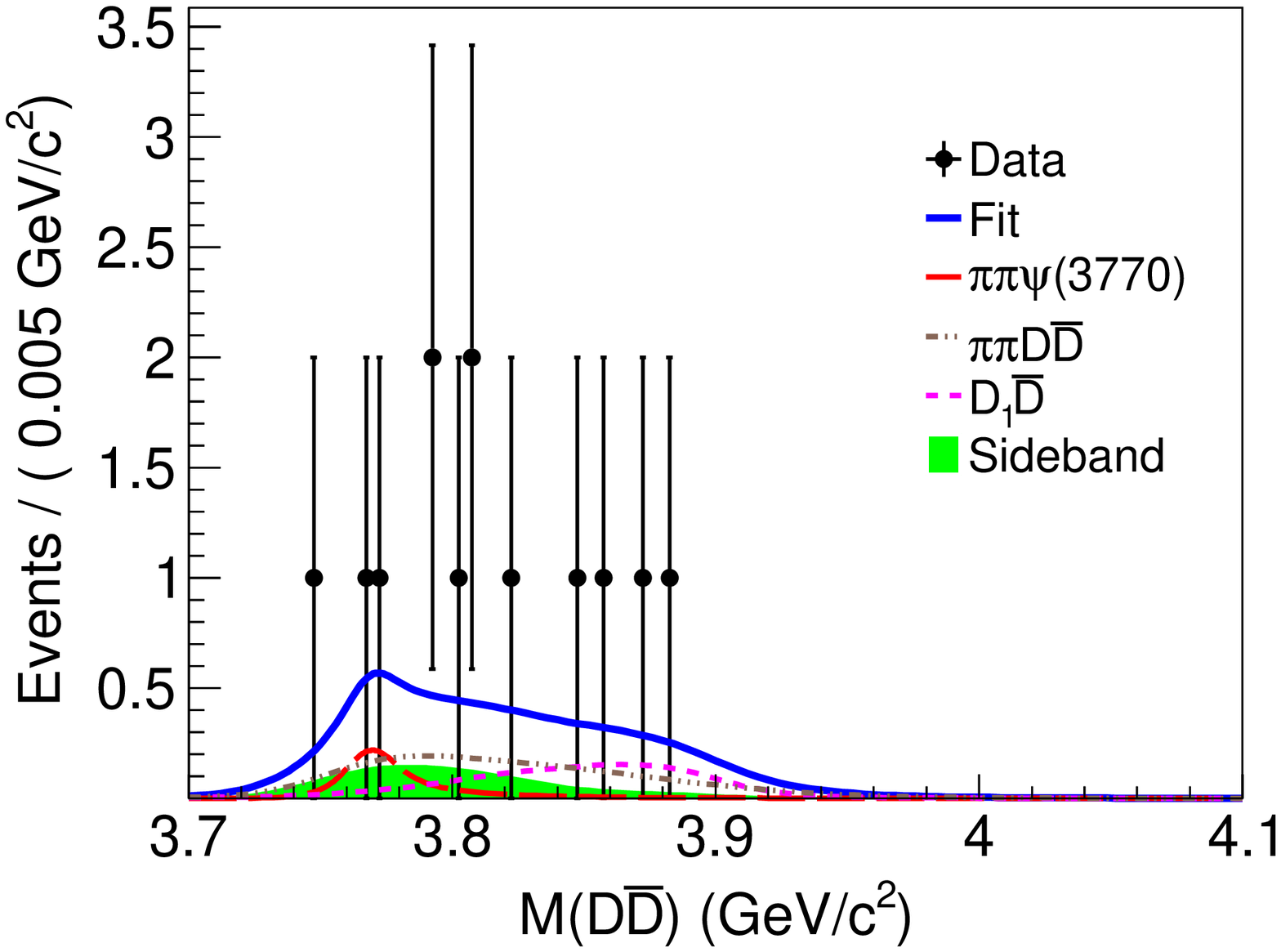}
   \put(80,64){(l)}
   \end{overpic}
   \begin{overpic}[width=0.329\textwidth]{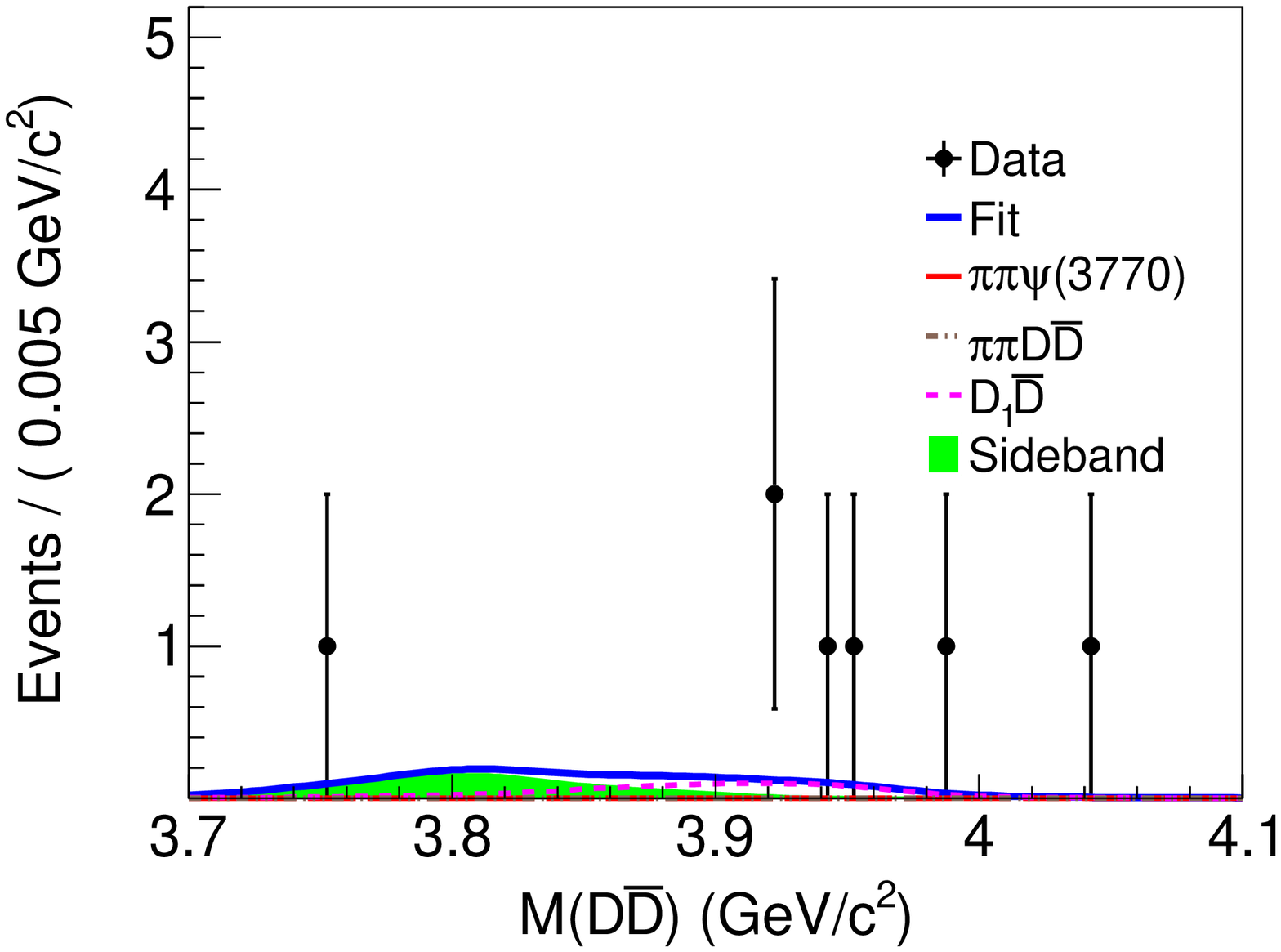}
   \put(80,64){(m)}
   \end{overpic}
   \begin{overpic}[width=0.329\textwidth]{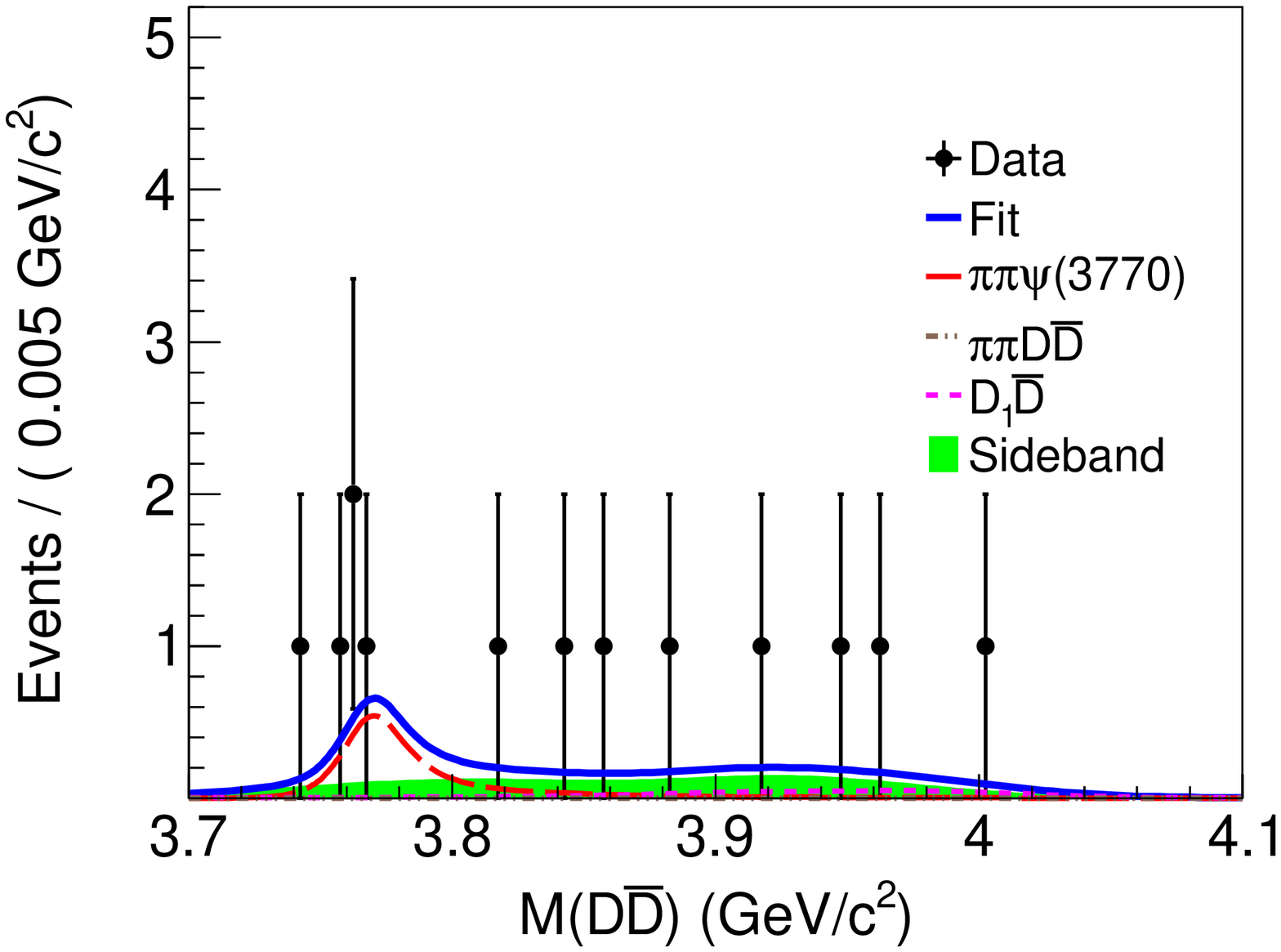}
   \put(80,64){(n)}
   \end{overpic}
   \begin{overpic}[width=0.329\textwidth]{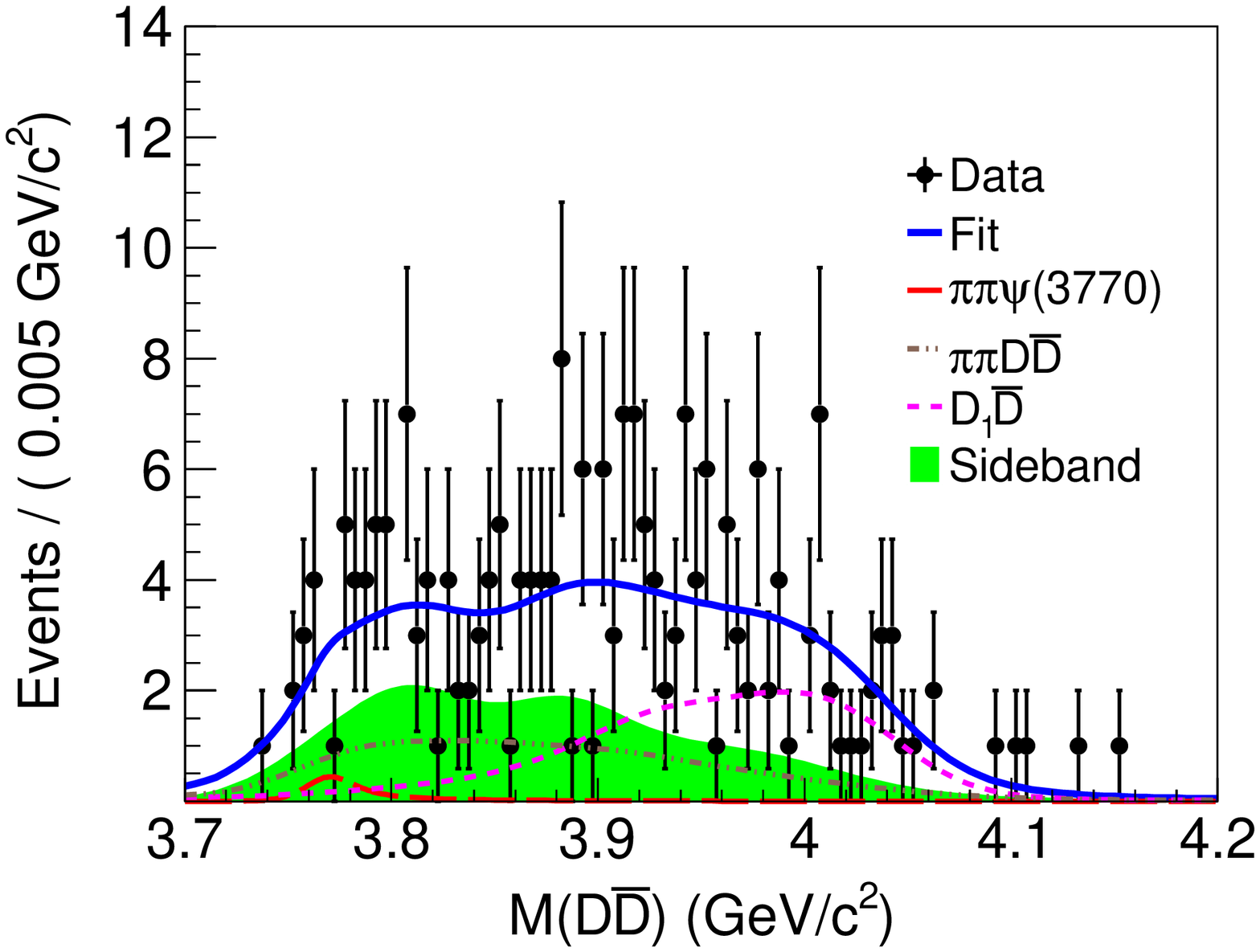}
   \put(80,64){(o)}
   \end{overpic}
   \caption{Fit to the  $D\bar{D}$ invariant
mass distribution at $\sqrt{s}$ = 4.09 (a), 4.19 (b), 4.21 (c), 4.22 (d), 4.23 (e), 4.245 (f), 4.26 (g), 4.31 (h), 4.36 (i), 4.39 (j), 4.42 (k),  4.47 (l),  4.53 (m), 4.575 (n) and  4.60 (o)~GeV.
The black dots with error bars
are data and the blue solid lines are the fit results. The red long-dashed lines indicate the contribution of the $\pspp$ and the pink
dashed lines the contribution of the $D_{1}(2420)\bar{D}$ final state. The brown dotted-dashed lines show the
$\pi^{+}\pi^{-} D\bar{D}$ background contributions and the green shaded histograms are the distributions from the sideband regions.}
  \label{FitDDbar_all}
\end{figure}

\newpage

\section{Fits to the  $D\pi^{+}\pi^{-}$ invariant mass distributions }

\begin{figure}[h]
  \centering
  \begin{overpic}[width=0.329\textwidth]{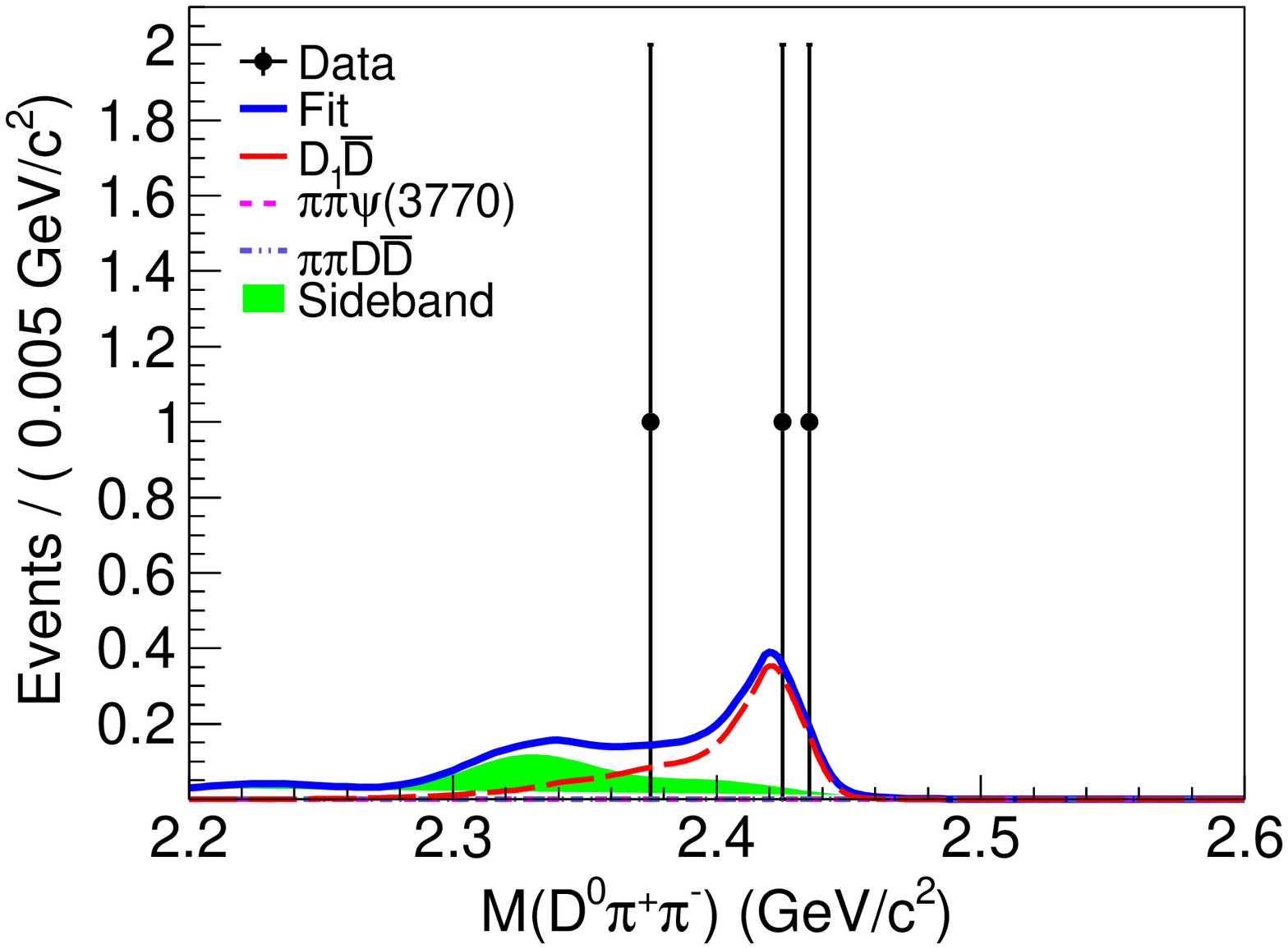}
   \put(80,65){(a)}
   \end{overpic}
   \begin{overpic}[width=0.329\textwidth]{Dpipifit_D0unbin_4360.eps}
   \put(80,65){(b)}
   \end{overpic}
   \begin{overpic}[width=0.329\textwidth]{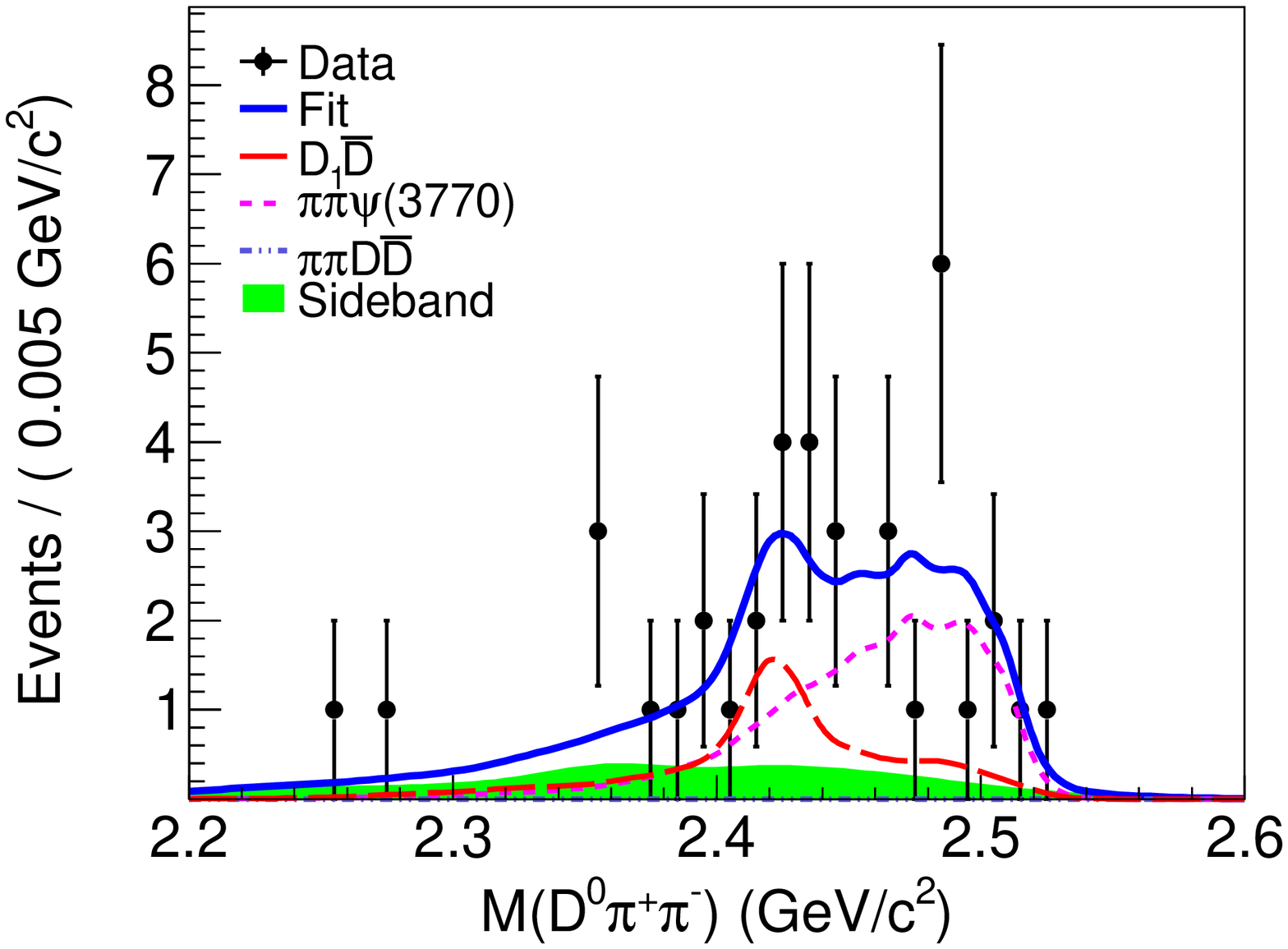}
   \put(80,65){(c)}
   \end{overpic}
   \begin{overpic}[width=0.329\textwidth]{Dpipifit_D0unbin_4420.eps}
   \put(80,65){(d)}
   \end{overpic}
   \begin{overpic}[width=0.329\textwidth]{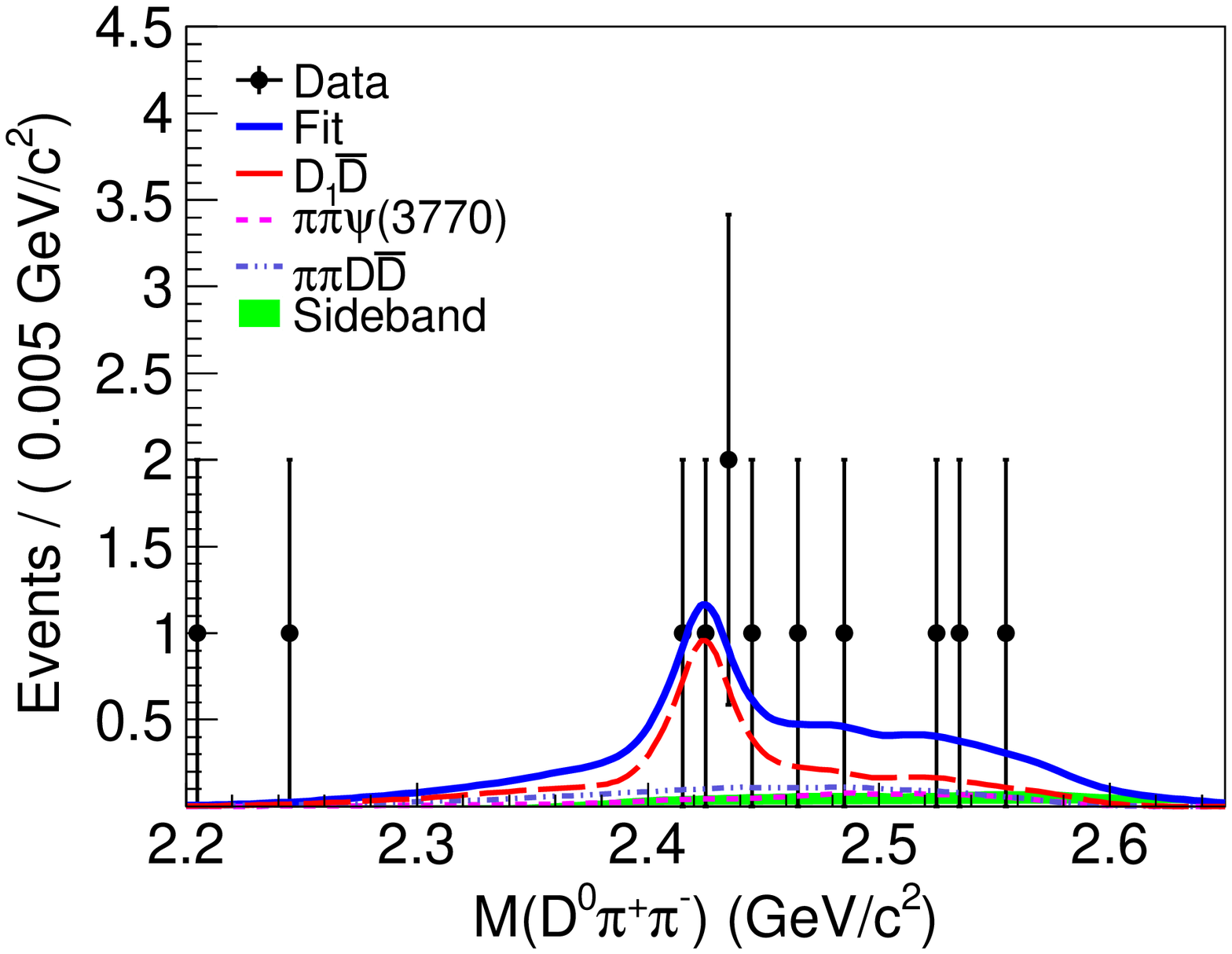}
   \put(80,65){(e)}
   \end{overpic}
   \begin{overpic}[width=0.329\textwidth]{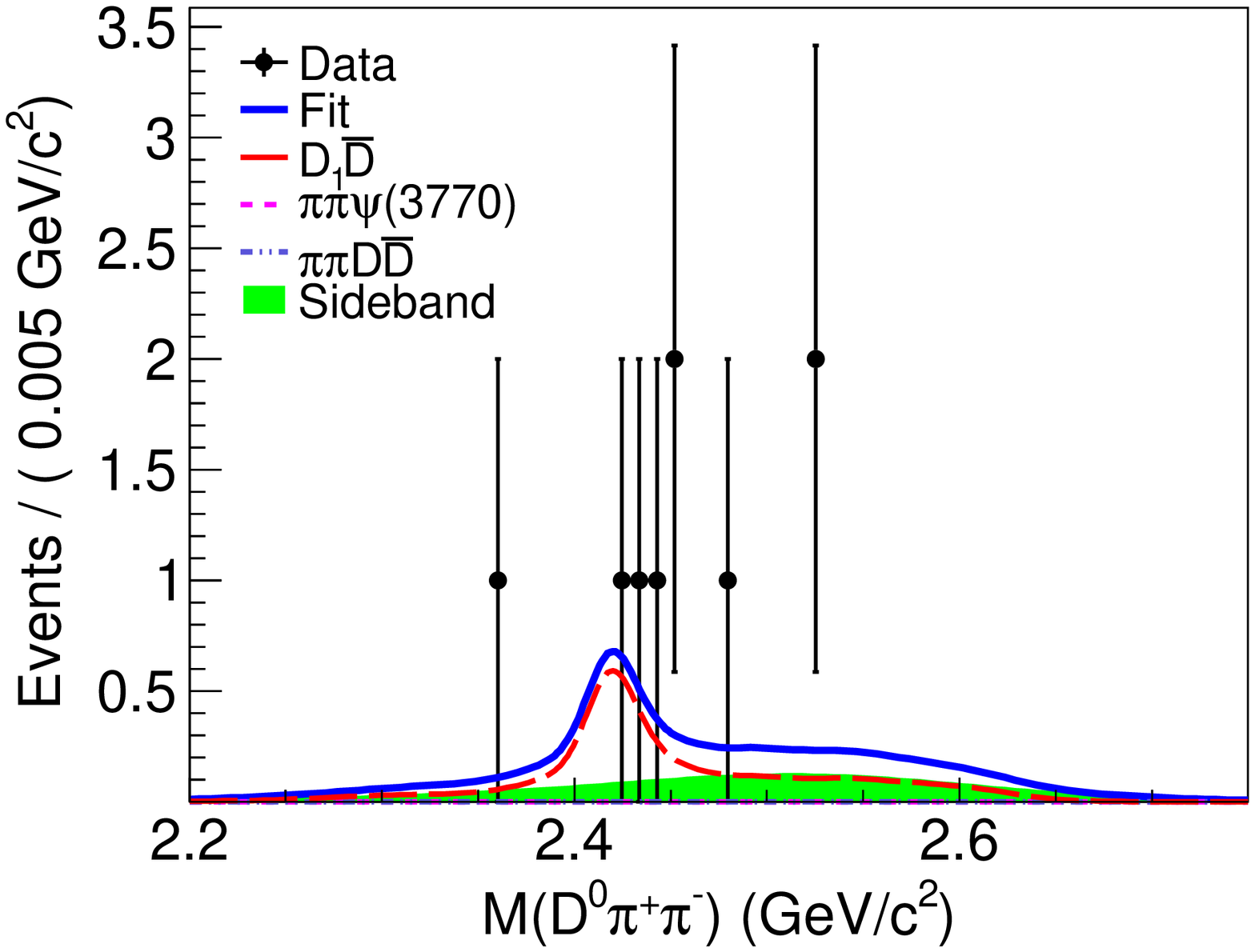}
   \put(80,65){(f)}
   \end{overpic}
   \begin{overpic}[width=0.329\textwidth]{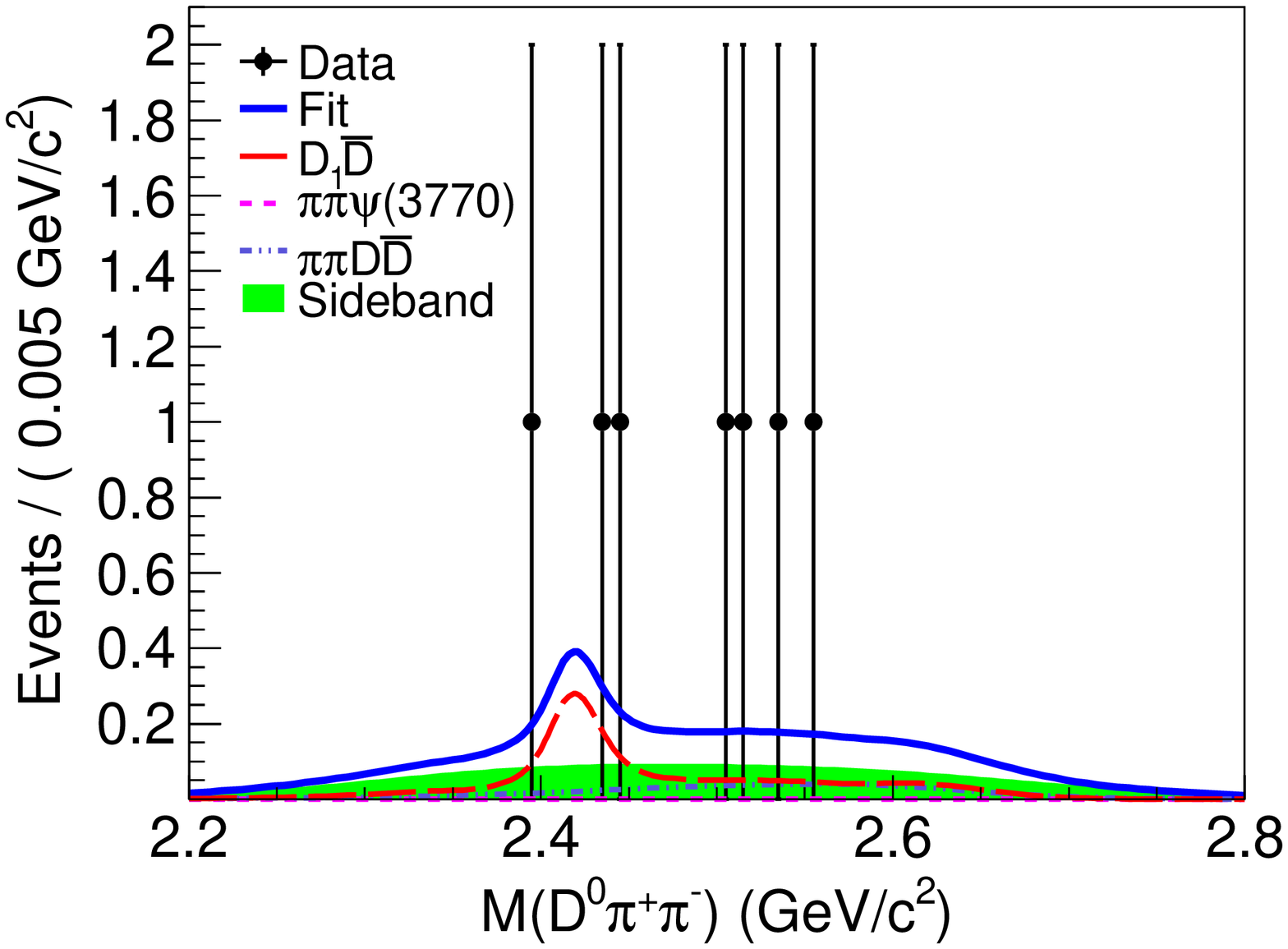}
   \put(80,65){(g)}
   \end{overpic}
   \begin{overpic}[width=0.329\textwidth]{Dpipifit_D0unbin_4600.eps}
   \put(80,65){(h)}
   \end{overpic}

\caption{Fit to the  $D^{0}\pi^{+}\pi^{-}$ invariant
 mass distribution at $\sqrt{s}$ = 4.31 (a), 4.36 (b),  4.39 (c), 4.42 (d), 4.47 (e), 4.53 (f), 4.575 (g) and  4.60 (h)~GeV. The black dots with error bars are data, the blue solid curves  the fits results, and the red long-dashed lines the $D_{1}(2420)$ signal contributions. The pink dashed lines are the contributions of the final state
$\pp\psi(3770)$ and the blue dot-dot-dashed lines these of the $\pp D\bar{D}$ final state, while the green shaded histograms are the distributions from the sidebands regions.}
  \label{FitD1_D0}
\end{figure}

\begin{figure}[H]
  \centering
   \begin{overpic}[width=0.329\textwidth]{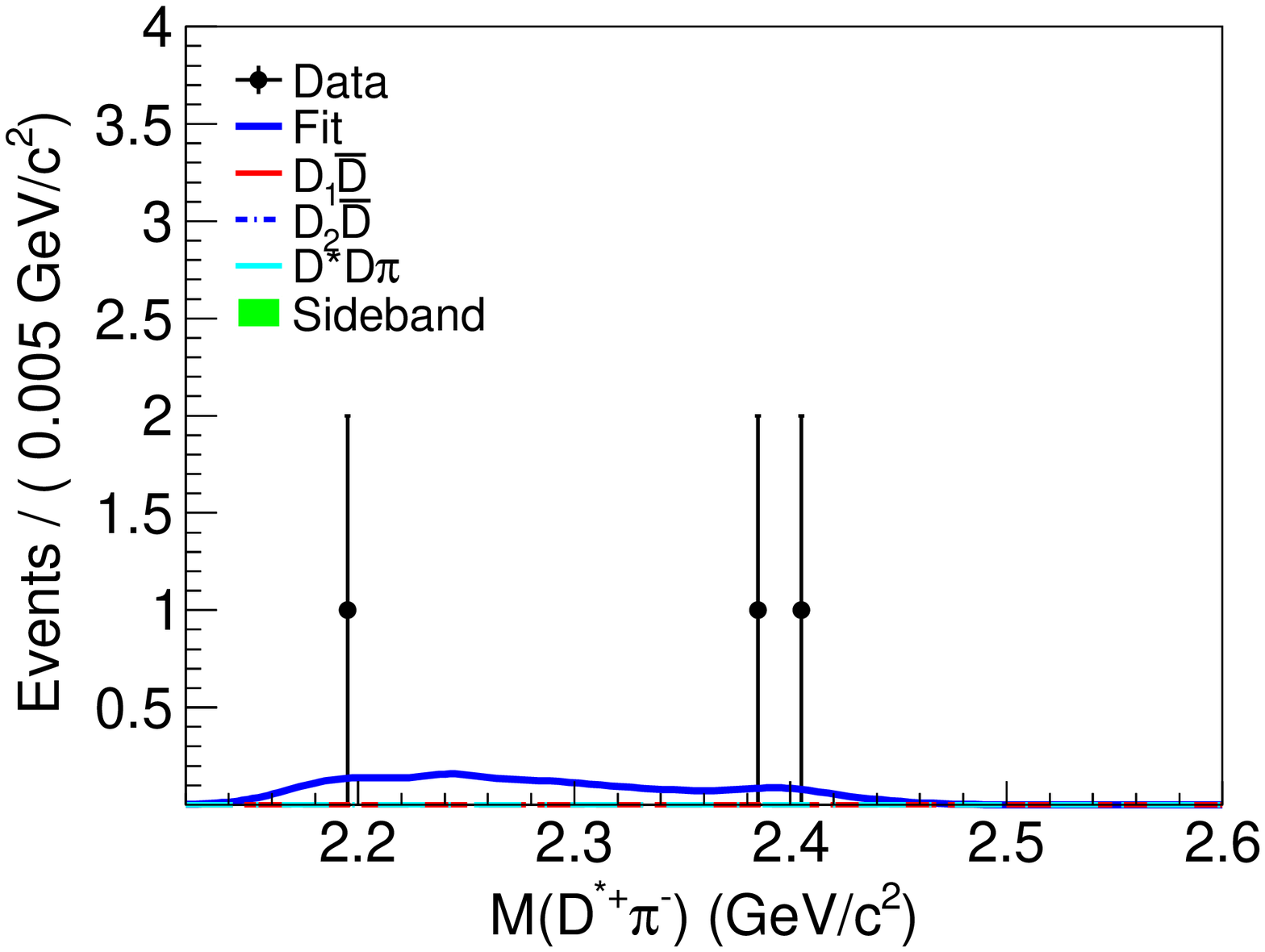}
   \put(80,65){(a)}
   \end{overpic}
   \begin{overpic}[width=0.329\textwidth]{Dpipifit_D0starunbin_4360.eps}
   \put(80,65){(b)}
   \end{overpic}
   \begin{overpic}[width=0.329\textwidth]{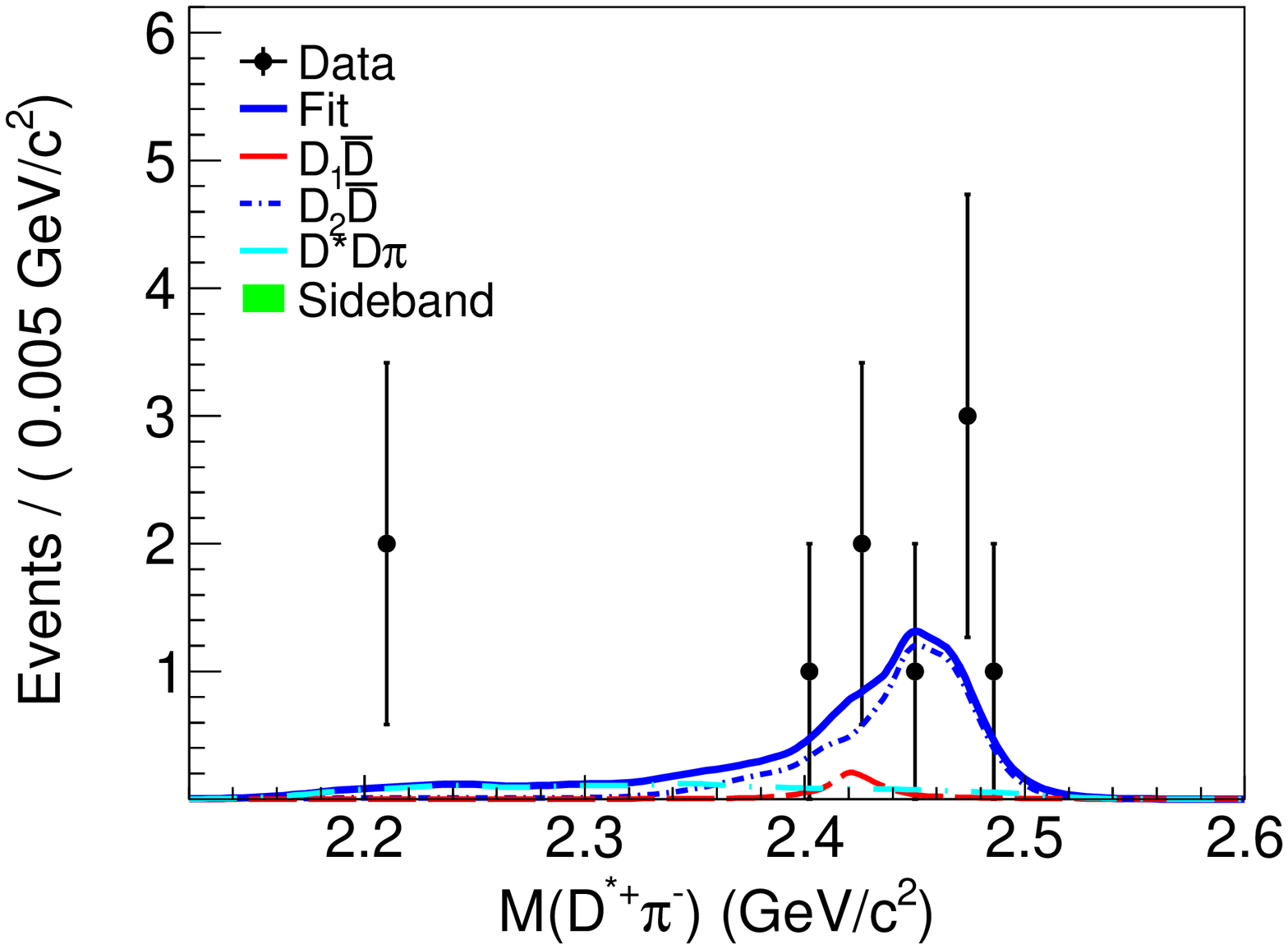}
   \put(80,65){(c)}
   \end{overpic}
   \begin{overpic}[width=0.329\textwidth]{Dpipifit_D0starunbin_4420.eps}
   \put(80,65){(d)}
   \end{overpic}
   \begin{overpic}[width=0.329\textwidth]{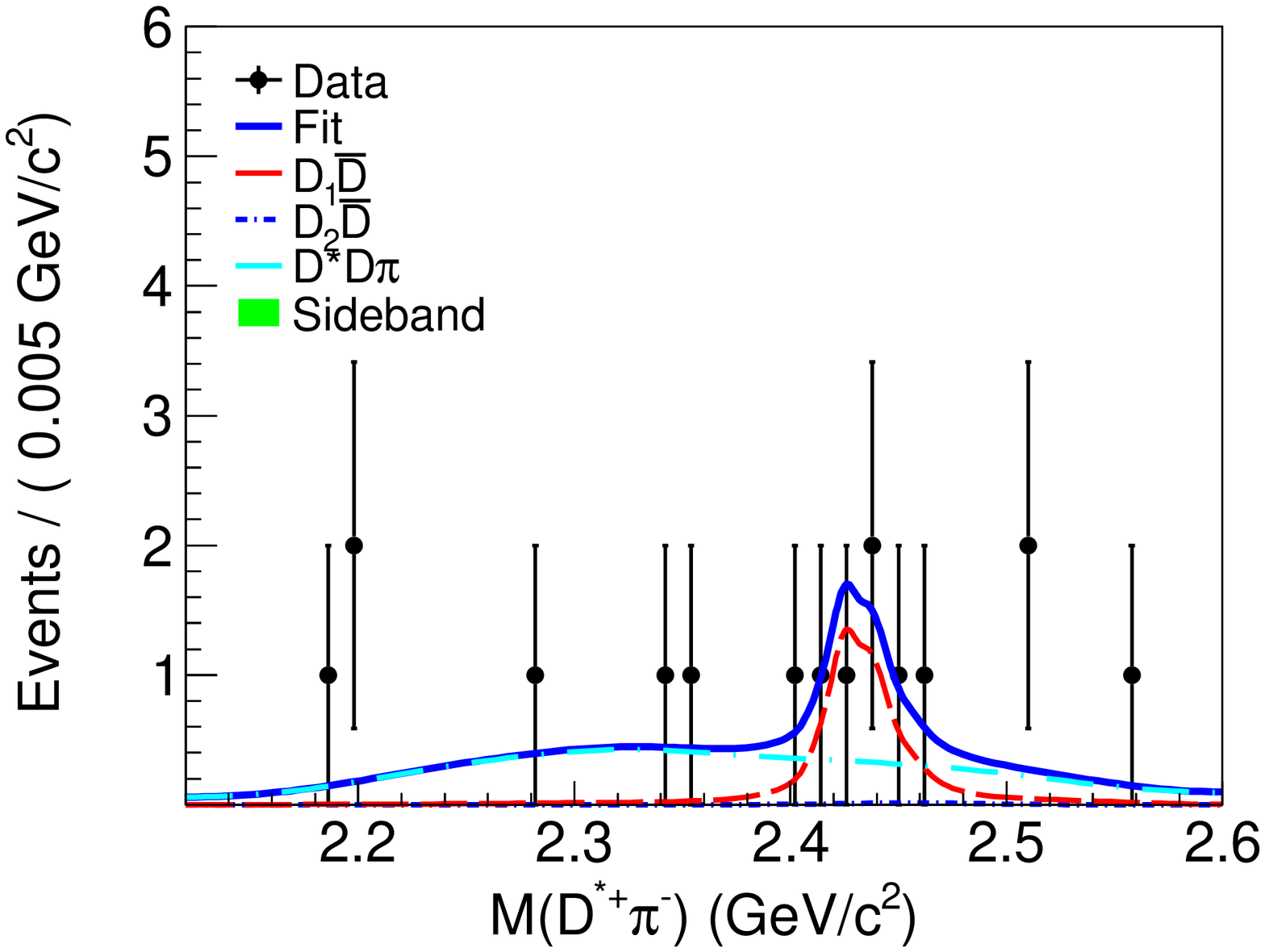}
   \put(80,65){(e)}
   \end{overpic}
   \begin{overpic}[width=0.329\textwidth]{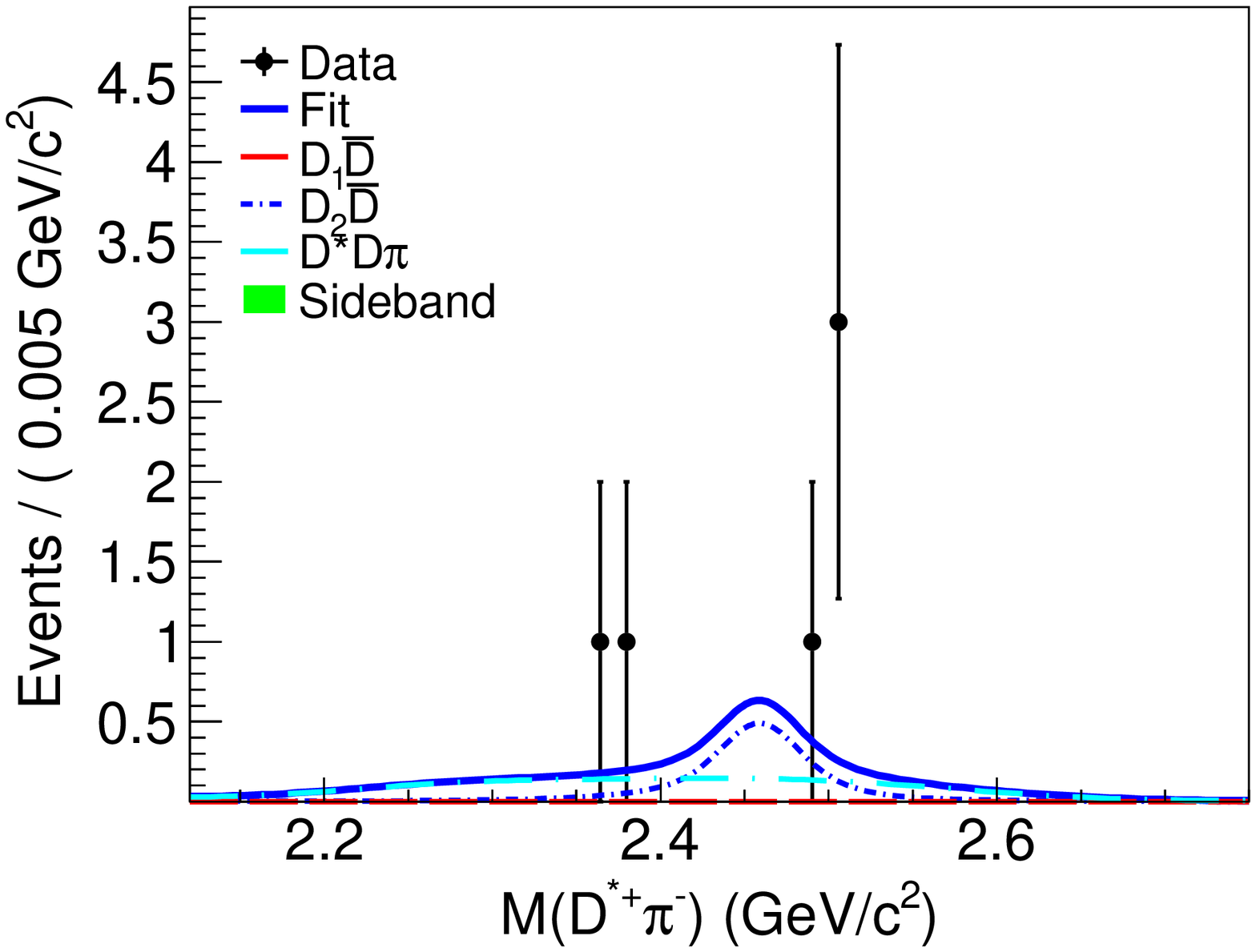}
   \put(80,65){(f)}
   \end{overpic}
   \begin{overpic}[width=0.329\textwidth]{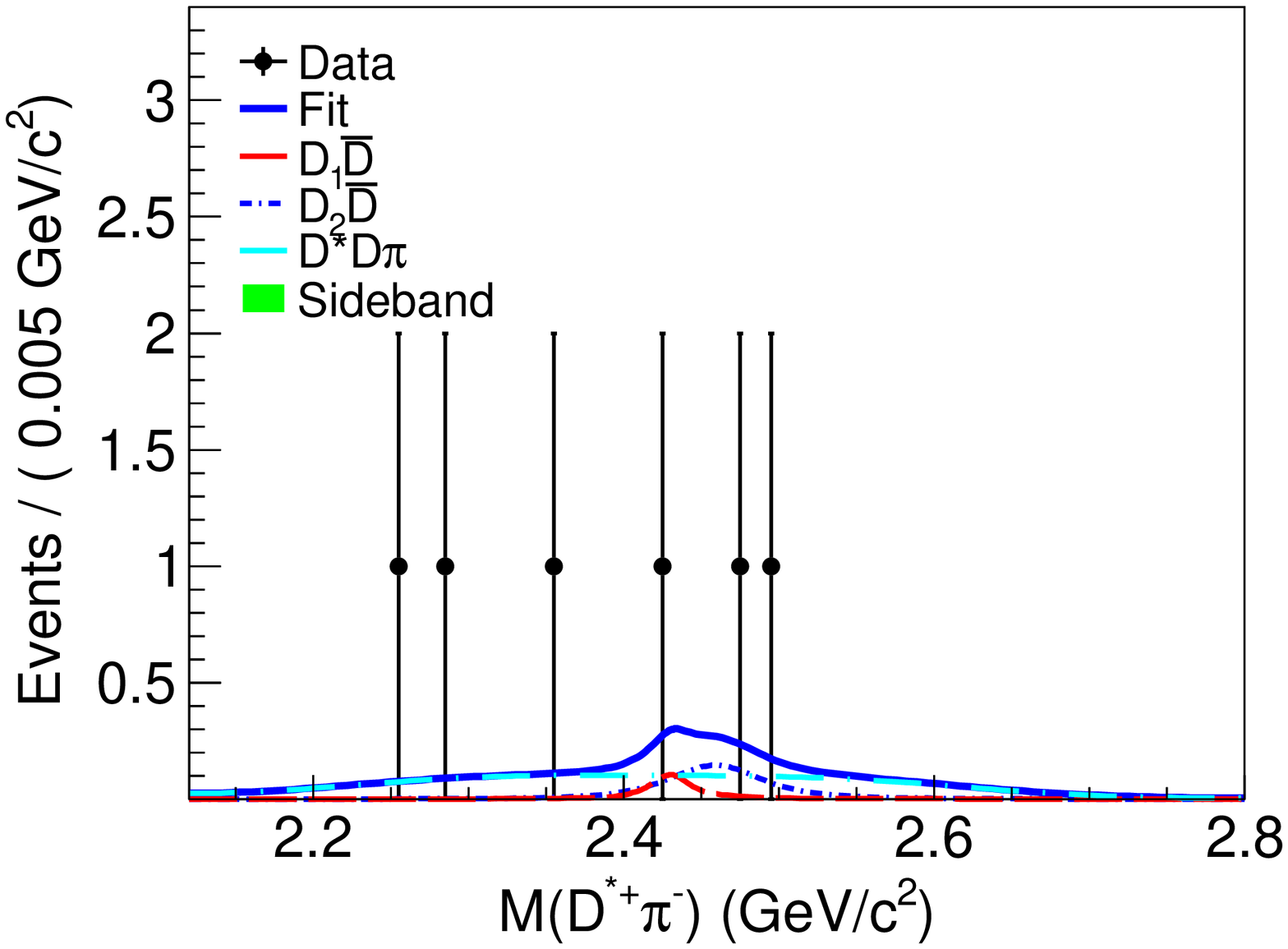}
   \put(80,65){(g)}
   \end{overpic}
   \begin{overpic}[width=0.329\textwidth]{Dpipifit_D0starunbin_4600.eps}
   \put(80,65){(h)}
   \end{overpic}
\caption{Fit to the  $D^{*+}\pi^{-}$ invariant
mass distribution at $\sqrt{s}$ =  4.31 (a), 4.36 (b), 4.39 (c), 4.42 (d), 4.47 (e), 4.53 (f), 4.575 (g) and 4.60 (h)~GeV. The black dots with error bars are data, the blue solid curves the fit results, and the red long-dashed lines the  $D_{1}(2420)$ signal contributions. The light blue dot-long-dashed lines are the $D^{*}\bar{D}\pi$ and the blue dot-dashed lines the $D_{2}^{*}(2460)^{0}\bar{D}$ background contributions, while the green shaded histograms are the distributions from the sideband regions.}
  \label{FitD1_D0Dstar}
\end{figure}

\begin{figure}[H]
  \centering
   \begin{overpic}[width=0.329\textwidth]{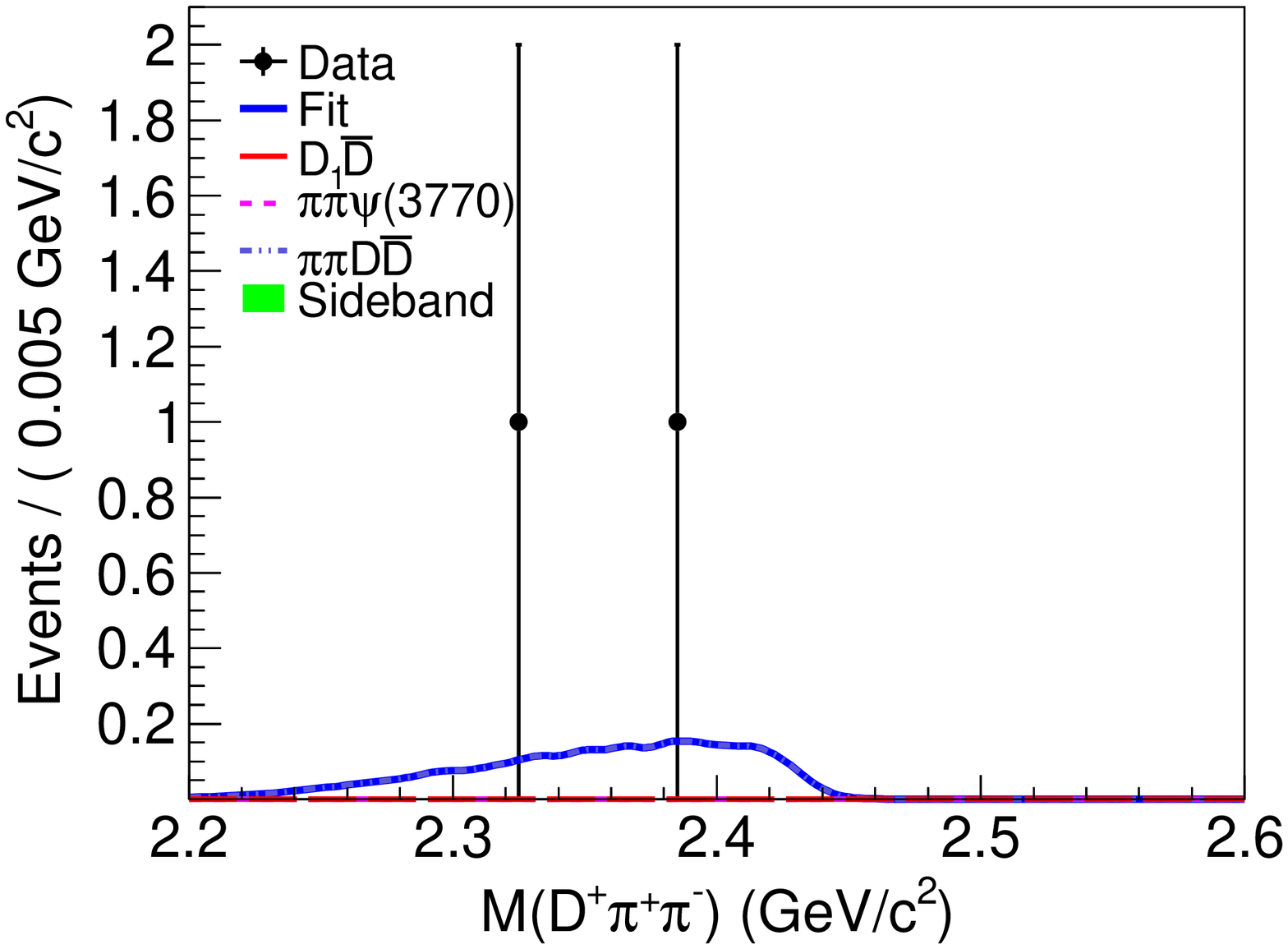}
   \put(80,65){(a)}
   \end{overpic}
   \begin{overpic}[width=0.329\textwidth]{Dpipifit_Dpunbin_4360.eps}
   \put(80,65){(b)}
   \end{overpic}
   \begin{overpic}[width=0.329\textwidth]{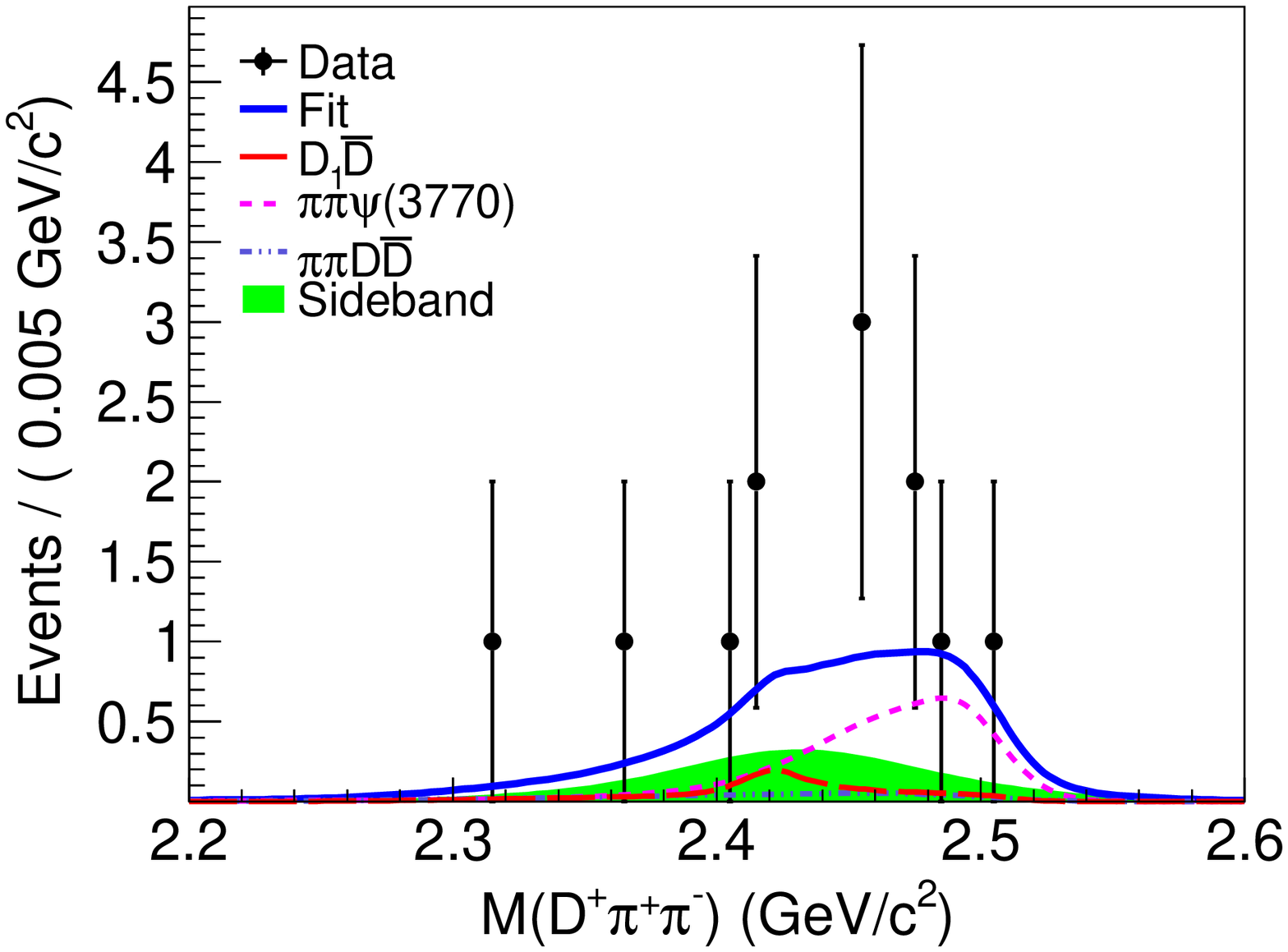}
   \put(80,65){(c)}
   \end{overpic}
   \begin{overpic}[width=0.329\textwidth]{Dpipifit_Dpunbin_4420.eps}
   \put(80,65){(d)}
   \end{overpic}
   \begin{overpic}[width=0.329\textwidth]{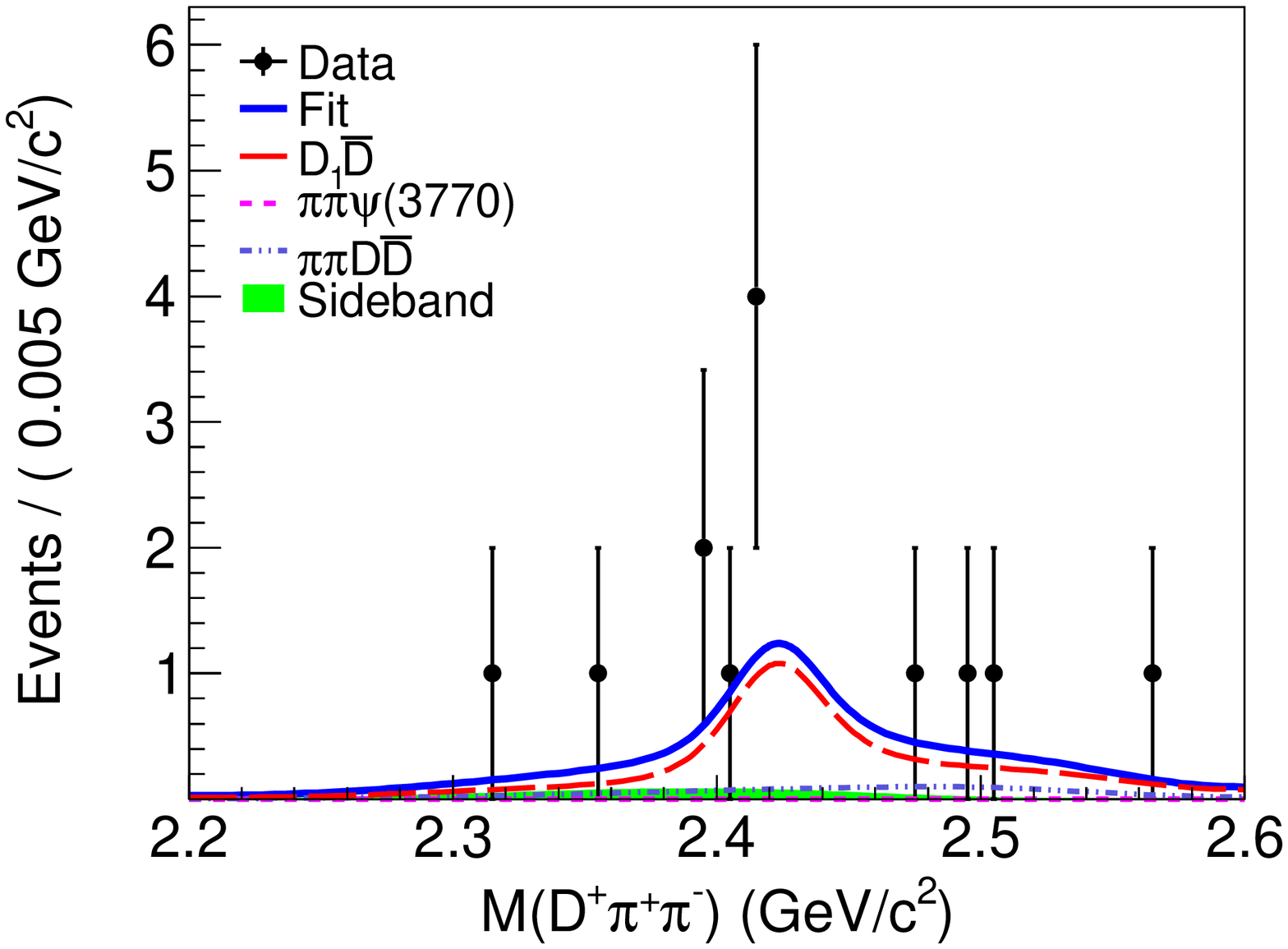}
   \put(80,65){(e)}
   \end{overpic}
   \begin{overpic}[width=0.329\textwidth]{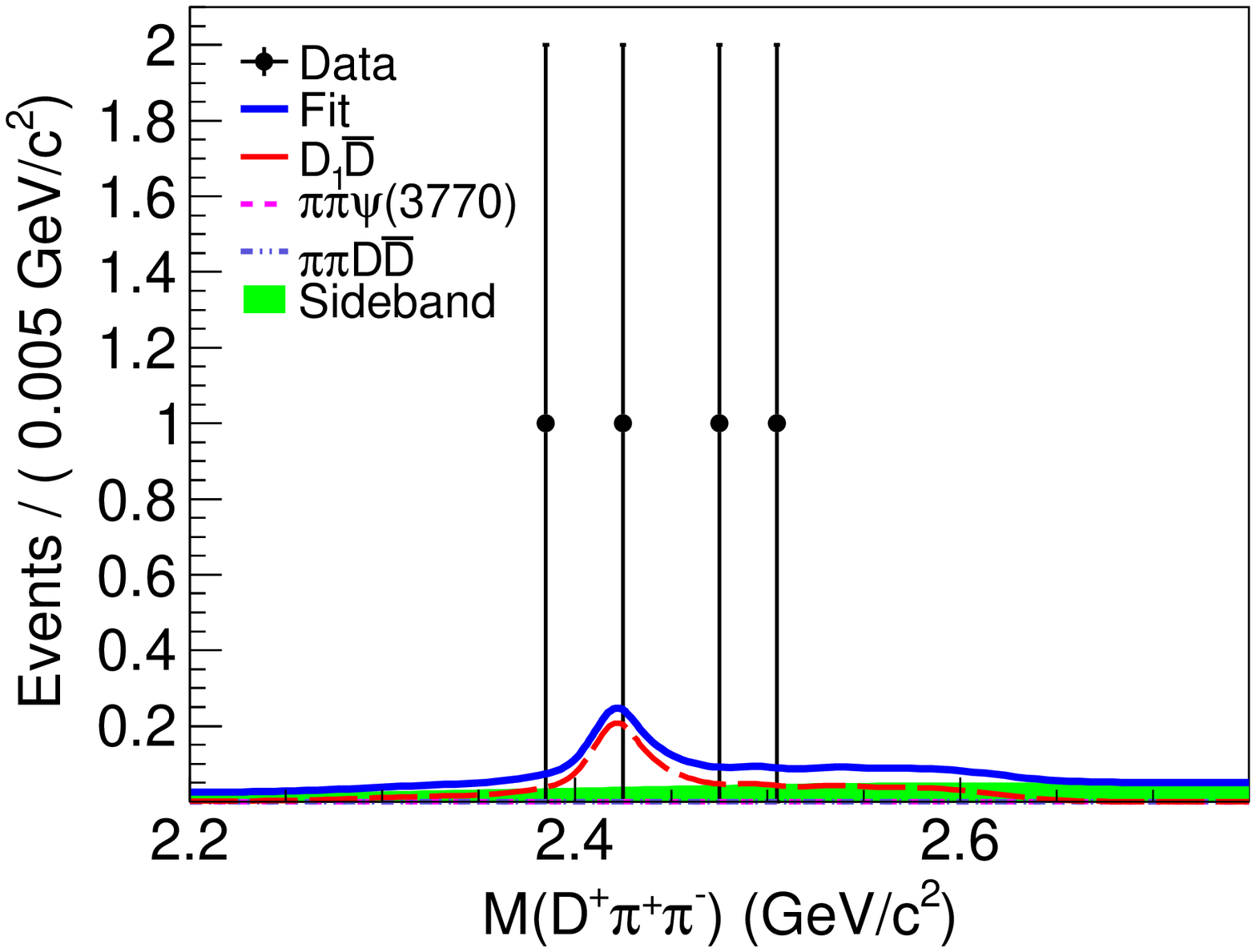}
   \put(80,65){(f)}
   \end{overpic}
   \begin{overpic}[width=0.329\textwidth]{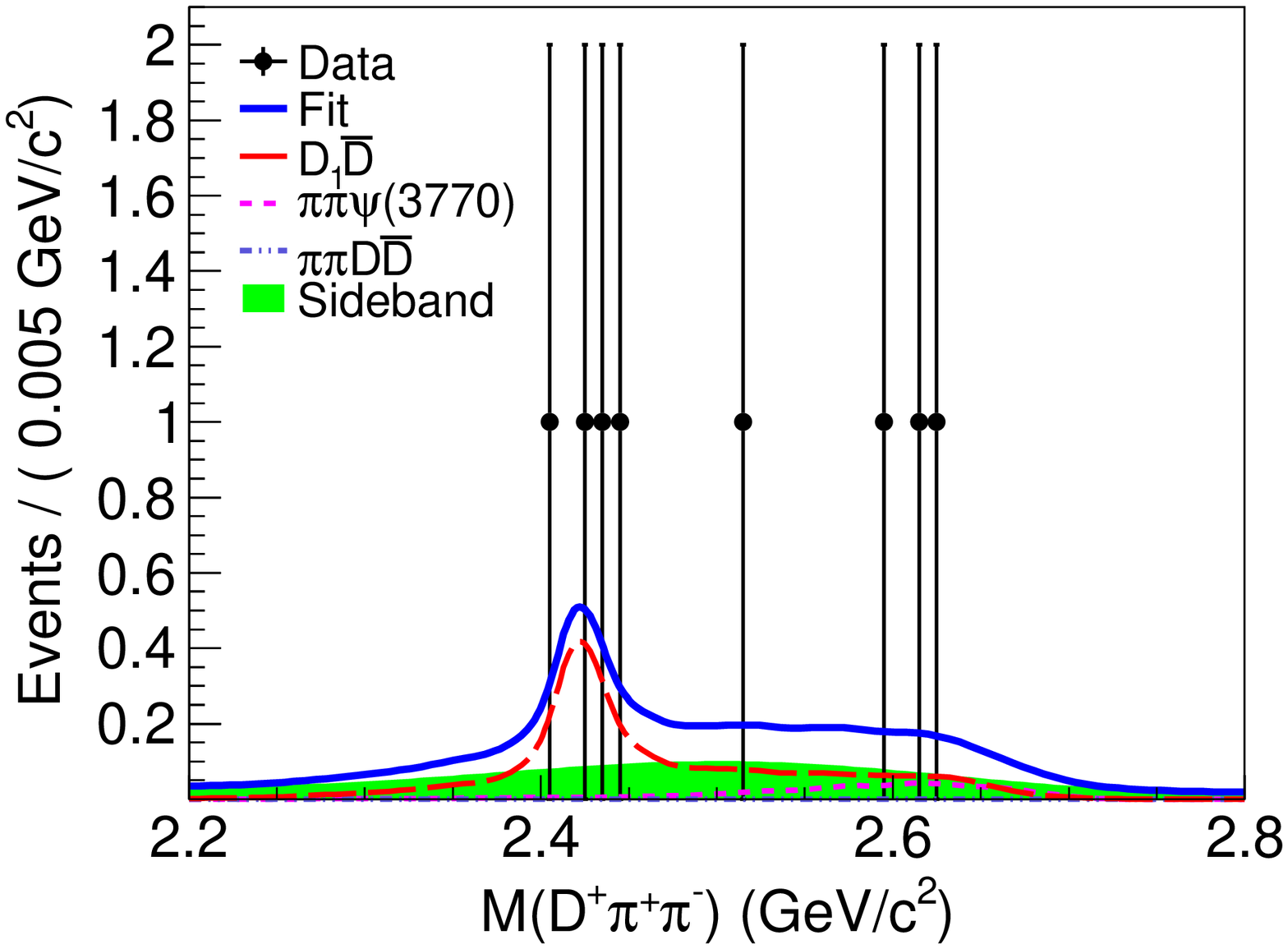}
   \put(80,65){(g)}
   \end{overpic}
   \begin{overpic}[width=0.329\textwidth]{Dpipifit_Dpunbin_4600.eps}
   \put(80,65){(h)}
   \end{overpic}
\caption{Fit to  the  $D^{+}\pi^{+}\pi^{-}$ invariant mass distribution at $\sqrt{s}$ = 4.31 (a), 4.36 (b), 4.39 (c), 4.42 (d), 4.47 (e), 4.53  (f), 4.575 (g) and 4.60 (h)~GeV. The black dots with error bars  are data, the blue solid curves  the fit results, and the red long-dashed lines the  $D_{1}(2420)$ signal contribution. The pink dashed lines are contributions of the final state $\pp\psi(3770)$ and the blue dot-dot-dashed lines these of the $\pp D\bar{D}$ final state,  while the green shaded histograms are the distributions from the sideband regions.}
  \label{FitD1_Dp}
\end{figure}

\section{\boldmath Systematic uncertainties for the measurements of  $\sigma(e^{+}e^{-}\to \pi^{+}\pi^{-}\psi(3770))$ and $\sigma(e^{+}e^{-}\to D_{1}(2420)^{0}\bar{D}^{0})$}

\begin{table}[H]
\centering
\setlength{\tabcolsep}{1.5mm}{
\scriptsize
\caption{Systematic uncertainties (in \%)
in $\sigma(e^{+}e^{-}\to \pi^{+}\pi^{-}\psi(3770))$ measurement.}
\label{sys_err_psipp_all}
\begin{tabular}{l r r r r r r r r r r r r r r r}
\hline \hline
     Sources / $\sqrt{s}$ (GeV) &4.0854 &4.1886 &4.2077  &4.2171  &4.2263  &4.2417 &4.2580  &4.3079   &4.3583  &4.3874   &4.4156  &4.4671   &4.5271  &4.5745   &4.5995\\
    \hline
    Integrated luminosity                  & 1.0 & 1.0  & 1.0  & 1.0  & 1.0  & 1.0  &1.0   &1.0    & 1.0  & 1.0   & 1.0  & 1.0   & 1.0   & 1.0   & 1.0  \\
    Efficiency related          &9.2 &9.3 &9.2 &9.2 &9.0 &9.0 &9.2 &9.0 &8.9 &8.8 &8.8 &8.7 &8.7 &8.7 &8.6     \\
    Radiative correction        &1.3 &1.5  &2.1  &1.5  &0.6  &4.4  &1.1  &6.3   &0.6  &1.8   &2.1  &1.8   &0.7   &2.5   &0.2  \\
    Signal shape                 &3.4  &3.4   &3.4   &3.4   &3.4   &3.4   &3.4   &3.4    &3.4   &3.4    &3.8   &3.8    &3.8    &3.8    &3.8   \\
    Background shape            &9.9  &9.9   &9.9   &9.9   &9.9   &9.9   &9.9   &9.9    &9.9   &9.9    &7.6   &7.6    &7.6    &7.6    &7.6   \\
    Fit range                   &2.0  &2.0   &2.0   &2.0   &2.0   &2.0   &2.0   &2.0    &2.0   &2.0    &2.4   &2.4    &2.4    &2.4    &2.4   \\
    Signal region of double tag &5.3  &5.3   &5.3   &5.3   &5.3   &5.3   &5.3   &5.3    &5.3   &5.3    &2.0   &2.0    &2.0    &2.0    &2.0   \\
    \hline
    Total                       &14.3  &15.2  &15.2  &15.1  &15.0  &15.6  &15.1  &16.2  &14.9  &15.0  &12.8  &12.7  &12.6  &12.8  &12.5      \\
    \hline
    \hline
\end{tabular}}
\end{table}

\begin{table}[H]
\centering
\setlength{\tabcolsep}{3.0mm}{
\caption{Relative systematic uncertainties (in \%) in $\sigma(e^{+}e^{-}\to D_{1}(2420)^{0}\bar{D}^{0}, D_{1}(2420)^{0}\to D^{0}\pi^{+}\pi^{-} + c.c.$) measurement.}
\label{sys_err_D0_all}
\begin{tabular}{l  r  r  r r r r r r}
\hline \hline
   Sources / $\sqrt{s}$ (GeV)    &4.3079   &4.3583  &4.3874   &4.4156  &4.4671   &4.5271  &4.5745   &4.5995\\
    \hline
    Integrated luminosity        & 1.0   & 1.0  & 1.0   & 1.0  & 1.0   & 1.0   & 1.0   & 1.0 \\
    Efficiency related           &12.4   &12.4   &12.3   &12.2   &12.1   &12.1   &12.1   &12.0 \\
    Radiative correction         &6.4   &0.6  &1.8   &2.1  &1.8   &0.7   &2.5   &0.2   \\
    Signal shape                 &7.2   &7.2  &7.2   &5.5  &5.5   &8.2   &8.2   &8.2  \\
    Background shape             &3.2   &3.2  &3.2   &1.2  &1.2   &7.4   &7.4   &7.4  \\
    Fit range                    &0.9   &0.9  &0.9   &1.0  &1.0   &0.9   &0.9   &0.9   \\
    Signal region of double tag  &2.6   &2.6  &2.6   &4.2  &4.2   &4.8   &4.8   &4.8  \\
    $\sigma(\pi\pi\psi(3770))$   &11.4  &11.4 &11.4  &0.5  &0.5   &1.1   &1.1   &1.1   \\
    \hline
    Total                        &19.9  &18.8 &18.8  &14.3  &14.2  &17.1  &17.3  &17.0   \\
    \hline
    \hline
\end{tabular}}
\end{table}

\begin{table}[H]
\centering
\setlength{\tabcolsep}{3mm}{
\caption{Relative systematic uncertainties (in \%) in $\sigma(
e^{+}e^{-}\to D_{1}(2420)^{0}\bar{D}^{0}, D_{1}(2420)^{0}\to D^{*+}\pi^{-} + c.c.$) measurement.}
\label{sys_err_D0star_all}
\begin{tabular}{l  r  r  r r r r r r}
\hline \hline
   Sources / $\sqrt{s}$ (GeV)    &4.3079   &4.3583  &4.3874   &4.4156  &4.4671   &4.5271  &4.5745   &4.5995\\
    \hline
    Integrated luminosity                  & 1.0   & 1.0  & 1.0   & 1.0  & 1.0   & 1.0   & 1.0   & 1.0 \\
    Efficiency related          &14.0   &14.0   &13.9   &13.8   &13.8   &13.7   &13.7   &13.6 \\
    Radiative correction        &6.3   &0.6  &1.8   &2.1  &1.8   &0.7   &2.5   &0.2   \\
    Signal shape                 &5.3   &5.3  &5.3   &2.8  &2.8   &4.9   &4.9   &4.9  \\
    Background shape            &0.2   &0.2  &0.2   &1.8  &1.8   &7.4   &0.1   &0.1  \\
    Fit range                   &2.7   &2.7  &2.7   &3.9  &3.9   &0.8   &0.8   &0.8   \\
    Signal region of double tag &4.9   &4.9  &4.9   &2.6  &2.6   &9.8   &9.8   &9.8  \\
    \hline
    Total                       &17.2  &16.0  &16.1  &15.1  &15.1  &19.1  &17.7  &17.5  \\
    \hline
    \hline
\end{tabular}}
\end{table}

\begin{table}[H]
\centering
\setlength{\tabcolsep}{3mm}{
\caption{Relative systematic uncertainties (in $\%$) in $\sigma(
e^{+}e^{-}\to D_{1}(2420)^{+}D^{-}, D_{1}(2420)^{+}\to D^{+}\pi^{+}\pi^{-}$ + c.c.) measurement.}
\label{sys_err_Dp_all}
\begin{tabular}{l  r r r r r r r r}
\hline \hline
    Sources / $\sqrt{s}$ (GeV)   &4.3079   &4.3583  &4.3874   &4.4156  &4.4671   &4.5271  &4.5745   &4.5995 \\
    \hline
    Integrated luminosity                   & 1.0   & 1.0   & 1.0   & 1.0   & 1.0   & 1.0  & 1.0   & 1.0 \\
    Efficiency related           &13.5   &13.4  &13.3  &13.2  &13.1  &13.1  &13.1  &13.0\\
    Radiative correction         &2.4   &0.2   &1.0   &1.0   &0.6   &3.4  &1.4   &1.5\\
    Signal shape                  &7.4   &7.4   &7.4   &4.2   &4.2   &4.1  &4.1   &4.1 \\
    Background shape             &1.6   &1.6   &1.6   &0.8   &0.8   &3.8  &3.8   &3.8  \\
    Fit range                    &1.7   &1.7   &1.7   &0.9   &1.0   &1.7  &1.7   &1.7  \\
    Signal region of double tag  &1.1   &1.1   &1.1   &1.3   &1.3   &5.2  &5.2   &5.2\\
    $\sigma(\pi\pi\psi(3770))$   &11.1  &11.1  &11.1   &0.3  &0.3   &1.7  &1.7   &1.7   \\
    \hline
    Total                        &19.3  &19.1  &19.1  &14.0  &13.9  &15.8  &15.5  &15.4          \\
    \hline
    \hline
\end{tabular}}
\end{table}

\end{appendix}

\end{document}